\pdfoutput=1 % arxiv: force pdflatex
\documentclass[aps,final,notitlepage,oneside,twocolumn,nobibnotes,nofootinbib,superscriptaddress,showpacs,showkeys]{revtex4}
\usepackage{graphicx}
\usepackage{amsmath}
\usepackage{amsfonts}
\usepackage{amssymb}
\usepackage{epigraph}
\usepackage{bbm}  % blackboard math für R, C, H
\usepackage[vcentermath]{youngtab}  % Young diagrams
\usepackage{yfonts}  % Fraktur für Complexe
\newcommand{\MOo}{\protect{M{\"o}bius}}

\newcommand{\PLu}{\protect{Pl{\"u}cker}}
\newcommand{\PLL}{\protect{Pl{\"u}cker }}
\begin{document}
\title{On Some Geometrical Aspects of Space-Time Description
and Relativity}
\preprint{bfIS-2019-002}
\author{\firstname{Rolf}~\surname{Dahm}}
\email{dahm@bf-IS.de}
\affiliation{Permanent address: beratung f{\"{u}}r 
Informationssysteme und Systemintegration,
G{\"{a}}rtnergasse 1, D-55116 Mainz, Germany}
\homepage{http://www.researchgate.net/profile/Rolf_Dahm}

\begin{abstract}
In order to ask for future concepts of relativity, one has
to build upon the original concepts instead of the nowadays 
common formalism only, and as such recall and reconsider 
some of its roots in geometry. So in order to discuss 3-space
and dynamics, we recall briefly Minkowski's approach in 1910 
implementing the nowadays commonly used 4-vector calculus
and related tensorial representations as well as Klein's 
1910 paper on the geometry of the Lorentz group. To include
microscopic representations, we discuss few aspects of Wigner's
and Weinberg's 'boost' approach to describe 'any spin' with
respect to its reductive Lie algebra and coset theory, and 
we relate the physical identification to objects in $P^{5}$
based on the case $(1,0)\oplus(0,1)$ of the electromagnetic 
field. So instead of following this -- in some aspects -- 
special and misleading 'old' representation theory, based 
on 4-vector calculus and tensors, we provide and use an 
alternative representation based on line geometry which
-- besides comprising known representation theory -- is 
capable of both describing (classical) projective geometry 
of 3-space as well as it yields spin matrices and the 
classical Lie transfer. In addition, this geometry is 
capable of providing a more general route to known Lie
symmetries, especially of the su(2)$\oplus$i~su(2) Lie
algebra of special relativity, as well as it comprises 
gauge theories and affine geometry. Thus it serves as 
foundation for a future understanding of more general 
representation theory of relativity based, however, on
roots known from classical projective geometry and $P^{5}$.
As an application, we discuss Lorentz transformations in
point space in terms of line and Complex geometry, where
we can identify them as a subset of automorphisms of the
\PLu-Klein quadric $M_{4}^{2}$ of $P^{5}$. In addition,
this description provides an identification as a special,
but singular parametrization of the tetrahedral Complex, 
too. As such, we propose to generalize and supersede the
usual rep theory of relativity by an embedding into the 
general geometry of $P^{5}$, and the use of appropriate
concepts of projective and algebraic geometry in \PLu's
sense by switching geometrical base elements and using 
transfer principles.
\end{abstract}

\pacs{
02.20.-a, %Group Theory
02.40.-k, %Geometry, differential geometry and topology
03.70.+k, %Theory of quantized fields
04.20.-q, %Classical general relativity
04.50.-h, %Higher-dim gravity and...
04.62.+v, %Quantum fields in curved spacetime
11.10.-z, %Field theory
11.15.-q, %Gauge field theories
11.30.-j, %Symmetry and conservation laws
12.10.-g  %Unified field theories and models
}

\keywords{relativity, unification, quantum field theory, Dirac theory, 
Lie algebras, Lie groups, geometry, projective geometry, line geometry, 
line Complex, Complex geometry, congruences, null systems}

\maketitle

\onecolumngrid
\epigraph{\textsc{Sommersprossen sind doch keine Gesichtspunkte!}}
{\textit{Ekkhard Verchau, Historisches Seminar,\\JGU Mainz, 1990}}
\vspace{4mm}
\twocolumngrid

\pagebreak

\section{Introduction}
Having been asked to contribute to this topical collection
on the future of relativity on occasion of Valeriy's $60^{th}$
birthday anniversary, it is a pleasure to appreciate the 
jubilarian and his long-term work and interest, and it is
-- not less -- an honour to contribute some aspects which 
might be interesting with respect to geometrical (and
possibly alternative) views on relativity which are lost
nowadays in the community's prevailing memory. While on 
the one hand, content being capable of addressing an 
alternative viewpoint grew and grew throughout working 
with classical projective geometry (below briefly 'PG')
during the last years, and even throughout writing this
text, we've experienced on the other hand the standard 
arguments with respect to relativity over and over again
during conferences, in papers, and especially in public 
networks to protect and shield often a naive textbook 
reasoning, sometimes even to avoid scientific\footnote{And
in this context, we do NOT address or reference the
exchange on some current 'social networks' where the 
signal-to-noise ratio with respect to discussions on
relativity, at least according to our perception, is 
close to zero.} discussion by claiming that 'everything
is well established and settled'. Now, it is NOT that 
we want to contradict on the subsequent pages the physical
aspects of relativity, however, we feel it necessary AND
worth to discuss several aspects of the various formalisms
in use and their associated interpretations. We think 
that those aspects are different facets of a more general
geometrical description which is known for more than 
two centuries now, and which -- formally starting with
Hamilton's formulations and his work on optics -- reappeared
from time to time in few aspects but is nowadays covered
by a patchwork of formal approaches and additional --
sometimes implicit -- assumptions. As such, this background
in geometry most often is not really visible on its own.
So we want to provide at next an extended outline, melting
some background with the organizational outline of the 
following pages towards a possible alternative view on
relativity.

\subsection{Outline}
As there is no 'future of \ldots' without founding on its 
past, we want to recall briefly in sec.~\ref{sec:known} 
two of the central 'classical papers' on the formalism 
of relativity --  Minkowski 1908/1910 and Klein 1910 --
as well as Weinberg's papers 1964/65 on 'quantum' reps
in terms of 'any spin' in order to explain where we want
to work the switches later on towards geometry. Minkowski's
and Klein's papers focus mainly on invariant theory of 
the 'Lorentz invariant' quadric $x^{2}+y^{2}+z^{2}-t^{2}$
in point space and related aspects, and they yield an
interpretation of coordinates and their transformations 
in point space, and appropriate rep theory. If we neglect 
for a moment the coordinate-oriented approaches closer 
to experimental observations like pursued by Maxwell, 
Lorentz, Poincar{\'{e}} and Einstein, then the standard 
reference with respect to 'classical relativity' nowadays
is Minkowski's paper\footnote{Originally published in 
1908 in the 'G{\"{o}}ttinger Nachrichten', p.~1, Klein 
being the editor of 'Mathematische Annalen' decided 
to republish it there again in 1910 after Minkowski's 
sudden death. Due to better availability of 
'Mathematische Annalen', we've decided to reference 
the latter publication, so the page references given 
below relate to \cite{mink:1910}.} \cite{mink:1910} 
(where he introduced the 4-vector calculus and some 
related tensorial representations\footnote{We suppress
the discussion of differential geometry and Riemannian
spaces here, as we have discussed already various of
those aspects thoroughly \cite{dahm:2008}, \cite{dahm:MRST1}
by the coset approach summarized by Gilmore \cite{gilmore:1974}
and detailed by Helgason's books \cite{helgason:1978},
\cite{helgason:GGA}. So we consider this discussion 
as a subsidiary concept, and focus instead on the 
synthesis of geometry and physics in the spirit of 
Einstein's efforts towards an unified concept comprising 
both. As before, we use the shorthand notation 'rep' 
to denote both representations as well as realizations
as long as the context is obvious.}) which marks the 
center of interest. Not less important to pave the 
way for Minkowski's 4-dim rep theory were Klein's 
papers 1910 \cite{klein:1910}, and 1872 \cite{klein:1872a}
-- the older one creating obviously the 4-dim metric 
foundations on 4-dim space concepts, however, founding
on a different but in the beginning of the $20^{th}$ 
century apparently forgotten or misinterpreted 
geometrical background.

Weinberg's papers are necessary in this context because
they attach Lie symmetries to a skew 6-dim 'tensor'
$\omega_{\mu\nu}$ and provide 'quantum' rep theory which 
we are going to discuss in sec.~\ref{sec:weinbergNEW} 
using a different interpretation of $\omega_{\mu\nu}$.

As such, after having sketched the standard trail very 
briefly\footnote{There is an enormous number of textbooks
floating around and treating the formalism, typically 
based on Weyl's affine thoughts \cite{weylRZM:1918}. 
However, we are not aware of even one focusing on the
{\it different characters} and {\it possible interpretations}
of space-time coordinates beyond typical Euclidean/metric
or affine -- and thus intrinsically linear -- concepts in 
the context of 'relativity'.} by discussing few geometrical 
aspects and identifications, we use sec.~\ref{ch:std-rev}
to extract the main ideas which we are going to use in 
order to rearrange and generalize the discussion. Thus,
we comment in sec.~\ref{ch:std-rev} also on Klein's review
1910 \cite{klein:1910} -- marked by both highlights and 
lowlights -- which seems to have emphasized the historical
mathematical discussion in favour of Minkowski's rep theory,
and as such also as a foundation of Weyl's concepts \cite{weylRZM:1918}.
There, we turn briefly back to Minkowski's paper \cite{mink:1910}
and recall our alternative identification \cite{dahm:MRST8}
in order to approach physics and geometry much closer. In 
\cite{dahm:MRST8}, it turned out that Minkowski's calculus
can be seen as a special case of line and Complex geometry,
i.e.~based on 5-dim objects $a_{\alpha} p_{\alpha}=0$ -- 
so-called 'line Complexe', or 'Complexe' for short, which
'live' in $P^{5}$ -- where $1\leq\alpha\leq 6$, 
$a_{\alpha}\in\mathbbm{R}$, and $p_{\alpha}$ denote six 
line coordinates of $P^{3}$. Very recently, we've indeed
found additional work published much later by von Laue 
\cite{vonLaue:1950} which emphasizes this concept, too,
and which discusses some consequences with respect to 
form and structure of the energy tensor. Moreover, our 
approach underpins Einstein's ongoing, lifelong quest for
a common geometrical description after having committed 
to Minkowski's 4-dim reps not earlier than 1912.

Projective line and especially Complex geometry allow 
to rearrange and supersede various 'well-known' formalisms
and symbolisms used throughout physics, and to put them 
at their places -- as the analytic counterpart of the 
geometry of 3-space. We address few aspects throughout 
sec.~\ref{sec:reinterpretation}. However, the description
we are going to use differs in the respective choices of 
representations by using six line coordinates $p_{\mu\nu}$
by means of ray coordinates, or using directly the six 
coordinates $p_{\alpha}$ of $P^{5}$ given above instead
of the usual sets of four point and four plane coordinates.
The six line coordinates, in order to work in $P^{3}$, 
have to obey the \PLL condition 
$p_{01}p_{23}+p_{02}p_{31}+p_{03}p_{12}=0$ (or 
equivalently $p_{1}p_{4}+p_{2}p_{5}+p_{3}p_{6}=0$)
as an additional quadratic constraint. Line coordinates
grant immediate use of their comprised planar and 
spatial geometry in $P^{3}$ or to its subsets, or of 
line Complexe and Complex geometry in $P^{5}$, and as 
such they induce quadratic constraints in $P^{3}$
which originate from the \PLu-Klein quadric $M_{4}^{2}$
in $P^{5}$ and additionally reflect in line-line 
duality in $P^{3}$. This, however, is helpful to 
understand physical aspects as we'll discuss in 
sec.~\ref{sec:quadratic}. Needless to mention, 
that due to the different objects and dimensions,
the respective transformation theories as well as 
the reps of their respective invariant theories 
differ, too! However, this aspect is nothing but 
one of the essential clues of Klein's 'Erlanger 
Programm', and shouldn't worry people.

The 'quantum' aspects, especially Bargmann's and Wigner's 
concepts and the various spin reps, can be related 
\cite{dahm:MRST7} to the Lie transfer of line geometry 
\cite{lie:1872} (or even using more sophisticated 
transfer principles like e.g.~Laguerre's geometry, 
or Fiedler's cyclography, and their related rep 
theories). Thus, it is straightforward to relate
\PLu's Euclidean line rep to spin reps \cite{dahm:MRST7}
and (projective) point reps to include absolute elements
(and spinorial reps). While emphasizing \PLu's fifth
line coordinate $\eta$ -- a determinant and formally
quadratic in the four original line coordinates $r, s,
\rho,$ and $\sigma$ -- we have introduced Pauli matrices 
and quaternions \cite{dahm:MRST7}, and constructed a 
parallel, matrix-based formalism to represent the 
original, projective geometry of 3-space in terms of 
commutators, and thus by means of a formal su(2) Lie 
algebra. However, the respective physical interpretations 
require sophisticated geometrical interpretations and 
appropriate care to proceed because we discuss objects 
and algebra of the transferred space \cite{dahm:MRST7}.
Using this identification as building block, common 
'quantum concepts' like the projective line rep in 
relation to a complex number (by means of an angle, 
or 'phase', i.e.~by pencil coordinates like visualized
in appendix~\ref{app:coordinatesystems}, figure~\ref{figA2}),
the construction of Clifford algebras by Pauli 
matrices, Heisenberg models, point identifications
on the line and Hesse's transfer principle vs.~binary
forms, etc. may be recast in terms of (or at least 
related to) classical line geometry, and as such 
to projective geometry (or subsequently for short
just 'PG') of 3-space.

Here, due to the celebration of Valeriy's anniversary 
and his great work on rep theories and equations of 
motion, we do not want to carry coals to Newcastle, 
so we skip most rep details here, but relate later 
(see sec.~\ref{sec:weinbergNEW}) few of our aspects 
and identifications to Valeriy's overviews \cite{dvoe:1993a},
\cite{dvoe:1993b}, \cite{dvoe:1993c}, \cite{dvoe:1994},
\cite{dvoe:2018} on different reps and formalisms, and
especially the 2(2S+1)-Formalism, or $(S,0)\oplus(0,S)$-reps,
with $S=1$.\\

Instead, in order to approach future aspects of relativity,
we want to pave the way for an alternative geometrical
rep. So by presenting some background and some examples,
we propose in sec.~\ref{sec:reinterpretation} to focus 
on $P^{5}$ as well as on projective and algebraic geometry.
After having recalled null systems in sec.~\ref{sec:nullsystems},
we have to discuss the Lie algebra approach as a subsidiary
concept of line geometry, and of projective geometry when 
generating surfaces by lines, and when generating higher 
order objects in general by means of projective mechanisms\footnote{There
are even much deeper background and possibilities to relate
projective geometry to the human observation ability, i.e.~to
neural, medical and psychological aspects and the adoption
of the brain and its recognition patterns to real physical
observations \cite{schmeikal:2018}, but further details are
off-topic here. Nevertheless, I'm very grateful to B.~Schmeikal
for pointing me to A.~Trehub's \protect{\cite{trehub:2019}}
work on the 'Retinoid System' and for helpful private 
discussions.}. So in sec.~\ref{sec:remarkstransformations},
we rewrite Lorentz transformations of point space by line
(or Complex) transformations, and show how they fit into
properties of tetrahedral Complexe and the \PLu-Klein 
quadric $M_{4}^{2}$ in $P^{5}$.

Especially Complexe and their geometry serve as unifying 
concepts as they are related linearly to null systems\footnote{
And as such they relate mathematically to correlations and
involutions!}, i.e.~to the very description of 6-dim forces
('Dynamen'), or their possible decomposition into 3-dim 'forces'
and 'moments' like e.g.~in the theory of the top \cite{kleinso:1897}
\cite{kleinso:1898}, as well as to the 'gauge description' 
$F_{\mu\nu}$ of the 'photon field' in terms of a special linear 
Complex\footnote{The parameter spaces of the transformations 
exhibit even more and higher-dimensional symmetries, as described
e.g.~in \cite{study:1903} by 'Somen', 'Protosomen', 'Pseudosomen',
etc.} (see \cite{dahm:MRST3} II.A, \cite{dahm:MRST5}, or 
\cite{dahm:MRST8}). More general, regular linear Complexe 
describe null systems and forces (see e.g.~sec.~\ref{sec:nullsystems}),
so our reasoning with respect to two Complexe (or a 'Congruence'
\cite{plueckerNG:1868}) as outlined and discussed in \cite{dahm:MRST3}
comprises von Laue's work, but one has to look much deeper 
into this kind of analytic reps of $P^{5}$, and -- even 
more important -- reconnect this rep theory to physics in
$P^{3}$ again in order to gain a deeper understanding 
after all the formalism as well as the, most often, empty
formal calculus of the past century since Einstein's work.
And as three Complexe describe ruled surfaces (reguli, 
or 'Configurations' in \PLu's original notion \cite{plueckerNG:1868}), 
we are thrown back onto the necessity to discuss surfaces
and even more intricate objects using lines as building 
blocks in $P^{3}$, or Complexe in $P^{5}$. A typical 
example based on three Complexe yields the two generator
sets of real lines of the one-sheeted hyperboloid, and
a special linear Complex can be used in this scenario 
to describe its symmetry axis by identification of the
axis/directrix\footnote{German: Treffgerade/Leitgerade}.
So there is immediately a deep connection of linear 
Complexe to surfaces of $2^{nd}$ order and surfaces 
up to $2^{nd}$ class, as well as obvious relations of 
Complexe to surfaces and their projective construction,
and to polar and focal theory.

However, in this context it is also immediately obvious, that 
complex numbers (as well as hypercomplex number systems in 
general) are nothing but an algebraical symbolism (and as 
such a mathematical, {\it not a physical} tool!) to represent
various geometrical cases analytically and algebraically in
a unified manner. So although we appreciate complex analysis
and function theory, differential geometry and Riemannian 
spaces -- as we want to focus here on some physical roots 
in order to propose an additional geometrical trail, we have 
to extend the usual discussion by additional facets. And
although our summary so far is even far from presenting {\it new}
knowledge, due to an exuberant technical focus in these days
on mathematical concepts like complex numbers, point spaces,
or differential geometry, people often use different understanding
and emphases, especially when applying Grassmann numbers and 
associated point space concepts\footnote{Note, that this 
differs considerably from \PLu's, Lie's (\cite{lie:1872},
p.~151), and Study's \cite{study:1903} reasoning in geometry!}.

On the other hand, we know of the central r\^{o}le of the
\PLu-Klein quadric $M_{4}^{2}$, i.e.~a quadratic constraint
in $P^{5}$ (the \PLL condition) which guarantees that points
on the quadric in $P^{5}$ map to lines in $P^{3}$. So whereas
this yields a mechanism {\it to derive} line geometry of $P^{3}$
as a subset of geometry of $P^{5}$ and related (quadratic)
constraints, this background emphasizes the existence and 
occurence of quadratic (and as such right from the beginning
nonlinear!) structures like known from involutions, or polar 
and focal theory related to $2^{nd}$ order surfaces which
reflects in quadratic algebras like known from Pauli or 
Clifford rep theory. Considering $P^{5}$ and the quadric
$M_{4}^{2}$, this automatically rises more questions beyond
just linear point reps of $P^{5}$ in that we have to consider
at least quadratic Complexe and their rep theory in $P^{3}$, 
too\footnote{The general (or pure) approach via $P^{5}$ 
will be addressed in sec.~\ref{sec:outlook}.}. As such, 
the tetrahedral Complex has played a prominent and important 
r\^{o}le throughout physics and geometry\footnote{See e.g.~Lie's 
discussion in \cite{lie:1896}, or related discussions in 
\cite{dahm:MRST4}.}, and for our discussion here -- besides 
being related to the quadratic structure of $P^{3}$ rep 
theory and yielding the Lorentz-invariant quadric 
$x^{\mu}x_{\mu}=0$ in point space \cite{dahm:MRST3}
-- the tetrahedral Complex yields an invariant theory 
in that it classifies all lines in $P^{3}$ by the 
anharmonic ratio of their four intersection points with
the planes of the fundamental coordinate tetrahedron.
Grouping the lines according to this ratio yields one
free parameter, the double ratio, and because we know 
that projective transformations preserve this ratio, we 
may immediately stress Klein's 'Erlanger Programm', and 
attach and apply Lie's theory of continuous groups and 
their algebras to these equivalence classes. The invariant 
theory, however, a priori is a geometrical one.

So on one hand, the tetrahedral Complex is deeply connected
to the very definition of point and plane coordinates 
of $P^{3}$ (and to polar theory, if we circumscribe the
fundamental coordinate tetrahedron by a sphere), on the
other hand, it induces structures on the lines in $P^{3}$,
or their different reps -- either in terms of complex 
numbers when we interpret the line as a special case of
a circle or when discussing point sets on such lines like
in Hesse's transfer principle (\cite{hesse:1866}, or
\cite{kleinHG:1926}, \S51) or using Clebsch's binary forms
\cite{clebsch:1872}.

Below we'll give some more arguments and discuss few more 
aspects. In general, we want to emphasize the idea that 
what we denote by 'relativity' in $P^{3}$ is part of 
Complex geometry in $P^{5}$, especially when represented
in terms of lines in $P^{3}$ and treated by classical 
projective line geometry. As an example, we discuss briefly
von Laue's identification, and relate those aspects to the 
tetrahedral Complex and 'field reps' in terms of point 
coordinates in that we construct field lines by an involution
on a line. This is well-known from classical projective 
geometry but here we can relate such field configurations
additionally to the tetrahedral Complex and appropriate 
classes of lines.

So thinking of 'relativity' in terms of transformations on 
point coordinates is much too short\footnote{\ldots~because
these symmetry properties are intrinsic properties of lines
and Complexe, see below!} and restricted to catch the background,
especially when attributing affine geometry, and its transport
and connection concepts, only. Accordingly 'miraculous effects' 
like Thomas precession, 'space-time mixture -- but only in 
velocity direction' --, etc.~appear which can be simply 
resolved by linear Complex geometry, and which thus can be 
understood as artefacts of $P^{5}$-geometry \cite{dahm:MRST8} 
\cite{dahm:MRST3} in $P^{3}$.

With respect to this background, it is important to recall
the quadratic character of line geometry. As necessary 
background, we discuss in sec.~\ref{sec:quadratic} senary
quadratic forms in line (or Complex) space $P^{5}$ (which
are associated quaternary forms in point space \cite{dahm:MRST3})
and two of their applications: the six Klein coordinates 
$\vec{E}\pm i\vec{B}$, and Klein's $3\oplus 3$ right- and 
left-handed linear fundamental Complexe.

In the last section \ref{sec:discrete}, we recall briefly
some related aspects of Hudson's book \cite{hudson:1905}
which besides discrete transformations based on point-plane
incidence, the Heisenberg group and K3-surfaces, connects
a wealth of additional aspects of projective and algebraic
geometry.

In order to prepare a future discussion of the associated 
physics and of the 'future of relativity', we close this 
document in sec.~\ref{sec:outlook} with an outlook by 
emphasizing the urgent necessity to 'reunify' geometry and
geometrical objects again with the respective rep theories
by means of Complexe, i.e.~we argue to focus again on Klein's
'Erlanger Programm' and the lessons learned from advanced 
projective geometry and invariant theory\footnote{In this 
context, Klein's footnote $***$ in \cite{klein:screws} on 
page~419 naturally comments on an ongoing simplification 
by pure formalisms and by shrinking concepts to subsidiary
problems for the sake of analytical presentability, only.}
by choosing lines as base elements of geometry \cite{plueckerNG:1868}
instead of points only.

\section{Known Aspects of Rep Theory}
\label{sec:known}
To discuss possible modifications, changes and reinterpretations
later on, we summarize throughout this section briefly some 
of the major aspects of the standard approaches to relativity
and some relevant aspects of rep theory. As such, this reflects
-- of course -- our personal reasoning and concentrates on aspects
which we want to emphasize with respect to the upcoming discussion
of geometrical identifications and reps. Especially, we do not 
want to discuss formalisms by themselves like Riemannian geometry
or Hilbert's axiomatization of Einstein's general relativity 
\cite{hilbert:1915}, \cite{hilbert:1916}, \cite{klein:1917},
\cite{hilbert:1924} here in depth. With respect to physical 
modeling and reasoning, we think it is at first necessary to
gain an impression on the physical objects and to find an 
adequate mathematical rep in order to model our observations
sufficiently. Only afterwards, we can go and see what we 
can get from the respective formalisms. A nice example is
Weinberg's statement (\cite{weinberg133:1964}, p.~B1319) 
where he refuses 'ab initio' to present field equations 
or Langrangians -- even nowadays a standard question on 
conferences and during discussions -- simply because they
are not needed. He emphasizes the evident fact that as 
soon as one has found reps which fulfil covariance and 
irreducibility, everything is known, and that there is 
no more need to suppress superfluous components of the 
reps. In other words, just consider symmetry and find 
suitable reps of objects and transformation groups.

So in this section, we address aspects of the coordinate
definition which can be seen as the fundamental rep for 
applications of the invariant and group theory of the 
Lorentz group, and as such, one has to address aspects 
of second order surfaces and especially the invariance 
of a quaternary quadric, too. As emphasized above, we 
discuss some aspects of the historical approach by means
of coordinate and physical identifications, on the one 
hand in order to keep track of associated physics, on 
the other hand to avoid pure mathematical formalism and
prevent the physical aspects from being buried by abstract
generalizations or axiomatizations like Berlin- or 
Bourbaki-type formalizations, or Hilbert's or Weyl's 
axiomatic approaches, which sometimes loose connection
to physics, or cover physical aspects by mathematical
formalism (and sometimes formal artefacts). It is 
sufficient to know that in case we have the need to 
calculate analytically, we can always find appropriate 
mechanisms from group theory, algebra or differential 
geometry to write things down. As such, we remember
Minkowski's paper(s) on 4-vector formalism, and Klein's
background, and with respect to discussions in 'quantum'
theories, we want to mention Weinberg's formalism 
because we can use it later to attach geometry.

\subsection{Standard Approach -- Minkowski 1908}
\label{sec:standard-minkowski}
%Usual Rep Theory
As with respect to Minkowski's paper \cite{mink:1910}, in his
introductory remarks he claims to derive the basic equations 
of motion from the 'principle of relativity'\footnote{German:
Prinzip der Relativit{\"{a}}t.} in a manner, determined uniquely
by this principle. To proceed, in \cite{mink:1910}, \S1, p.~475,
he {\it defines} a coordinate system $x$, $y$, and $z$ without
explicitly claiming the {\it type} of the chosen coordinates, 
and in addition, he defines 'time' in terms of a fourth and
a priori {\it independent} rectangular coordinate $t$. As 
we'll see later, although this sounds familiar today, in terms
of Euclidean (or 'metric') coordinates, it is ambiguous. 
Implicit later use of the coordinates in his paper indicates
that he uses an Euclidean coordinate interpretation\footnote{See
beginning of \S3, p.~477, and beginning of \S4, p.~480, where 
he argues with rotations of the three rectangular space axes.}.

After having defined the projections of the point 'vector' 
$\vec{r}$ onto a general vector $\vec{v}$, $|\vec{v}|=q<1$,
his eqns.~(10) and (12) in \S4,
\begin{equation}
\vec{r}_{||}\,=\,\frac{\vec{r}_{||}-qt}{\sqrt{1-q^{2}}}\,,\quad
t'\,=\,\frac{-q\vec{r}_{||}+t}{\sqrt{1-q^{2}}}\,,
\end{equation}
yield what he denotes as 'special Lorentz transformation' 
while the orthogonal components of $\vec{r}$ with respect 
to the velocity $\vec{v}$ remain invariant. $q$ at that 
time has been defined in a complicated manner related to
a transformation parameter $\psi$ according to $q=-i\,\tan i\psi$
(\cite{mink:1910}, eq.~(2)).

However, Minkowski's reasoning so far is based on some kind
of 'cut and paste' transfer of electromagnetism which he derived
after identifying (\S2, p.~476, bottom) the six skew symmetric 
components $f_{\alpha\beta}$, $1\leq\alpha,\beta\leq 4$, with 
electromagnetic field components by
\begin{equation}
\label{eq:ComplexComponents}
\begin{array}{rl}
& f_{23},f_{31},f_{12},f_{14},f_{24},f_{34}\\
\cong & m_{x},m_{y},m_{z},-ie_{x},-ie_{y},-ie_{z}
\end{array}
\end{equation}
without explaining the background of $f$. Then, by a 
lot of arguments, he derives the transformation equations
\begin{equation}
e'_{x'}\,=\,\frac{e_{x}-qm_{y}}{\sqrt{1-q^{2}}},\,
m'_{y'}\,=\,\frac{-qe_{x}+m_{y}}{\sqrt{1-q^{2}}},\,
e'_{z'}\,=\,e_{z}\,,
\end{equation}
\begin{equation}
m'_{x'}\,=\,\frac{m_{x}+qe_{y}}{\sqrt{1-q^{2}}},\,
e'_{y'}\,=\,\frac{qm_{x}+e_{y}}{\sqrt{1-q^{2}}},\,
m'_{z'}\,=\,m_{z}\,,
\end{equation}
of the fields (see \cite{mink:1910}, eqns.~(6) and (7)). 
Although he remarks that the structure of these two 
equation sets may be superseded by vectorial 
representations\footnote{His symbol $[\cdot \cdot]$
denotes the 3-dim vector product, where $\vec{v}=q\hat{z}$.}
according to $\vec{e}+[\vec{v}\vec{m}]$, and 
$\vec{m}-[\vec{v}\vec{e}]$, even here he doesn't discuss
the higher and distinct background structure of $f$.

Instead, Minkowski discusses the quadratic invariant 
$x^{2}+y^{2}+z^{2}-t^{2}\longleftrightarrow x'^{2}+y'^{2}+z'^{2}-t'^{2}$
in point coordinates $(x,y,z,t)$, some issues of rep theory,
and after having introduced total differentials $\mathrm{d}x$,
$\mathrm{d}y$, $\mathrm{d}z$, and $\mathrm{d}t$ of the point 
coordinates, he identifies velocities and some physics.

It is only in \S5, p.~484, eq.~(23), 
\begin{equation}
\label{eq:MinkComplex}
\begin{array}{rl}
& f_{23}(x_{2}u_{3}-x_{3}u_{2})
+ f_{31}(x_{3}u_{1}-x_{1}u_{3})\\
+ & f_{12}(x_{1}u_{2}-x_{2}u_{1})\\
+ & f_{14}(x_{1}u_{4}-x_{4}u_{1})
+ f_{24}(x_{2}u_{4}-x_{4}u_{2})\\
+ & f_{34}(x_{3}u_{4}-x_{4}u_{3})
\end{array}
\end{equation}
that Minkowski begins to construct 'space-time vectors of 
2$^{nd}$ kind'\footnote{German: Raum-Zeit-Vektoren II.~Art} 
$f$ with six components out of two 'space-time vectors of 
1$^{st}$ kind'\footnote{German: Raum-Zeit-Vektoren I.~Art} 
$x$ and $u$ which he requires to transform invariantly, 
i.e.~eq.~(\ref{eq:MinkComplex}) has to transform into
\begin{equation}
\label{eq:MinkComplexNew}
\begin{array}{rl}
& f'_{23}(x_{2}'u_{3}'-x_{3}'u_{2}')
+ f'_{31}(x_{3}'u_{1}'-x_{1}'u_{3}')\\
+ & f'_{12}(x_{1}'u_{2}'-x_{2}'u_{1}')\\
+ & f'_{14}(x_{1}'u_{4}'-x_{4}'u_{1}')
+ f'_{24}(x_{2}'u_{4}'-x_{4}'u_{2}')\\
+ & f'_{34}(x_{3}'u_{4}'-x_{4}'u_{3}')\,.
\end{array}
\end{equation}
In addition to this observation, it is important for our 
later use that in his eqns.~(25) and (26), he claims that
the {\it two} quantities
\begin{equation}
\label{eq:MinkInvariante1}
\vec{m}^{\,2}-\vec{e}^{\,2}\,=\,
f_{23}^{2} + f_{31}^{2} + f_{12}^{2} + f_{14}^{2} + f_{24}^{2} + f_{34}^{2}\,,
\end{equation}
and
\begin{equation}
\label{eq:MinkInvariante2}
\vec{m}\vec{e}\,=\,i\,\left(
f_{23}f_{14} + f_{31}f_{24} + f_{12}f_{34}\right)\,,
\end{equation}
derived from components which constitute the 'space-time 
vectors of 2$^{nd}$ kind' $(\vec{m},-i\vec{e})$, {\it are 
invariant under Lorentz transformation} (\cite{mink:1910},
p.~485).

Here, we do not follow his further and sometimes weird 
formal constructions and discussions as he obviously 
didn't notice\footnote{We leave a thorough discussion of 
these strange historical circumstances to more qualified 
people like science historians. Indeed, it is {\it very 
strange} to see that Minkowski either didn't notice or 
even desperately avoided the geometrical background of 
Complexe (or '6-vectors') -- after having studied in 
K\"{o}nigsberg under Weber and especially Voigt, having
studied in Berlin under Kummer, having been advised by
von Lindemann during his PhD thesis, having worked in 
Bonn -- \PLu's long-term domain -- for seven years, 
having been professor of {\it geometry} in Zurich during 
Einstein's studies there, and -- last not least -- 
having been in permanent local contact with Klein in 
G{\"{o}}ttingen since his professorship started there 
in 1902 -- Klein, who himself acted as editor and 
contributing author to publish \PLu's heritage on line
geometry and Complexe in conjunction with Clebsch 
\cite{plueckerNG:1868}, and who published additional
basic work on line geometry, Complexe and transformation
groups with Lie in the early 1870s (see e.g.~\cite{klein:1871}
or \cite{klein:1872a}). However, it's a fact that 
Minkowski didn't use the existing, very elegant line
and Complex geometry established back since the 1860's
and 1870's, which moreover had been emphasized much 
stronger by Study's exhaustive contemporary work 
\cite{study:1903} on Dynamen! This lack is underpinned
by his statement \cite{mink:1910}, p.~499, \#6, where 
he constructs line coordinates out of his two vectors
$w$ and $s$ -- whether point or plane coordinates -- 
without mentioning the line geometrical background
or null systems. However, he noticed that he had 
to introduce a dual $[w,s]^{*}$ being relevant for 
subsequent physical discussions. We have given a more
detailed (but still introductory) treatment and partial
transfer to line and Complex geometry in \protect{\cite{dahm:MRST8}},
however, based on formal and algebraic arguments only.}
(or didn't want to notice\footnote{see \cite{dahm:MRST8},
footnote 10.}) that eq.~(\ref{eq:MinkComplex}) relates 
to the very definition of a line Complex\footnote{The 
3-vectors $\vec{E}$ and $\vec{B}$ may be formally 
arranged as linear Complex, or special 'Dyname' when
interpreted as 'force'.} according to \PLL (see 
\cite{plueckerNG:1868}, or eq.~(\ref{eq:complexinhomogen})
here, or e.g.~\cite{lueroth:1867}, or \cite{clebsch:1870}).
Moreover, according to his introduction of complex 
phases in eq.~(\ref{eq:ComplexComponents}) above, 
he didn't use real Complex components, but switched
implicitly to Klein coordinates instead of \PLL 
coordinates\footnote{\label{fn:so6}We postpone the 
phase discussion and the reality considerations of 
the fields -- due to Minkowski's artificial introduction
of an imaginary 4-component of the point reps $x$ 
and $u$ -- to a later stage, see sec.~\ref{sec:quadratic}.
As a consequence, in \protect{eq.~(\ref{eq:MinkComplex})} 
the coefficients $f_{14}$, $f_{24}$ and $f_{34}$ 
have to be imaginary, too, to represent meaningful
geometry which would convert the signature of the
squares in \protect{eq.~(\ref{eq:MinkInvariante1})}
to SO(3,3) instead of their formal SO(6) invariance.
Alternatively, instead of real coefficients with 
Klein coordinates, one could choose imaginary 
coefficients with \PLL coordinates, or -- as the 
best and most transparent approach -- start from 
scratch with line and Complex geometry in terms 
of tetrahedral homogeneous coordinates which 
fixes the phases and the coordinate sets {\it uniquely}.
Moreover, as within this context it is necessary 
to introduce conjugation of Complexe as well, like 
(at least formally) performed e.g.~in von Laue 
'reprise' \cite{vonLaue:1950}, one thus approaches
the background concepts of $P^{5}$, too. However, 
both authors -- Minkowski and Klein -- do not 
create the necessary relationships to an underlying
Complex geometry, and thus miss this original 
background of $P^{5}$ and the r\^{o}le of the 
\PLu-Klein quadric.}. Minkowski's invariance 
requirement with respect to eq.~(\ref{eq:MinkComplex})
and the transformation into eq.~(\ref{eq:MinkComplexNew})
constrains the objects\footnote{Being a priori a 
constraint on the form, in other words with respect
to irreducibility of the chosen reps in eqns.~(\ref{eq:MinkComplex})
and (\ref{eq:MinkComplexNew}), one can, of course, 
identify the expressions in parenthesis there 
as well as their accompanying coefficients $f$ 
geometrically. If we identify later the coordinate
expressions in parenthesis as reps of line 
coordinates in terms of point reps, i.e.~as
'ray' reps, their invariance forces the linear
Complex to remain invariant, too. We have discussed 
special cases already in \cite{dahm:MRST5}, the 
general discussion is given in \ref{sec:remarkstransformations}.}.
He identifies this Complex by its six components
('coordinates') $(\vec{m},-i\vec{e})$, and if we
rewrite eqns.~(\ref{eq:MinkInvariante1}) and 
(\ref{eq:MinkInvariante2}) according to
\begin{equation}
\label{eq:ComplexInvariante1}
\vec{m}^{\,2}+(-i\vec{e})^{\,2}\,=\,
f_{23}^{2} + f_{31}^{2} + f_{12}^{2} + f_{14}^{2} + f_{24}^{2} + f_{34}^{2}\,,
\end{equation}
and
\begin{equation}
\label{eq:ComplexInvariante2}
\vec{m}(-i\vec{e})\,=\,f_{23}f_{14} + f_{31}f_{24} + f_{12}f_{34}\,,
\end{equation}
we are led automatically into Complex geometry of $P^{5}$
and the related invariants\footnote{This relation is discussed
in sec.~\ref{sec:quadratic} in more detail.} of a linear 
Complex $(\vec{m},\vec{e})$. As such, the rhs of eq.~(\ref{eq:ComplexInvariante2})
is the so-called 'parameter'\footnote{With respect to textbooks, 
both notations with $x_{4}$ and $x_{0}$ are in use, however,
the rhs of \protect{eq.~(\ref{eq:ComplexInvariante2})} differs 
by an overall minus sign when one switches to $x_{0}$
instead of $x_{4}$ according to the usual ordering of the 
coordinates and the antisymmetry of the line coordinates.
This has consequences with respect to respective physical 
interpretation(s).} of the linear Complex which in the 
case of a special linear Complex $\vec{m}(-i\vec{e})=0$
yields the \PLL condition related to the line coordinates
$f_{\alpha\beta}$ of an axis/a directrix\footnote{German:
Treffgerade}. In this case, the special linear Complex
$f \sim (\vec{m},\vec{e})$ is described by a line/axis
hit by the lines of the Complex. Physically, we have used
this picture to describe a linearly/uniformly moving point
emitting light rays (see \cite{dahm:MRST3}, sec.~2.2, 
and ibd.,~appendix~A, and \cite{dahm:MRST4}, sec.~3).

The rhs of eq.~(\ref{eq:ComplexInvariante1}) has its deeper
background directly in $P^{5}$ and possible coordinatizations
which we discuss later in vicinity of Klein's remarks in
\cite{kleinHG:1926}, especially \S\S 22 and 23. Here, 
by referring also to footnote~{\ref{fn:so6}}, it 
is obvious that according to complexifications/phases of
the underlying $P^{3}$ point/plane coordinates one can 
treat the usual quadratic invariant in 4 variables by 
considering the whole 'family' of symmetry groups 
SO($\alpha$,$\beta$), $0\leq \alpha,\beta\leq 4$, 
$\alpha+\beta=4$ in point as well as in plane (or 
'momentum') space. However,
related to their associated six line coordinates, we 
have to discuss also the accompanying 'family' SO($n$,$m$),
$0\leq n,m\leq 6$, $n+m=6$, \cite{dahm:MRST3}, \cite{dahm:MRST4},
as well as quadratic mappings like the \PLu-Klein quadric, 
the Veronese mapping, birational maps, etc. 

\subsection{An Early 'Relativistic' Example}
\label{sec:relativisticexample}
For now, we want to mention (as the first and as an early
example) only the relation to generating lines of the 
one-sheeted hyperboloid\footnote{This case of this second 
order surface yields the most practical access to discuss 
real lines within a surface, and by departing from this 
special surface -- which is analytically easy to handle 
and geometrically easy to understand for being a ruled 
surface in 3-space -- we can relate further discussions 
of general $2^{nd}$ order or class surfaces by complexifying
individual point/plane or line coordinates.}. Each generating
family of lines of the hyperboloid can be derived from 
three Complexe, so by Lie transfer \cite{dahm:MRST7} or 
by the differential rep discussed in appendix~\ref{app:polarity},
we may relate as well the two operator sets $L_{i}$ and 
$K_{j}$, each comprising three generators\footnote{Due to
possible alternative identifications and ambiguities, we 
have summarized some details in appendix~\ref{app:polarity}}.

Thus, we may use the operator discussion in \cite{alfaro:1973},
ch.~1, sec.~11.3, and especially in \cite{alfaro:1973}, ch.~1, 
appendix~III, as well as the rep theory discussed there in 
terms of su(2)$\oplus$i~su(2) by means of the 3-dim operator
sets $J_{i}$ and $K_{i}$, and the associated invariants
$F$ and $G$, where
$F=\frac{1}{2}(\vec{J}^{\,2}-\vec{K}^{\,2})
=\frac{1}{4}M_{\mu\nu}M^{\mu\nu}
=\frac{1}{2}\left(N(N+2)+M^{2}\right)$, and 
$G=\vec{J}\cdot\vec{K}
=\frac{1}{4}\epsilon_{\mu\nu\rho\sigma}M^{\mu\nu}M^{\rho\sigma}
=iM\,(N+1)$.
$M$ and $N$ denote 'quantum numbers' (see \cite{alfaro:1973},
ch.~1, appendix~III, eq.~(III.14), or see sec.~\ref{sec:weinberg}
and compare to eqns.~(\ref{eq:ComplexInvariante1}) and 
(\ref{eq:ComplexInvariante2})), and the 'translation'
of $\vec{K}=i\vec{B}$ (see \cite{alfaro:1973}, ch.~1, 
appendix~III, eq.~(III.5), or sec.~\ref{sec:weinberg})
yields the complexification of the original su(2) operators,
and thus to line geometry when recalling the line generators
of second order surfaces (see appendix~\ref{app:polarity}). 
Moreover, it connects to Minkowski's artificial complexification
of the fourth space-time coordinate above, and as such of 
the second 3-dim generator set $K_{i}$ (the 'boosts') which
involves the zero- or four-coordinate. So the reference to 
Gegenbauer polynomials as 'basis functions' of such polynomial
reps is evident, however, one should take care of the projective
character and an appropriate identification of the coordinates
involved in such a rep theory. This reflects once more the 
relation of the quadrics in terms of four homogeneous point
or plane coordinates, and in terms of their six homogeneous
line coordinates, i.e.~$P^{3}$ and $P^{5}$, or SO(4) and SO(6)
(and their respective reps) when discussing transformation
groups and invariants. The real (\PLu) case relates SO(3,1)
and SO(3,3) (see e.g.~\cite{dahm:MRST4}, sec.~1.6, and the
discussion there).

An analogous reasoning can be extracted from \cite{joos:1962},
sec.~2.1, where a skew $4\times 4$ Lie algebra rep $M_{\mu\nu}$
is mapped (or decomposed by separating space coordinates 
1,2,3 from a 'time' coordinate 0) to two operator triples 
$\vec{M}=(M_{23},M_{31},M_{12})$ and $\vec{N}=(M_{01},M_{02},M_{03})$.
Both are usually treated in terms of su(2) Lie algebras and 
attached to operator reps of the inhomogeneous Lorentz group
on a Hilbert space\footnote{Not only by this approach, it 
is evident that people wanted to construct Hilbert space 
reps in order to handle the skew 6-dim object $M_{\mu\nu}$, 
however, the formal treatment so far ended in su(2)$\oplus$su(2)
or su(2)$\oplus$i~su(2) discussions while neglecting the 
geometrical background of the 6-dim Complex rep. As this 
is the 'classical' decomposition of the line rep resp.~the
organization of the line coordinates by point coordinates,
for us the only open issue is to discuss the occurrence 
of the skew $4\times 4$ rep and the differential reps in 
appendix~\ref{app:polarity}. In a naive approach, in order 
to switch from classical geometry to differential reps, one 
can replace the line coordinates $p_{\mu\nu}$ by operators
$M_{\mu\nu}$ (and vice versa) and check the consequences 
with respect to linear reps and second order surfaces.}.
 
The important issue in this context for now is Minkowski's
claim that for $\vec{m}^{\,2}-\vec{e}^{\,2}=0$ and $\vec{m}\vec{e}=0$,
i.e.~according to our 'new' notation due to a singular 
Complex, these two properties remain invariant under every
Lorentz transformation (\cite{mink:1910}, p.~485). So here,
with respect to our Complex approach \cite{dahm:MRST8},
we've recovered the two Lorentz invariants related formally 
to su(2) algebras, which e.g.~in the su(2)$\oplus$i~su(2) 
Lie algebra rep approach \cite{alfaro:1973} are related to 
the invariant 'quantum numbers' of the reps. We discuss in 
appendix~\ref{app:polarity} briefly the necessary operator
representation derived by classical polarity and line 
geometry, whereas the 'quantum' notion can be introduced
according to \cite{dahm:MRST7}, \cite{dahm:MRST9} based 
on Lie transfer of lines to spheres and their reps in 
terms of the Pauli algebra. However, already here, it is 
immediately evident that the linearization and the rep
by infinitesimal su(2) generators have to be traced back 
to their origins and background in PG in order to understand
more details of relativity, symmetry and 'quantum' reps.
In other words, the example rises the question on how to
find linear reps, given a quadric, which can -- of course 
-- be answered immediately from the viewpoint of PG (at
least for elements of low grades). Later, we can enhance
this discussion on generating higher order (or class) 
elements.

\subsection{Further Aspects}
\label{sec:furtheraspects}
As is nowadays common textbook knowledge, one can develop 
both theories of relativity -- special and general\footnote{see
e.g.~Einstein's papers (or the notes in appendix~\ref{app:einstein},
or Hilbert's papers cited above.} -- based on Minkowski's
space-time (or point-geometrical) approach and related differential
geometry, although in cosmology there are various models with
symmetry groups SO($n$,$m$) under discussion where $n+m\leq 6$.
Whereas the Poincar\'{e} group -- which is also often mixed 
up with transformation groups acting on homogeneous coordinates
-- can be obtained {\it only after contraction} from SO(3,2) 
or SO(4,1)\footnote{For a detailed discussion of contractions,
their geometrical background and the relevant references, see 
e.g.~\cite{gilmore:1974}, especially ch.~10. Gilmore's detailed
approach by coset spaces as well as Helgason's marvelous work 
\cite{helgason:1978}, \cite{helgason:GGA} can be seen as a 
mechanism and framework to comprise, generalize and supersede
Wigner's approach by boosts \cite{wigner:1962} which has been
used by Weinberg's formalism (\cite{weinberg133:1964}, 
\cite{weinberg134:1964}, \cite{weinberg138:1965}, and see also
\cite{dvoe:1993a}, \cite{dvoe:1993b}, \cite{dvoe:1993c},
\cite{dvoe:1994}, and \cite{dvoe:2018}). We have summarized
some aspects in \ref{sec:weinberg} later in order to identify
the operators and the reductive algebra structure.}, its rep
theory accordingly requires Euclidean/metric coordinates and 
semidirect products. 

Beginning with Bateman's and Cunningham's work in 1909 on 
Maxwell's equations, the discussions of conformal symmetries 
related to SO(4,2) -- and also to invariant 6-dim quadrics 
with different signatures -- showed up and entered the field
and the physical discussions. In our understanding, this 
topic is completely resolved and superseded by line geometry
and Complexe as soon as one accepts the $P^{5}$ background
represented in terms of {\it six homogeneous} coordinates,
their related symmetry groups and subgroups, and -- last 
not least -- various transfer principles and projections 
techniques to lower dimensional spaces (see e.g.~\cite{kleinHG:1926}
'Zweiter Hauptteil'). In other words: just strictly apply 
Klein's 'Erlanger Programm' from scratch!

Classical relativity has been elaborated further by Einstein
using Minkowski's formalism, emphasized by Hilbert's
(re-)formulation of Einstein's general theory in a short
series of papers \cite{hilbert:1915}, \cite{hilbert:1916},
\cite{hilbert:1924} in exchange with Klein \cite{klein:1917},
and also by Weyl with his axiomatization and the affine 
approach \cite{weylRZM:1918}. The standard textbook 
procedure treating and handling electromagnetism in 
terms of 4-vectors and -tensors starts here, too. Lots
of textbooks in mathematics and physics so far followed 
this trail, always accepting and using those coordinate 
and differential definitions without deeper investigations.
Now, we agree, of course, with the fact that this formalism
works in point space while using 4-vectors (or 'space-time 
vectors of 1$^{st}$ kind') only, and by specifying circumstances
as well as obeying obvious and usual restrictions. However, 
answers with respect to transformation properties of 
extended physical objects, even with respect to basic 
mathematical objects like lines or planes, most often 
are not treated at all, but the focus is set to point 
transformations only\footnote{Even in this context, 
people focus on point coordinates instead of recalling 
that e.g.~length contraction or time dilation intrinsically
argue with {\it differences} of points (or their coordinate
reps). So it is not sufficient to include an 'origin' of 
a coordinate rep by the null {\it vector} which is neglected
but one has to recall that the very coordinate definition 
of such 'vectors' already includes geometrical assumptions
e.g.~on linearity as in the case of Euclidean or affine 
coordinates, and on the metric, i.e.~in these cases on 
the polar system in the absolute plane. So one should 
consider naive generalizations from 3 to 4 coordinates
very carefully, like e.g.~in the case of gauge theories.}.

Moreover, the general background of Complex geometry and
its relation to physics, and especially to relativity and
electromagnetism, has been forgotten and has been lost in 
literature throughout the years. So for us, the milestone 
of the Minkowski paper \cite{mink:1910}, and especially 
von Laue's paper \cite{vonLaue:1950} much later, show how
to attach the differential representation of relativity 
and electromagnetism to the geometry of (singular) linear
Complexe and their invariant theory \cite{dahm:MRST8}. 
However, from the viewpoint of Complex geometry (or PG
in general), this is a singular and very special case,
so the main switch we are going to work later is to 
generalize special/singular Complexe to general Complexe 
-- regular/general linear Complexe as well as higher 
order Complexe and their respective geometrical properties,
instead of generalizing the dimensions of the irreps or 
'quantum numbers' of the SU(2) subgroups as usual.

The 'twofold antisymmetric tensor' $F^{\mu\nu}$, of course,
occurs in various formalized contexts in physics like 
classical field theory, affine connection concepts, 
covariant derivatives and differential geometry, even 
in quantum field theory when formulating gauge theories. 
However, its foundation in null systems and the analytic 
description of forces which yields the physical background 
and -- according to our opinion -- the correct trail to 
unify representation theory by geometry, has been forgotten
and suppressed by applying tensor calculus, formal algebra 
and differential geometry, although the framework of 
(skew) Lie algebra reps and Lie symmetries in general 
is a common tool-set in these days.

Also in Einstein's papers, we've found\footnote{So far, 
we've checked mainly publications in 'Annalen der Physik'
and the published/printed versions of Einstein's academy
contributions in Berlin, (re-)printed in 'Sitzungsberichte
der Preu\-{\ss{}}i\-schen Akademie der Wissenschaften', 
1914-1932 \cite{einstein:1914-32}. We've given a brief 
summary of what we've found in appendix~\ref{app:einstein}.}
only few statements and contexts mentioning the 'six-vector
formalism', and only one of them \cite{einstein:1916} 
addressing the electromagnetic field rep in conjunction
with the gravitational field in detail. However, all of
his approaches in appendix~\ref{app:einstein} are based 
on Minkowski's paper \cite{mink:1910} by using only one
six-vector and its dual, however, with major use of 
4-vector and various derived tensorial reps. So we see
von Laue's paper \cite{vonLaue:1950}, by now introducing
{\it two} 'six-vectors' to describe field {\it and} matter,
as the next relevant step towards Complex geometry. The
associations of the 3-vector fields $-i\vec{E}$, $\vec{B}$
to a 'six-vector' $\textfrak{M}$, and $-i\vec{D}$, $\vec{H}$
to a 'six-vector' $\textfrak{B}$ is used \cite{vonLaue:1950}
to derive Maxwell's equations, and matter properties are
introduced by coupling the two 'six-vectors' by
\begin{equation}
\label{eq:mattercoupling}
\textfrak{B}_{\alpha\beta}Y_{\beta}
=\epsilon\,\textfrak{M}_{\alpha\beta}Y_{\beta}\,,\quad
\textfrak{M}^{*}_{\alpha\beta}Y_{\beta}
=\mu\,\textfrak{B}^{*}_{\alpha\beta}Y_{\beta}\,,
\end{equation}
where von Laue used the 4-velocity $Y_{\beta}$, and conjugation
of the 'six-vectors' denoted by ${\cdot}^{*}$. Details can
be found in \cite{dahm:MRST6}.

Moreover, in all of these references, we've found no remark
or discussion of the original background of null systems 
and forces. In contrary, all authors base their reasoning
and their calculations upon the rep incidence of antisymmetric
twofold tensor reps with the ray rep of singular linear 
Complexe in terms of $4\times 4$-matrices, however, they
subsume this rep as part of a tensorial calculus based 
on 4-vector reps. This resembles using the 4-vector as 
a fundamental (ir-)rep\footnote{Which, of course, works
only severely limited in this context!} and constructing
higher order reps of a Lie algebra or group, and they 
proceed performing formal algebra.

Thus, a lot of physical background has been overlooked 
in favor of technocracy and stiff algebra applications,
and was lost throughout the last 100 years for general 
discussion(s) of the underlying physics and geometry.
% formalized from geometry, geometry forgotten
% Bourbaki style algebra, far from physics

\subsection{Standard Approach -- Klein 1910}
\label{sec:standard-klein1910}
Although we have already discussed the major switch 
-- Minkowski's 1910 paper \cite{mink:1910} -- which we 
want to work below in that we want to use general line 
and Complex geometry to comprise and supersede this paper
and its concept, it is necessary to consider another 
important paper in 1910 which emphasized Minkowski's 
coordinate identifications above: Klein's paper \cite{klein:1910}
on the Lorentz group! To forestall one of the major 
results -- his arguments have not only emphasized 
Minkowski's 4-vector formalism, but in addition they 
have forced the coordinate discussion to generalize 
this formalism to '5-dim' point reps by expressing the
four coordinates of space and time in terms of five 
homogeneous point coordinates. This has had dramatic
impact on the formulation of various 5-dim theories 
(Kaluza-Klein, or see e.g.~Pauli \cite{pauli:1933a}, 
\cite{pauli:1933b}) as well as on approaches in projective 
(differential) geometry like Veblen's efforts (see 
e.g.~\cite{veblen:1930}, \cite{veblen:1933a}, or 
\cite{veblen:1933b}, and the associated literature 
and discussion). In order to gain more control on the 
context, we want to recall and discuss briefly some 
aspects of Klein's paper \cite{klein:1910} on the 
geometry of the Lorentz group.\\

So whereas in the first part Klein summarizes the historical
development so far, beginning with Cayley's $6^{th}$ 
memoir on quantics in 1859, and Klein's personal geometrical
approach towards his 'Erlanger Programm' in 1872, he 
focuses there mainly on collineations and invariant 
theory. However, in the first part, he keeps track on
the correct identification of the {\it different types}
of coordinates and their respective transformation groups
which we see as a real highlight of the paper! There, 
he distinguishes Euclidean ('metric') point coordinates
explicitly from (ordinary) homogeneous coordinates\footnote{Klein
doesn't mention tetrahedral homogeneous coordinates in
this paper at all!}, he discusses some aspects of order
vs.~class from classical projective geometry of 3-space,
he addresses some issues related to conics as an application
to discuss the Cayley-Klein approach to metrics, and -- 
last not least -- he focuses on affine transformations.

However, on \cite{klein:1910}, p.~293, and this -- in the
light of his achievements -- is from our viewpoint not 
one of the highlights of \cite{klein:1910}, he begins 
'to generalize' point coordinates, and it is here, where
we'll have to switch later on to another reasoning.

Following Klein's arguments, he interprets the four space
and time coordinates $x$, $y$, $z$, and $t$ as Euclidean
('metric') coordinates which can be seen in his next 
paragraph where he introduces five homogeneous coordinates
$x_{1}$, $x_{2}$, $x_{3}$, $x_{4}$, and $x_{5}$. He relates
both sets by
\begin{equation}
x=\frac{x_{1}}{x_{5}},\,
y=\frac{x_{2}}{x_{5}},\,
z=\frac{x_{3}}{x_{5}},\,
t=\frac{x_{4}}{x_{5}},
\end{equation}
in the 'classical manner' to define ordinary homogeneous
coordinates, and as such naively recalls and transfers the
affine approach of 3-space, or $P^{3}$, to be applied to 
some kind of affine $P^{4}$, represented inhomogeneously 
by the coordinates $x$, $y$, $z$, and $t$, and by the 
homogeneous coordinates $x_{1}$, $x_{2}$, $x_{3}$, $x_{4}$,
and $x_{5}$. So, in addition, he gains direct contact to 
his old 4-dim metric representation of line geometry in 
the early 1870s, however, the need for a satisfactory 
interpretation of the four 'metric' coordinates appears
immediately and should have been addressed there already.

We do not want to discuss his reasoning with respect to 
his 4-dim 'metric' space-time spanned by $x$, $y$, $z$, 
and $t$ further because we have given arguments already
(see e.g.~\cite{dahm:MRST3}, \cite{dahm:MRST4}, \cite{dahm:MRST8},
or appendix~\ref{app:coordinatesystems} here) that one 
should consider 'time' in a different manner\footnote{This
is especially important from the physical point of view 
because one has to keep track of physical dimensions. 
It is NOT straightforward to set a 'velocity' $c=1$ 
without caring about the consequences. Already physical
dimension considerations require to treat {\it all} 
the coordinates $x$, $y$, $z$, and $t$ either as space 
coordinates, however, while taking care of their 
{\it identification and interpretation} (see below), 
or to change the interpretation and introduce {\it a 
time} conceptually into the picture which is typically
done by restriction to a Euclidean point setup or to 
a setup compatible with special relativity, i.e.~requiring
uniform and as such linear motion. This, however, 
introduces linear velocities and their relations, 
appropriate transformations, as well as 'an upper 
bound' in order to handle absolute elements. So 
implicitly this corresponds to the axioms of standard 
projective geometry of 3-space. It is then evident 
that 'time' has to connect the point and the velocity
picture \cite{dahm:MRST3}, \cite{dahm:MRST4}, 
\cite{dahm:MRST5}, and we find projective relations
e.g.~between the velocities to justify $\beta$ or 
the Lorentz factor $\gamma$.} than just as an additional
'coordinate' mixed up with point space. Moreover, 
by assuming the simple case of uniform motion, we've
shown (see e.g.~\cite{dahm:MRST3}, sec.~3.5, or 
\cite{dahm:MRST4}, or \cite{dahm:MRST5} sec.~2) 
that it is projective geometry which -- represented
in point space -- allows to understand the Lorentz 
factor just by relations of velocities\footnote{If 
expressed by anharmonic ratios, or 'W{\"{u}}rfe', 
the assumption of a common 'time' (or 'time' 
difference, or projection) allows to relate points 
and velocities, and to transfer projective relations.}
when identifying $ct$ with the absolute element. 
In appendix~\ref{app:coordinatesystems}, we discuss
some more examples which show the relation of 
point space and metric versus the identification
of 'time'. From within this context, the important 
message is that point reps and velocities (considering
uniform motion and an origin on the line of the 
motion, or the trajectory, to simplify the vector
description) are proportional due to choosing the
usual linear time rep, so time represents a line 
parameter as soon as one introduces additionally 
individual points on lines.

In the derivation of the tetrahedral Complex (see 
\cite{dahm:MRST4}, sec.~3, especially sec.~3.3), and in 
appendix~\ref{app:coordinatesystems}, it is easy to 
construct cases where the 'time' is derived from other 
coordinatization, e.g.~only using line geometry and line 
intersections. It is just the geometrical picture 
which changes the interpretation of the parameter, and 
especially of 'a time' associated to points and their 
motion. So in sec.~\ref{sec:reinterpretation}, we'll 
try to adjust Klein's problematic conception of the 
five homogeneous coordinates founding on the point 
interpretation and a linear absolute element above, 
and instead we argue in favour of Complex geometry 
of $P^{5}$ in terms of {\it six} homogeneous coordinates.
The projection to usual space-time can be realized 
by the \PLL condition, i.e.~by a quadratic constraint,
so whereas in 3-space it is a generalization to switch
from point/plane to line representations, the physical
background discussion should reconsider Complex geometry
in $P^{5}$ as is known from mechanics (null systems/forces),
optics (rays), or the electromagnetic description (see 
e.g.~von Laue \cite{vonLaue:1950}). Note already here,
that this framework generalizes the idea of forces by
3-dim line sets distributed over the whole 3-dim space,
however, respecting a certain geometrical distribution
\cite{plueckerNG:1868}. So whereas on the one hand this
is an alternative picture of 'fields' and potential 
theory by identifying the individual members of such
line sets in 3-space which are passing through the 
individual point coordinates\footnote{In detail, given
a point $x$, one can ask which lines of a Complex, a 
Congruence, a Configuration (or ruled surface), or 
other sets of lines, pass this point. This is answered
immediately in the case of a Complex by an associated 
planar line pencil, and in the case of a Congruence 
(or a 'ray system (German: 'Strahlensystem erster Ordnung
und erster Classe') by a 2-dim set of lines in space 
where one space-point is passed by exactly one line
of the set. In this case, there is only one line of
the set within each spatial plane. An example is the
set of lines being incident with two skew lines. 
Moreover, one can consider cones or general second
order surfaces -- having Monge's equation, Lie's 
sphere geometry, or second order partial differential
equations in mind -- which are related to the point
$x$ in the respective geometry.},
on the other hand, one has to switch notion from 
points (without extension) and 'point-based' 
interactions into a picture which includes spatial
extension (or non-locality) right from the beginning.
This framework allows for a plethora of different 
reps, only one being points and their accompanying
planes in 3-space in order to complete dualism 
as transfer principle. Accordingly, we've found
a wealth of 'transfer principles' between different
reps and different interpretations, and one should
be careful to attach physics naively to only one 
special rep or picture, especially one -- the 'point'
-- which is the perfect candidate to prescind. So 
a 'wave-particle dualism' is far from being mystical,
but it reflects only one out of several transfer 
possibilities between lines and spheres (like 
e.g.~in Lie transfer), and each of the patterns 
holds as long as the underlying transfer principle holds mathematically (i.e.~the
rep 'exists' faithfully). Coined in \PLu's, early 
Klein's, Lie's, and Study's words, we may change 
the base elements of the respective geometry 
appropriately.

An immediate identification within Lie's description
of line elements (\cite{lie:1896}, Abschnitt~II), 
however, is evident where according to Lie we may 
use ('metric') point coordinates and the two ratios
$p$ and $q$ of their differentials $\mathrm{d}x:\mathrm{d}y:\mathrm{d}z$,
so Lie has presented a 5-dim rep $(x,y,z,p,q)$ which
he uses to develop his rep theory of line and area
elements (and which, of course, appears in related
scenarios like dynamics and differential geometry 
having his contact geometry as background).

\subsection{Standard Approach -- Klein before 1910}
\label{sec:standard-kleinbefore}
In order to gain some more insight into possible origins
of a 4-dim 'metric' concept of special relativity, its
embedding into PG and of the discussion of the 'well-known'
invariant quadric $x^{2}+y^{2}+z^{2}\pm t^{2}=0$, it is 
worth to mention some earlier aspects of Lie's and Klein's 
work on transfer principles as well as Klein's and Lie's 
work of sustaining and summarizing \PLu's ideas on line 
and Complex geometry, both a vista of years before 1908/1910.\\

As such, we can start our discussion with \cite{klein:1872a},
footnote $**$ on page 257, in 1872. There, Klein claims in 
the text his goal to generalize Lie's transfer principle 
\cite{lie:1872} -- which maps two 3-dim spaces $r$ and $R$
of lines and line Complexe to spheres and sphere Complexe 
under special metric restrictions -- to a 4-dim {\it metric}
space because line geometry itself is 4-dim\footnote{It is 
restricted by the additional \PLL condition, i.e.~the 
\PLu-Klein quadric in $P^{5}$, to 3-dim, and it is thus
suited to describe 3-space geometrically besides using 
points and planes only.}. In the accompanying footnote 
$**$, he introduces four rectangular point coordinates 
$x$, $y$, $z$, and $t$, which according to his statement
define an absolute quadric $x^{2}+y^{2}+z^{2}+t^{2}=0$ 
and thus provide the basis of the metric in this 4-dim 
{\it metric} space. So, vice versa, the only possible, 
valid coordinate interpretation in this scenario is to 
understand the four rectangular point coordinates $x$, 
$y$, $z$, and $t$ in his text as {\it ordinary homogeneous}
point coordinates within a set of {\it five} coordinates 
where the fifth coordinate has already been set to 0 in 
order to realize the absolute quadric in the absolute 
'plane', and accordingly the 'metric' by means of a 
direct generalization of the 3-dim case. So although 
Klein's explanation -- even in german -- is difficult 
to read and understand, he obviously\footnote{See also 
the discussion in the next paragraph based on his later
publication \cite{kleinHG:1926}!} discusses a $P^{4}$ 
where in the philosophy of the Cayley-Klein mechanism
the metric properties have to be derived by projective 
relations of objects with respect to the absolute 
quadric $x^{2}+y^{2}+z^{2}+t^{2}=0$. In other words,
the conceptual focus is set on {\it linear} elements 
when defining coordinates and attached to a special
identification of absolute elements\footnote{In 
\cite{dahm:MRST3}, we have 'derived' the hyperbolic 
quaternary quadric in point space from its association
to the senary quadric in line geometry. In other words,
the 'light cone' may also be associated with a certain
quadratic Complex, or an element of the \PLu-Klein 
quadric in $P^{5}$. This, at the same time has some 
influence on the coordinate definition by self-polar
fundamental tetrahedra, however, that's beyond scope
here. Another well-known relation of hyperbolic 
geometry is known by considering three linear Complexe
which constitute a Configuration, i.e.~a ruled surface
and as such an hyperboloid \cite{plueckerNG:1868}.
In elementary geometry, this is reflected e.g.~by
generating the hyperboloid by lines, and discussing
the two families of generating lines.}.

By the way, having discussed this construction scheme
with respect to coordinate definitions and the 
construction of the metric via a quadric now for a 
second time, we want to rephrase this context with
respect to rep theory, and especially to linear reps.
As such, it is interesting to ask how 'to resolve' a
quadric by identifying two appropriate linear reps
(like the rep and its 'adjoint' or 'dual', or the 
roots of a square as a special case of relating 
linear reps and their adjoints). Especially the way
around to construct quadrics from linear reps, or in
general the generation of elements of higher grade
(order or class) can be performed using PG. In the
case of linear elements, prominent examples are the
generation of conics by line pencils in the plane
(see e.g.~\cite{doehlemann:1905}) or the generating
line families of second order surfaces (see e.g.~\cite{kleinHG:1926}),
but this can be extended, of course, to schemes using
higher order elements right from the beginning and
classifying e.g.~the intersection results, or unions.

Klein's coordinate identification as discussed above
and the related reasoning can be verified also in another
context -- although published by Klein more than 50 years
later -- if we refer once more to Klein's book on advanced
geometry \cite{kleinHG:1926}. There in \S\S37--39, Klein 
develops the four historical stages of the foundations of 
projective geometry. Especially what he calls 'the third 
stage' in \S39, it is where he comments on the historic 
approach to derive metric properties from projective geometry 
in terms of homogeneous coordinates. There, he uses appropriate 
analytic examples to justify the coordinate identification/interpretation 
we have discussed already above.

His planar example uses analytic point coordinates $x$, $y$, 
and $t$, whereas his spatial example uses point coordinates 
$x$, $y$, $z$, and $t$. In the first case, he derives the two
absolute points by starting from the quadric $x^{2}+y^{2}+t^{2}=0$,
so the absolute line $t=0$ yields $x^{2}+y^{2}=0$ which results 
in two conjugate imaginary point solutions.

Then, he has related the 'metric' 3-dim case to $x^{2}+y^{2}+z^{2}+t^{2}=0$
in terms of four ordinary homogeneous coordinates, and he thus
obtains the absolute conic section $x^{2}+y^{2}+z^{2}=0$ in the
absolute plane $t=0$. His reasoning is attached to the projective
arguments that if the analytic expression 
$(x^{2}+y^{2}+z^{2})+2\alpha x+2\beta y+2\gamma z+\delta =0$ of
a sphere is given in terms of Euclidean coordinates $x$, $y$, 
and $z$, and four parameters $\alpha, \beta, \gamma, \delta$, 
the related expression in {\it homogenous coordinates}\footnote{It 
is important to note that Klein thus implicitly switched the 
{\it character} and {\it interpretation} of the originally 
Euclidean point coordinates $x$, $y$, and $z$ to the set 
$(x,y,z,t)$ of ordinary homogeneous coordinates without 
changing notation appropriately!} reads as 
$(x^{2}+y^{2}+z^{2})+2\alpha xt+2\beta yt+2\gamma zt+\delta t^{2}=0$.
Its intersection with the absolute plane $t=0$ thus yields the
conic section $x^{2}+y^{2}+z^{2}=0$ which is common to {\it all}
spheres because the set of equations $x^{2}+y^{2}+z^{2}=0$, $t=0$,
evidently doesn't depend on the parameters $\alpha, \beta, \gamma, 
\delta$ comprising the sphere description. So this conic section
is suitable to classify spheres as those second order surfaces 
which leave the absolute conic invariant. Note, that throughout
all those discussions, $t$ is a homogeneous coordinate and NOT
'time'. Further contemporary examples of this notation of the 
fourth homogeneous variable can be found e.g.~in \cite{hudson:1905},
or \cite{hilbert:1909}, so it is necessary to distinguish standard
reasoning in PG from a physical 'time' identifications of 't' 
as done by Einstein\footnote{Klein attributes this identification
of 't' and time to Einstein (\cite{klein:1917}, p.~474) when 
writing letters to Hilbert on occasion of Hilbert's papers on
'Die Grundlagen der Physik' \cite{hilbert:1915}, \cite{hilbert:1916},
\cite{hilbert:1924}. Extracts taken from the letters exchanged
between Hilbert and Klein were published in Klein's paper 
\cite{klein:1917}.}.

As such, the case of four 'metric' coordinates expressed by 
means of five ordinary homogeneous point coordinates $x$, $y$,
$z$, $t$, and $s$ thus would give rise to the related set of
corresponding equations in $P^{4}$, i.e.~$x^{2}+y^{2}+z^{2}+t^{2}=0$
where $s=0$. Note however, that in order to obtain 'metric' 
coordinates in an 'affine picture', i.e.~fixing a {\it linear}
'plane' in $P^{4}$, one has to divide the four coordinates by 
$s$ linearly, and thus absorb the singularity of the absolute 
element already in the 'metric' (or Euclidean) coordinate 
definition\footnote{We have started this discussion in \cite{dahm:MRST3}
as we see this as a possible mechanism as long as one considers 
linear transformations and related invariants, however, we have 
given there an alternative (and to our opinion more general) 
definition which has to be extended if we discuss non-linear 
transformations and spaces but it includes the linear case 
(see \cite{dahm:MRST3}, sec.~3.5). So it can serve as basis 
to generalize the coordinate description and related rep 
theory.}.\\

Throughout all this reasoning, the related concepts are driven 
by an invariant {\it linear} absolute element so that e.g.~in 
the case of spatial affine transformations the absolute plane 
$t=0$ remains invariant under linear transformations\footnote{This,
recalling duality vice versa as a correlative mapping, involves
the origin of the coordinate system, too.}. This requirement, 
however, is a strong restriction which e.g.~in case of non-linear 
transformations in general can be fulfilled only in infinitesimal
limits or by special considerations. And this aspect yields exactly
an approach towards future discussions of space-time and relativity 
in that e.g.~line geometry is a priori quadratic. So in order
to discuss finite transformation and geometry, we necessarily 
have to consider non-linear transformations, and as well 
invariants of higher order or class. Although invariant theory
and form systems underwent mathematical and physical research 
and extensions over the decades since Clebsch, Gordan, Study,
Hilbert, and Weitzenb{\"{o}}ck, the necessary quaternary or
even senary formalism is tedious and error-prone. Various
algebraic approaches introduce formal problems, or physical
and mathematical ambiguities, or even lack physical evidence
completely. Typical examples are large Lie groups, or associated
algebras of various kinds, where final SU(2) reductions 
(e.g.~by using cosets or coset chains) and the necessary 
physical identifications are ambiguous\footnote{The various
possibilities can be 'counted' phenomenologically e.g.~in 
the Dynkin diagram.}, supersymmetry which so far seems to 
have its place in nuclear physics according to Iachello's 
work and research, however, NOT in the description of 
elementary particles \cite{catto:2019}, or string (or 
'superstring') models which remind of some aspects of
\PLu's reasoning when interpreting geometry and their 
base elements, however, in different clothing. Last not
least, so far we are not aware of significant evidence 
in physics.

To summarize, Klein's reasoning introduces the concept of a 
$P^{5}$ (or four 'metric' coordinates) which has been 
reused\footnote{As of today, we cannot verify whether 
Minkowski used the same underlying geometrical concept while
discussing the quadric $x^{2}+y^{2}+z^{2}-t^{2}=0$.} by Klein
in 1910 and in \cite{kleinHG:1926}. However, it is important
to emphasize Klein's identification of the two 4-dim spaces,
i.e.~whereas one can of course start in the first space by 
using ray/point or axis/plane coordinates to introduce 
antisymmetric line coordinates there, he identifies the 
second space again with a metric space which is relevant 
to physical observations, too. As we don't want to discuss 
related aspects and problems in detail in this section,
but only want to report the approaches, we close here and
address in the next subsection \ref{sec:weinberg} one more
common rep which yields some insight into the origin of 
the appearance of the Lie algebras su(2)$\oplus$i~su(2)
and su(2)$\oplus$su(2). Please note right here, that the
derivation is valid also from the classical viewpoint, 
and that -- in contrary to common belief -- there is 
no need at all to discuss this rep solely in terms of
'quantum' theory or a 'quantum' interpretation although
this framework has been developed there due to the common 
use of the theory of Lie groups and Lie algebras\footnote{Or
as hostile researchers named it: 'Gruppenpest'.}. \\

\subsection{Feynman Rules and Spin}
\label{sec:weinberg}
A lot of work has been done on 'quantum' reps and concepts 
to treat relativity on the quantum level. Especially 
Valeriy has published a lot of great work on various 
aspects of the different rep theories, so we feel free
to skip most of the discussion here. Instead, we focus
in this subsection on the overlapping conceptual aspects
which are relevant for our later discussion here, and as
such, feeling guided by Valeriy's papaers \cite{dvoe:1993a},
\cite{dvoe:1993b}, \cite{dvoe:1993c}, and \cite{dvoe:1994},
we want to refer here only to Weinberg's presentations 
\cite{weinberg133:1964}, \cite{weinberg134:1964}, and 
\cite{weinberg138:1965} on spin reps. Why?

On the one hand, they give a strong hint with respect to the
identification of the two su(2) algebras\footnote{More precise:
with respect to the association of the skew $4\times 4$ matrix
rep of real transformation parameters $\omega_{\mu\nu}$ and
the differential reps of the operators by means of two su(2)
algebra reps.} related to his 'infinitesimal Lorentz 
transformation',
\begin{equation}
\label{inflortrans}
\Lambda^{\mu}_{\hphantom{\mu}\nu}\,
=\,\delta^{\mu}_{\hphantom{\mu}\nu}+\omega^{\mu}_{\hphantom{\mu}\nu}\,,\quad
\omega_{\mu\nu}=-\omega_{\nu\mu}
\end{equation}
as given by \cite{weinberg133:1964}, eqns.~(2.18) and (2.19),
and 'quantum' reps in this Hamiltonian approach\footnote{Moreover,
it seems to be the major source of \cite{alfaro:1973}, ch.~1, 
appendix~III, cited in our first example above.}. His 'proper
homogeneous orthochronous Lorentz transformations' were 
defined\footnote{Compared to our picture, this Ansatz describes
a subset of collineations whereas correlations are handled only
implicitly here in the context of the adjoint rep, or conjugation,
raising and lowering indices, or in some implicit aspects of 
invariant theory.} by \cite{weinberg133:1964}, eq.~(2.1),
\begin{equation}
\label{eq:pholtFinite}
\begin{array}{c}
x^{\mu}=\Lambda^{\mu}_{\hphantom{\mu}\nu}x^{\nu}\,,\quad
g_{\mu\nu}\Lambda^{\mu}_{\hphantom{\mu}\lambda}\Lambda^{\nu}_{\hphantom{\mu}\rho}=g_{\lambda\rho}\\[2mm]
\mathrm{det}\,\Lambda=1\,,\quad \Lambda^{0}_{\hphantom{0}0}>0\,,
\end{array}
\end{equation}
which can be understood as well as a covariance requirement 
of the 'metric tensor', 
$\Lambda^{\mu}_{\hphantom{\mu}\lambda}g_{\mu\nu}\Lambda^{\nu}_{\hphantom{\mu}\rho}=g_{\lambda\rho}$.

On the other hand, Weinberg identifies physics in terms of the 
electromagnetic field reps within his $2(2j+1)$-dim reps, here 
with $(1,0)\oplus(0,1)$ while $j=1$. Because this physics is a 
direct and a really well-established and accepted showcase by 
identifying and relating the different rep concepts, we can use
it later 'vice versa' to break the geometrical aspects down 
to 'known' rep theory\footnote{\protect{\ldots}~although, in 
consequence, we'll see how the need arises to modify some of 
those reps and interpretations.}.\\

Now, addressing the first aspects of 'infinitesimal Lorentz 
transformations' in eq.~(\ref{inflortrans}), Weinberg decomposes 
the transformation into a symmetric and antisymmetric part, 
$\delta^{\mu}_{\hphantom{\mu}\nu}$ and $\omega^{\mu}_{\hphantom{\mu}\nu}$.
Whereas within the notion and terminology used nowadays in
Lie theory, people usually attribute this to the exponential
and 'infinitesimal' displacements, implicitly this comprises
already the background of Lie's logarithmic mapping and the 
polar system, i.e.~the metric properties of the coordinates.
The feature addressed by Weinberg, however, is the antisymmetric
part $\omega^{\mu\nu}$ which he denotes\footnote{Throughout 
his paper series cited above, the notation occurs twice without
any comment on the background. So it is not clear whether 
this 'six-vector' is named according to counting just the
six degrees of freedom of the skew \protect{$4\times 4$} 
transformation rep \protect{$\omega_{\mu\nu}$} or as a 
reference to the old notion 'six-vector' as used e.g.~by
Minkowski and other people.} as 'an infinitesimal 'six-vector',
and which he interprets by associating {\it unitary} operator
reps (\cite{weinberg133:1964}, eqns.~(2.20) and (2.21)) 
according to
\begin{equation}
\label{eq:hermitisierung}
U[1+\omega]=1+\frac{i}{2}J_{\mu\nu}\omega^{\mu\nu}\,.
\end{equation}
There, due to having introduced the additional complexification
$i$, he extracts the skew and hermitean operators $J_{\mu\nu}$ 
which he decomposes into two operator triples ('Hermitean 
three-vectors')\footnote{Whereas Wigner \cite{wigner:1962}
argues with the Poincar\'{e} group and relativistic angular
momentum $M_{\mu\nu}$, Joos decomposes $M_{\mu\nu}$ into two
operator triples $\vec{M}$ and $\vec{N}$ \cite{joos:1962}. 
With respect to a quantum-mechanical interpretation, Joos 
associates $\vec{M}$ with operator reps of angular momentum,
and the triple $\vec{N}=(M_{01},M_{02},M_{03})$ with a 
'barycenter of energy' \cite{joos:1962}, sec.~2.1, p.~69.},
\begin{equation}
J_{i}=\frac{1}{2}\epsilon_{ijk}J_{jk}\,,\quad 
K_{i}=J_{i0}=-J_{0i}\,,
\end{equation}
i.e.~in the spirit of requiring {\it unitary} reps, he sticks 
to {\it Hermitean} generator triples $J_{i}$ and $K_{j}$.
Please note, that at this point this rep theory maps Hermitean
generators by using the mapping $\exp i(\cdot)$, i.e.~by using
an additional $i$ to map Hermitean operators to unitary reps
of the group, contrary to standard use in mathematics where 
it is easier to map anti-Hermitean operators by $\exp$, thus
avoiding excessive phases $i$ without specific meaning\footnote{Therefore,
we've included some remarks in sec.~\ref{sec:weinbergNEW} 
to achieve a more transparent approach to reps of the algebras 
involved by avoiding superfluous phases, and moreover to 
attach directly to Helgason's work \cite{helgason:1978}, 
\cite{helgason:GGA}, and benefit from its enormous power.}.
It is especially the requirements of {\it unitary} reps due
to comparison within eq.~(\ref{eq:pholtFinite}) of the finite
Lorentz transformations which induces signs and phases like
e.g.~in \cite{weinberg133:1964}, eq.~(2.26) of the 'boost'
commutator.

Once more in other words: We are acting with unitary (or 
'compact') reps on complex rep spaces, and it is the 
$4\times 4$ skew matrix rep $\Lambda$ in eq.~(\ref{eq:pholtFinite}) 
acting as a collineation on space-time points -- this time 
$\omega_{\mu\nu}$ if we associate the parameters with physics
-- which is the origin of a certain model or rep theory. 
In Weinberg's case, he has chosen an su(2) based approach 
with complex coefficients to reflect the skewness and dim~6
of the generator algebra. Whereas this 'decomposition' of
so(4)$=$so(3)$\oplus$so(3) fulfils the algebraic and formal
requirements, and yields a formal $4\times 4$ skew matrix
rep, throughout his papers \cite{weinberg133:1964}, 
\cite{weinberg134:1964}, \cite{weinberg138:1965}, there 
is only one strict and physically accepted identification 
when treating the source-free electromagnetic field, 
i.e.~in the 6-dim case where the adjoint Lie algebra
rep(s) map directly to a special linear Complex using 
Klein coordinates\footnote{For our purposes and the 
following reasoning, this is sufficient, especially 
since the skew transformation reps of the Complex and
the generator set of the Lie algebra coincide.}. As 
we'll see below, this causes a lot of trouble with 
the correct identification of phases due to a lot of
additional $i$s appearing throughout rep (and as such
perturbation) theory.

He then discusses some operator algebra of $J_{i}$ and
$K_{j}$, and commutators similar to reductive algebra 
commutators, and he decouples them by linear combinations 
$\vec{A}=\frac{1}{2}(\vec{J}\pm i\vec{K})$ into two 
commuting su(2) (or $A_{1}$) algebras,
\begin{equation}
\label{eq:reductiveWeinberg}
\begin{array}{cclclcl}
\left[A_{i}, A_{j}\right] & = & i\epsilon_{ijk}A_{k}\,, & & 
\left[B_{i}, B_{j}\right] & = & i\epsilon_{ijk}B_{k}\,,\\
\left[A_{i}, B_{j}\right] & = & 0\,. & & 
\end{array}
\end{equation}
Decades ago, we've covered that formalism by our effective
SU(4) approach\footnote{However, covering at first compact
SU(2)$\otimes$SU(2) from Chiral Symmetry, or Chiral Dynamics,
by compact SU(4) \cite{dahm:diss}.}, because right from the
Dynkin it can be seen that $A_{3}$ comprises $A_{1}\otimes 
A_{1}$ \cite{dahm:diss}. Further reasoning lead us to consider
and discuss linear rep theory and the inclusion of axial 
charges \cite{dkr:1992}, \cite{dkr:1993}, \cite{dk:1994},
\cite{dk:1995}, to find suitable irreps. This led us later
to the non-compact group SU$*$(4) (which is isomorphic to
SL(2,$\mathbbm{H}$) and describes projective transformations
of quaternions) as well as to some associated real forms 
of SO(6,$\mathbbm{C}$) \cite{dahm:mex}, \cite{dahm:2008},
\cite{dahm:MRST1}, but details are beyond scope here.

For now, in order to stay in touch with Weinberg's notation 
and the reps within perturbation theory, it is important to 
remember the first 'overall' $i$ which has been introduced
into the exponential mapping (\ref{eq:hermitisierung}) in 
order to have unitary operator reps $U$ expressed by means
of the skew, but hermitean 'operators' $J_{\mu\nu}$,
despite of representing a non-compact group. Then, we 
have to remember the second, relative $i$ between the 
3-dim operator sets $\vec{J}$ and $\vec{K}$ in order 
to obtain commuting sets $\vec{A}$ and $\vec{J}$ in 
eq.~(\ref{eq:reductiveWeinberg}). Whereas formally this
is almost straightforward, we have postponed the discussion
of problems to a later stage when we have to identify 
the operators physically and when we have to take care 
of compact vs.~non-compact group transformations despite
the unitary character of the reps. We'll discuss a 
typical example later in terms of the field components 
$\vec{E}\pm i\vec{B}$ by \PLL versus Klein coordinates, 
and their relation to the 3$\otimes$3 $\pm$-handed 
fundamental Complexe introduced by Klein which have 
been 'reused' and reinterpreted in both relativistic 
theories as well as chiral SU(2)$\times$SU(2) symmetry
discussions during the late 1960s.\\

The second useful aspect is Weinberg's identification of 
the electromagnetic field within the $2(2j+1)$-dim reps 
$(j,0)\oplus(0,j)$. Choosing $j=1$, he associates the 
field components within the 6-dim rep $(1,0)\oplus(0,1)$,
and discusses various aspects throughout his paper series
(see \cite{weinberg133:1964}, sec.~VII, for general properties
and e.g.~\cite{weinberg134:1964} (especially the end of 
sec.~IV and of sec.~XI), and \cite{weinberg138:1965} sec.~II
ff.). Finally, he associates certain field combinations
$\vec{E}\pm i\vec{B}$ with the rep $(1,0)\oplus(0,1)$. 
However, as we've pointed out already a couple of times,
we are discussing the invariance condition of the quaternary
quadric and its related symmetry group SO(3,1) here in 
terms of {\it unitary} reps, so Weinberg's additional 
requirement of {\it unitary} reps $\Lambda$ by 
eq.~(\ref{eq:pholtFinite}) and the additional phases $i$ 
introduce additional complications. We postpone this phase 
discussion to sec.~\ref{sec:weinbergNEW}.

For our considerations later, the second aspect of Weinberg's 
'idea' has already been re-presented in Valeriy's overviews
appropriately\footnote{Although -- based on the idea of two
fundamental SU(2)s and 'spinors', obviously induced by 'quantum' 
notion -- Weinberg has considered 'any spin'. For our purposes, 
especially because the free electromagnetic field is the only 
case of this spin approach without additional assumptions when 
relating physics and spin/spinor reps, we prefer to fix $S=1$
in order to compare. We've discussed the fundamental spinor 
rep and its origin in PG sufficiently in \cite{dahm:MRST7}, 
\cite{dahm:MRST9} whereas we postpone the discussion of
'higher spins' to PG, e.g.~in the case of twisted cubics and
third order reps.} (\cite{dvoe:1993a}, \cite{dvoe:1993b}, 
\cite{dvoe:1993c}, and \cite{dvoe:1994}) by rearranging the 
field components $\vec{E}$ and $\vec{B}$ into a 'six-spinor' 
$\Psi$ according to
\begin{equation}
\Psi\,=\,\left(
\begin{array}{c}
\chi \\ \phi
\end{array}
\right)
\end{equation}
(see \cite{dvoe:1993b}, eqns.~(1) and (2), or \cite{dvoe:1994}, 
eqns.~(2.1) and (2.2)), and by acting with transformations 
$\exp{\pm\theta\,\hat{\vec{p}}\cdot\hat{\vec{J}}}$ on the two 3-dim
subcomponents of $\Psi$. For $S=1$, the operators
$\hat{\vec{J}}$, of course, have to be associated in this approach
to the adjoint rep of so(3), or su(2), and the subcomponents take 
the form $\chi=\vec{E}+i\vec{H}$, $\phi=-\vec{E}+i\vec{H}=-(\vec{E}-i\vec{H})$.
Some more aspects discussed in \cite{dvoe:1993b} are helpful to guide
and compare the different reps:
\begin{itemize}
\item[-] Valeriy mentions the fact that, according to Weinberg,
\cite{dvoe:1993b}, eq.~(3), $\Psi$ can be transformed to the
equations for left- and right-circularly polarized radiation 
in the case $S=1$ of massless fields\footnote{Whereas in the 
2(2S+1)-formalism this relates to the identification of $\chi$ 
and $\phi$ with electromagnetic field components as given above,
in our picture when considering the geometrical base objects 
this relates to Klein's fundamental Complexe of $P^{5}$ 
\cite{dahm:MRST8} with 3$\oplus$3 handedness \cite{dahm:MRST8},
or \cite{kleinHG:1926}, p.~99, with respect to involutions of
linear Complexe.}. See also his discussion in \cite{dvoe:1994}, 
eqns.~(2.24) and (2.25), where $\vec{E}$ and $\vec{H}$ are 
Pauli vectors, and we may stress Lie transfer \cite{dahm:MRST7}
to locate the six components in line space.
\item[-] The 4-vector Lagrangean density is constructed by means
of the field tensor and its dual (\cite{dvoe:1993b}, eq.~(6)),
and by using a 'current' $j^{\mu}$ which has to be discussed 
separately.
\item[-]The energy-momentum tensor is given by the vector 
Lagrangean and the 6-dim rep $\Psi$, only, and with respect
to the current tensor, there exist Sudbery's 'duality rotations' 
which mix the field strength tensor and its dual (\cite{dvoe:1993b},
eq.~(15))\footnote{In other words, we have to discuss the 
position of two dual lines given by a special linear Complex
and represented by $F_{\mu\nu}$ and its conjugate, and we 
thus enter the geometrical discussion of Congruences/ray 
systems and of conjugation with respect to polar and null
systems, in $P^{3}$ as well as in $P^{5}$.}. In the source-free
case, the action is invariant under this 'duality transformation'
(\cite{tiwari:2012}, eqns.~(36) and (37)), and by Noether's 
theorem one obtains the conservation laws for a symmetric 
energy-momentum tensor (\cite{tiwari:2012}, p.~7).
\item[-]One can find new 'gauge' transformations 
$F_{\mu\nu}\longrightarrow F_{\mu\nu}
+\partial_{\nu}\Lambda_{\mu}-\partial_{\mu}\Lambda_{\nu}$
(\cite{dvoe:1993b}, eq.~(31)), i.e.~by adding an antisymmetric
tensor $A_{[\mu\nu]}=\partial_{\nu}\Lambda_{\mu}-\partial_{\mu}\Lambda_{\nu}$.
Later, this addition of a second Complex rep $A_{[\mu\nu]}$
yields\footnote{See also the related aspects discussed in
sec.~\ref{sec:quadratic}.} the discussion of pencils of 
linear Complexe, finding lines within the pencil or line 
sets of quadratic Complexe, or analyzing Congruences.
\item[-] In \cite{dvoe:1993c}, eq.~(3), the decomposition 
of $\Psi$ is associated with vector and pseudo-vector parts. 
Here, we want to mention only the geometrical limit used in 
sec.~\ref{sec:minkrev}, or discussed in depth in \cite{kleinHG:1926},
\S16, which decomposes the antisymmetric line coordinates
{\it only after} having performed the affine or Euclidean 
limit. So it is the transition from homogeneous coordinates
to affine or Euclidean coordinates, which introduces the 
notion of vector and pseudo-vector, or of parity, which 
is subsidiary in homogeneous and projective theories 
because projective transformations in general mix the
coordinates. There is no 'absolute element' yet, or a
distinguished coordinate $x_{0}$ or its dual plane to 
introduce parity, or a linear orientation. In other 
words, as long as we perform PG, we benefit from pretty
'symmetric' line coordinates (because twofold in point
reps) with respect to the order of composition, i.e.~using
{\it two} homogeneous point coordinates and their antisymmetry
when exchanging the points. It is only {\it after} having 
fixed an absolute plane (or even after having defined 
an absolute circle within this plane) that we have to 
take care of polar and axial vectors with their 'new' 
asymmetry of being linear or quadratic in point coordinates
because now the absolute element is fixed and absorbed
in the coordinate definition (i.e.~by formally setting
$x_{0}=1$ in the affine case, or by switching to Euclidean
coordinates which 'absorb' the linear singularity $x_{0}=0$
in the fraction by their very definition).
\item[-] \cite{dvoe:1994}, eq.~(3.31), defines a 'new magnetic 
momentum vector' which is traced back to another skew tensor
$\Sigma_{(\mu\nu)}$, and the parameter triples $\vec{\Theta}$,
$\vec{\Xi}$ of $\hat{H}$ are derived from the three 0- or 
4-components on the one hand, 
$\vec{\Theta}=(\Sigma_{(41)},\Sigma_{(42)},\Sigma_{(43)})$,
and from 
$\vec{\Xi}=(\Sigma_{(23)},\Sigma_{(31)},\Sigma_{(12)})$ on the
other hand. This time, this skew structure is constructed by 
means of the Pauli-Lubanski-vector $W_{\mu}(\vec{p})$, and
$\Sigma_{(\mu\nu)}$ reads as
$\Sigma_{(\mu\nu)}(\vec{p})=\frac{1}{2}\left(
W_{\mu}(\vec{p})W_{\nu}(\vec{p})-W_{\nu}(\vec{p})W_{\mu}(\vec{p})
\right)$
\end{itemize}

We'll address both major aspects from above -- the Lie 
algebra and the 6-dim rep of electromagnetic field -- 
later after having established more geometrical and 
analytical background in sec.~\ref{sec:weinbergNEW}.
For now, by summarizing the standard reps, it is 
important to note that a generalization in these days
is usually achieved by generalizing the 'quantum numbers'
of the reps associated to the spin algebras. However, 
we'll see later that identifying the 6-dim structure 
$\Psi$ geometrically leads back to PG and yields 
generalizations in multi-linear algebra and algebraic
geometry, and it is thus more interesting for us with
respect to physics to follow the geometrical trail, 
parts of which we've addressed here and in the cited
publications, than generalizing the rep dimensions 
of su(2) reps, or switching to irreps with higher 
multiplicities and 'quantum numbers'. We have given
examples in the physics of the pion-nucleon-delta 
system \cite{dahm:diss} which due to use of SU(4) 
can be traced back to spatial geometry \cite{dahm:MRST4},
\cite{dahm:MRST6}, and there are lots of further 
examples where current formalisms may be derived 
easily from classical geometry of $P^{3}$ and $P^{5}$.

\section{Standards revisited}
\label{ch:std-rev}
%Original Geometry

Throughout sec.~\ref{sec:known}, we've presented the major
foundations of the usual and known formalism to treat relativity
(without claiming completeness of the presentation!). And
so far, we have introduced only few comments with respect
to this description of physics by means of the point picture
during this overview on 'standards'. However, before revisiting
some central aspects of this rep theory (expressed in point
coordinates) and working the switches as we've announced in
the beginning, in order to be not misunderstood or misinterpreted, 
we want to emphasize that we do {\it NOT} want to argue or even
reject the wealth of experimental evidence so far on relativity
confirming Einstein's physical picture! In contrary, we argue 
that switching to a description and treatment in terms of PG 
and Complex geometry yields the observed effects more naturally,
and that we may realize common textbook notation just by switching
to antisymmetric point or plane reps of lines (or Complexe) 
in 3-space at any time. Our following arguments relate to the
nowadays usual rep theory in terms of point reps, the 4-vector
calculus and differential geometry which cover only limited 
aspects of a full-fledged physical description.

From the viewpoint of projective geometry, it is well-known 
that a point description without plane reps is far from 
complete to describe even 3-space and its physics because
it lacks the oldest known transfer principle(s): duality,
or reciprocity. Moreover, in order to introduce the metric
or an absolute quadric, one usually needs additional 
assumptions e.g.~on absolute elements or on the polar system,
which are typically put in by additional requirements, and 
they do not result a priori from geometry like in the case 
of projective geometry when using the Cayley-Klein mechanism. 
However, here it is not our line of reasoning that switching 
to line and Complex geometry cures {\it all} the open questions
but first of all, we see this geometry as a very efficient 
ordering and rep scheme to collect well-known point descriptions,
and which allows to simplify, complete and generalize then
geometrically. Moreover, a description of the observation 
process is automatically included if we think of optical 
lines ('rays') connecting observer and observation in 
order to represent physical observations \cite{dahm:MRST3}, 
\cite{dahm:MRST4}. In linear scenarios, this concept comprises
or even requires classical projective geometry; in non-linear
scenarios, one has to include objects of higher orders\footnote{An
example we've mentioned already a couple of times, is the
coupling of the mesons to the nucleon-delta systems where
it is the 20-dim, $3^{rd}$ order symmetric rep which serves
as a basis for vector {\it and} axial currents which we 
have shown to be necessary in order to calculate the axial 
coupling constant. Other examples like skew cubics, \PLu's
\cite{plueckerNG:1868} or Kummer's (see e.g.~\cite{kummer:1866}
or \cite{hudson:1905}), or K3 surfaces are well-known as well.}
and classes as well as non-linear transformations like 
in advanced geometry or in algebraic geometry.

To approach an alternative -- and more general -- description
without getting lost in the enormous world of projective, 
or algebraic geometry, we use the preceding section as a
certain guideline to transfer the aspects here in sec.~\ref{ch:std-rev}
step by step and subsume them into the more general framework.
In sec.~\ref{sec:reinterpretation} -- after having switched 
to the more general framework -- we'll discuss some more 
aspects, however, without claiming completeness of the
outline.

\subsection{Minkowski Revisited}
\label{sec:minkrev}
As a first example, we may rewrite the linear Complex 
$\sum a_{\alpha}p_{\alpha}$, $1\leq\alpha\leq 6$, from 
above in terms of homogeneous point coordinates with 
$x_{\mu}$ and $y_{\mu}$, $1\leq\mu\leq 4$, and we associate
the $P^{5}$ coordinates $p_{\alpha}$ to \PLu's line coordinates
in $P^{3}$ by\footnote{In order to compare directly to Minkowski's
notation, eq.~(\ref{eq:staudemapping}) shows the mapping due 
to Staude \cite{staude:1905}, \S59.1, p.~322, which differs 
from our usual notation in e.g.~\cite{dahm:MRST3} or later,
where we typically use the point coordinate $x_{0}$ instead 
of $x_{4}$. This change introduces a sign in three of the 
line components $p_{4}$, $p_{5}$, and $p_{6}$ by reordering
the indices of the antisymmetric line rep in ray coordinates.
Whereas the \PLL condition itself is not altered, one has to 
keep track when working with Complex parameters, signatures 
and polar conditions in $P^{5}$, etc., and related linear 
coordinate sets. Note also, that Klein sometimes varies his 
notation at all in that he uses also other coordinates than
$x_{0}$ or $x_{4}$ as linear absolute elements. This notation 
in terms of $P^{5}$ coordinates $p_{\alpha}$ also introduces 
the notation of two coordinate triples, usually denoted by 
$k=1,2,3$ and $\overline{k}=4,5,6$ which comprise 3-dim forces
and moments in the force picture, or relate to axial and polar
'vectors' in the point picture. Please note also, the connection
to the decomposition of $M_{\mu\nu}$ into $\vec{M}$ and $\vec{N}$
as discussed in \cite{joos:1962}, or \cite{alfaro:1973}, see 
also sec.~\ref{sec:standard-minkowski}.}
\begin{equation}
\label{eq:staudemapping}
\begin{array}{l}
p_{1}\longleftrightarrow p_{23},\,
p_{2}\longleftrightarrow p_{31},\,
p_{3}\longleftrightarrow p_{12},\\
p_{4}\longleftrightarrow p_{14},\,
p_{5}\longleftrightarrow p_{24},\,
p_{6}\longleftrightarrow p_{34}
\end{array}
\end{equation} 
A general linear Complex $\textfrak{C}$ in the ray representation
may then be written according to
$\textfrak{C}=\sum a_{\alpha}p_{\alpha}\,\longrightarrow\,\sum f_{\mu\nu}p_{\mu\nu}$
with line coordinates $p_{\mu\nu}=x_{\mu}y_{\nu}-x_{\nu}y_{\mu}$, $1\leq\mu,\nu\leq 4$,
and $a_{\alpha}\in\mathbbm{R}$, i.e.~by
\begin{equation}
\begin{array}{rl}
\textfrak{C}=
& f_{23}(x_{2}y_{3}-x_{3}y_{2})
+ f_{31}(x_{3}y_{1}-x_{1}y_{3})\\
+ & f_{12}(x_{1}y_{2}-x_{2}y_{1})\\
+ & f_{14}(x_{1}y_{4}-x_{4}y_{1})
+ f_{24}(x_{2}y_{4}-x_{4}y_{2})\\
+ & f_{34}(x_{3}y_{4}-x_{4}y_{3})
\end{array}
\end{equation}
which we can compare to eqns.~(\ref{eq:MinkComplex}) and 
(\ref{eq:MinkComplexNew}). It is evident that eq.~(\ref{eq:MinkComplex})
above (up to the additional phase $i$ of the fourth component 
which has been introduced by Minkowski artificially into
point space to change the signature of the quadric) yields
exactly the very definition of a line Complex using ray 
coordinates\footnote{The linear Complex expressed in \PLL
coordinates with {\it real} coefficients/coordinates 
corresponds to an SO(3,3) invariant quadric and to an 
associated SO(3,1) invariant quadric in quaternary 
('4-dim') point coordinates \cite{dahm:MRST3}, sec.~1.6.
So with respect to introducing additional phases $'i'$ 
one should be careful about interpretation and use of 
coordinates! Moreover, we are obviously involved once 
more in a phase discussion like in the case of Weinberg's
'unitary reps' of the Lorentz group which we've discussed
in sec.~\ref{sec:weinberg}. This emphasizes the need 'to 
straighten' and unify this rep theory as we'll try to do 
in sec.~\ref{sec:weinbergNEW}.} if we replace $y\longleftrightarrow u$.
Therefore, Minkowski's claim of form invariance when 
rewriting the result {\it after} Lorentz transformations
in terms of primed components like in eq.~(\ref{eq:MinkComplexNew})
reflects nothing but the requirement of the invariance
of the respective linear Complex $\textfrak{C}$. So 
according to the components (or the coordinates) 
$f_{\alpha\beta}$ of the linear Complex $\textfrak{C}$,
we have to consider transformations which leave this 
Complex invariant. This can be accomplished by 
transformations leaving $\textfrak{C}$ invariant as 
a whole e.g.~by interchanging the line coordinates 
and thus constraining the components $f_{\alpha\beta}$,
or by leaving line coordinates individually invariant, 
etc. Minkowski's requirement thus has to be associated
with the second, restricted case which results in the 
stricter constraints
$f_{\alpha\beta}\longrightarrow f'_{\alpha\beta}$, and 
$p_{\alpha\beta}\longrightarrow p'_{\alpha\beta}$.

In all cases, 
we may use Klein's 'Erlanger Programm' as the appropriate 
framework (or tool-set) for continuous as well as for
discrete symmetries. One example is $r+\sigma=0$, $r$ and
$\sigma$ denoting inhomogeneous \PLL coordinates \cite{dahm:MRST7} 
\cite{lie:1872}, which yields Lie transfer when stabilizing
the linear Complex $r+\sigma=0$. In terms of homogeneous
coordinates (\cite{lie:1896} p.~283), $r+\sigma=0$ can be 
represented by 
$r+\sigma=\frac{p_{41}}{p_{43}}+\frac{p_{23}}{p_{43}}
=\frac{p_{01}}{p_{03}}+\frac{p_{23}}{p_{03}}=0$
which -- besides the special r\^{o}le of $p_{01}+p_{23}$ 
-- also emphasizes the special r\^{o}le of the individual
line coordinate $p_{43}=p_{03}$ with respect to the 
definition of inhomogeneous line coordinates $(r,s,\rho,\sigma,\eta)$.
This can be seen using Lie's rep of inhomogeneous versus 
homogeneous line coordinates (\cite{lie:1896}, p.~283),
\begin{equation}
\label{eq:inhomoglinecoord}
\begin{array}{ccc}
r=\frac{p_{41}}{p_{43}}=\frac{p_{01}}{p_{03}} & \,,
& s=\frac{p_{42}}{p_{43}}=\frac{p_{02}}{p_{03}}\,,\\[1mm]
\rho=-\frac{p_{31}}{p_{43}}=-\frac{p_{31}}{p_{03}} & \,,
& \sigma=\frac{p_{23}}{p_{43}}=\frac{p_{23}}{p_{03}}\,,\\[1mm]
\eta=\frac{p_{12}}{p_{43}}=\frac{p_{12}}{p_{03}} & \,, & 
\end{array}
\end{equation}
where we've expressed the coordinates in addition by using
the 0-component-notation instead of 4-components. It is
noteworthy, that the six coordinates $p_{\alpha\beta}$ 
transform homogeneously whereas the five coordinates 
$(r,\rho,s,\sigma,\eta)$ undergo projective transformations
if space itself is transformed by projective transformations
(see \cite{lie:1896}, p.~285, Satz~5). Moreover, 
eq.~(\ref{eq:inhomoglinecoord}) visualizes perfectly that 
the line coordinates involving $0$-components of point reps
determine the slopes $r$ and $s$ of the lines in terms of
inhomogeneous coordinates, i.e.~we may associate $p_{0\alpha}$
(or $r$ and $s$) with the direction (information) of the 
line.

Lie gives the corresponding definitions in terms of point
coordinates by
\[
\begin{array}{ccl}
p_{12}=x'y''-y'x'' & \longrightarrow & p_{12}=x_{1}y_{2}-x_{2}y_{1} \\
p_{23}=y'z''-z'y'' & \longrightarrow & p_{23}=x_{2}y_{3}-x_{3}y_{2} \\
p_{31}=z'x''-x'z'' & \longrightarrow & p_{31}=x_{3}y_{1}-x_{1}y_{3} \\
p_{41}=x''-x' & \longrightarrow & p_{01}=y_{1}-x_{1}=x_{0}y_{1}-x_{1}y_{0} \\
p_{42}=y''-y' & \longrightarrow & p_{02}=y_{2}-x_{2}=x_{0}y_{2}-x_{2}y_{0} \\
p_{43}=z''-z' & \longrightarrow & p_{03}=y_{3}-x_{3}=x_{0}y_{3}-x_{3}y_{0}\,.
\end{array}
\]
Whereas on the lhs we've used Lie original notation, we've 
re-expressed the rhs using $()'\longrightarrow x$ and $()''\longrightarrow y$
with quaternary homogeneous coordinates of the points $x$ 
and $y$, and we've used 0-components with $x_{0}=y_{0}=1$ 
like in the affine picture. So on the one hand, it is evident
that line coordinates $p_{\alpha\beta}$ comprise Minkowski's
transformation theory. On the other hand, triples $\sim$ 
$p_{0j}$ seem to be related to 'polar vectors' whereas 
triples $\sim$ $(p_{12}, p_{23}, p_{31})$ seem related to
'axial vectors', or moments. This decomposition $3\oplus 3$
of the 6-dim homogeneous line rep, however, occurs only AFTER
having performed the transition from homogeneous to affine,
or Euclidean, coordinates. So with respect to discussions
of 'parity' symmetry, they have to be applied (or located) 
in rep theories only AFTER having performed this transition,
and they are misleading in homogeneous rep theory. This can 
be seen easily if we exchange opposite lines (or edges) of
the fundamental tetrahedron by simple coordinate exchange
without altering the geometry. Moreover, linear combinations
of the edges like $(p_{01}\pm p_{23})$, $(p_{02}\pm p_{31})$,
or $(p_{03}\pm p_{12})$ which we've used with respect to
Congruences or Onsager theory (see e.g.~\cite{dahm:MRST9},
sec.~4.3), remain invariant by themselves, and thus yield
additional symmetries.

Earlier, we've re-expressed Lie's transfer principle already
in terms of Pauli matrices and thus attached the usual 
quantum notion to line geometry of $P^{3}$ \cite{dahm:MRST7}
\cite{dahm:MRST9}. Moreover, we've thus founded the constructing
scheme of Clifford algebras $\sigma^{(1)}\otimes\sigma^{(2)}\otimes\ldots\sigma^{(n)}$
to projective constructions based on lines (and Complexe)
instead of point notion and Euclidean concepts as usual.

In terms of transformation or group theory, we are thus 
at the very origin (or at least close to the heart) of 
symplectic transformations and of quadratic constraints 
in $P^{3}$ which originate e.g.~from the \PLu-Klein
quadric of $P^{5}$. 

So the major guideline of our reasoning within the next 
sections is the idea to switch back to line and Complex 
geometry in order to catch the complete background instead
of discussing only certain aspects of point reps and their
differential geometry. This automatically reestablishes 
the almost lost connection of formal and technical algebra
to physics in terms of null systems and forces, and thus 
interconnects experiments of the real world and well-known
geometry. As such, recalling the two {\it different} approaches
by Grassmann and \PLL to higher dimensional reps of geometry,
we want to underpin Lie's summary \cite{lie:1896}, p.~274/275,
and emphasize \PLu's achievements, i.e.~to express and understand
higher-dimensional 'spaces' and their respective coordinate 
reps by appropriate $n$-dim base elements of the respective 
geometry. Practically, this is established by transfer principles
connecting the different reps while changing the coordinate
{\it interpretations} and usually the dimension of the reps,
too. A simple example can be taken from Lie transfer above, 
or already from \PLu's observation \cite{pluecker:1838} when
connecting line reps with three types of associated second 
order surfaces which naturally explains otherwise mystified
rep relations like the wave-particle dualism. We think, this
identification and reasoning as well as the use of transfer 
principles is much closer to physical reasoning than using 
$n$-dim {\it point} sets in abstract spaces as exercised 
nowadays e.g.~in string or gauge theories.\\

\subsection{Line and Complex Geometry Revisited}

To depart from this introductory example in sec.~\ref{sec:minkrev},
featuring coordinate properties and the assembly of lines 
and linear Complexe, we now want to collect at least few 
of the technical debris floating around and relate them 
briefly to known geometrical concepts, whereas in the next 
section -- based on projective and especially on line and 
Complex geometry -- we want to catch up with the guiding 
topic of this book, with the future of relativity, however,
based on old physical principles and geometrical roots.\\

So after having identified Minkowski's 'six-vector' (or 'vector
of second kind') geometrically in sec.~\ref{sec:minkrev} with
a special line set in 3-dim space (a special linear Complex, 
or -- equivalently -- with a special physical system, a null 
system), it is obvious that people have introduced a special 
representation theory by means of 4-vector calculus (and 
related tensor calculus) to treat a small (sub-)aspect of line
geometry by an independent, different and a priori disconnected
formalism. Geometry and the ray representation of lines in terms
of homogeneous fourfold point coordinates have been replaced 
by 4-vector calculus, tensor algebra and differential geometry,
and we are faced with 'special relativity' comprising implicit
geometrical assumptions and singularities which cannot be 
resolved from within the 4-vector rep theory without additional
and non-trivial assumptions. We've stressed in sec.~\ref{sec:weinberg}
already Weinberg's rep approach and emphasized the case $j=1$ 
because we are going to compare this with line geometry and 
the case of a special linear Complex. However, as we have 
discussed already above, it is mandatory to take care with
respect to {\it the interpretation} of the point coordinates,
and in the same manner, we have discussed Minkowski's paper 
and central aspects of our reasoning in more detail elsewhere
\cite{dahm:MRST8}.

In another paper \cite{dahm:MRST5}, sec.~3.2, we have given 
analytic proofs of the invariance properties even of individual
line coordinates $p_{\mu\nu}$ which comprise the 'Lorentz 
transformations' in point space known nowadays from textbooks.
As such, by using homogeneous point coordinates, the usual 
Lorentz transformation 'leaving $x$ and $y$ invariant, and 
mixing $z$ and $t$' corresponds to the two individual invariance
properties of the line coordinates $p_{12}$ and $p_{03}$ (or 
$p_{34}$) which according to their definition above by 
$p_{\mu\nu}=x_{\mu}y_{\nu}-x_{\nu}y_{\mu}$ may be represented
as 2$\times$2-determinants, too\footnote{See eq.~(\ref{eq:detLinien})
in sec.~\ref{sec:remarkstransformations} on transformations 
later on. In this section, we provide the full discussion 
because up to this section, we'll have developed the context
to discuss first aspects of Complex geometry beyond the old
formalism and phenomenology.}. However, the 2$\times$2-determinants
-- besides exposing further invariance properties -- do 
not change under identity transformations or hyperbolic 
mixture within the 2$\times$2 blocks \cite{dahm:MRST5}.
And because geometrically the two line coordinates $p_{12}$
and $p_{03}$ build two opposite, non-intersecting sides 
of the fundamental coordinate tetrahedron, each invariance
corresponds to an invariant coordinate in line geometry, 
or to the invariance of a certain Congruence\footnote{More
precise: a 'ray system' ('Strahlensystem erster Ordnung 
und erster Classe').} in Complex geometry. As such, a 
switch to rep theory in terms of line geometry yields 
the additional benefit to avoid artificial invariance
requirements in point space and yields an easier and 
more evident treatment \cite{dahm:MRST9}, sec.~4.C. 
It is much easier to discuss invariance properties such 
as $(p_{12}, p_{03})\longrightarrow (p'_{12}, p'_{03})$ 
in terms of line coordinates than figuring out the antisymmetric
(and invariant) rep parts out of an arbitrary point rep 
tensor (or event point space dependent field reps) from
a given point space object. Last not least, according 
to their very construction, the line reps are the reps
most close to {\it linear} rep theory, and we may use 
Hesse transfer \cite{hesse:1866} and binary forms 
\cite{clebsch:1872} to treat various point or spinor 
reps, and various related aspects of PG. Although 
Poincar{\'{e}}'s/Minkowski's 4-vector calculus yields
a symbolism and tensorial rules on how to proceed and
calculate, it is a symbolism attached to the special
identifications above, and although it is capable to 
represent special geometrical aspects by the 4-dim rep,
it is not capable to cover the geometrical aspects of 
null systems and Complex geometry of $P^{5}$ completely.
It is stuck in the affine picture using point-based algebra 
reps and an appropriately related formalism (and symbolism).
As such we think, moving back to line and Complex geometry
yields the profound background especially when generalizing
later to higher orders or classes, and to transfer principles
and non-linear behaviour, i.e.~to the general aspects of 
projective and algebraic geometry. So the usual discussion
of 'Lorentz transformations' or special relativity can be
subsumed without loss of generality under the invariance 
requirement of linear Complexe, and as such, it is part 
of Complex geometry and projective geometry which provides 
a much larger mathematical framework. So a necessary 
generalization -- or the discussion of the future of 
relativity -- needs this more profound basis to overcome
and supersede the concepts of affine and differential 
geometry.

\subsection{Affine Picture Revisited}
It is also evident from the physical viewpoint, that one
shouldn't naively put $c=1$ (and neglect the physical 
dimension of this velocity!) in order to obtain a 'time'
and some 'space-with-time' mixture, or nowadays often 
for short 'space-time'. $c$ has physical dimensionality,
and if we want to interpret $c$ phenomenologically as 
an upper bound on velocities (which paves the path to 
hyperbolic models), the only consequence we see is that
if we interpret\footnote{With respect to the use of the
Hamilton formalism \cite{study:1905}, we feel free to 
do so at least with respect to equal time considerations
and the usual physical reasoning.} the three space
coordinates $x_{i}$ in the same manner by $x_{i}=v_{i}t$,
one can establish {\it for the same time interval $t$} 
a comparison of the velocities $v_{i}$ with their upper
bound $c$. However, in order to compare, one has to 
compare quantities of the same physical dimension! We 
have discussed some of those aspects in \cite{dahm:MRST3},
sec.~3.5, and derived the Lorentz factor. 

Moreover, we've pointed out \cite{dahm:MRST1}, \cite{dahm:MRST6},
that we thus catch up with Smorodinskij's beautiful treatment
of velocities and the geometry of velocity space \cite{smorod:1965}.
What is open today for discussion is a certain restorative 
return on the occurrence and interpretation of second order
surfaces which Study has summarized \cite{study:1905} in 
relation with normal congruences and focal theory although
they are deeply related to the underlying concept of Hamilton's
formalism. This can be addressed here only very briefly but 
has to be worked out in detail later on due to its importance
with respect to its universal applications throughout most 
fields of physics and Lie theory formalizing invariant theory
and symmetry transformations. Nevertheless, we've discussed 
the 'corrected' interpretation of 4-dim 'momentum' reps 
already in terms of plane coordinates of the Hesse form
\cite{dahm:MRST6}, i.e.~by '4-dim' normals, and it is the
picture of normals of planes that suggests to put focus on 
geometry and the related tetrahedral Complex of surfaces. 
Last not least, considering $ct$ as '0-component' with 
respect to Lorentz transformations, this coordinate is 
{\it mixed} with one of the spatial coordinates related 
to the axis or direction of motion, usually $z=x_{3}$. 
So we do {\it not} encounter an affine transformation 
(which should keep $x_{0}$ invariant), but we need projective 
transformations and PG to access the picture\footnote{See
sec.~\ref{sec:nullsystems}, and there especially figure~\ref{fig:complex}!}
of an invariant normal plane $x$, $y$ and mixture of 
$x_{0}$ and $z=x_{3}$ with respect to the direction of
motion. There is even the possibility to introduce new
coordinates with quadratic dependence \cite{dahm:MRST3}
and switch to a consistent formalism.\\

\subsection{Coordinate Notation and Terminology}
This forwards our discussion to Klein's paper \cite{klein:1910} 
featuring the remarkable break in terms of the transition from
'ordinary projective geometry' of 3-space -- according to Klein's
use of ordinary homogeneous coordinates mostly from the viewpoint
of an affine geometry -- to the identification of four 'space-time
coordinates' in the second part of \cite{klein:1910} and by 
introducing five homogeneous coordinates $x_{\beta}$, $1\leq\beta\leq 5$.
Now, of course, it is formally and algebraically/analytically possible
to start from four rectangular coordinates and by generalizing the 
{\it linear} affine approach like discussed in sec.~\ref{sec:standard-klein1910}, 
i.e.~by thus introducing a {\it linear} absolute element $x_{5}=0$
into this space. However, this opens a lot of room to figure out the 
additional {\it implicit intrinsic} assumptions of these definitions
and to discuss their relation to physics, the more as they impose 
further constraints and restrict objects and transformations.

Although we have already discarded to pursue this trail of using
$P^{4}$ further, we nevertheless want to point to some historical
but contemporary aspects of the coordinate notation $t$ in 
conjunction with the three space coordinates $x$, $y$, and $z$
in literature\footnote{We are not enough qualified to discuss 
this aspect exhaustively, and moreover, here, we think it's 
not the place to discuss all the historical and especially the
important sociological aspects of that time and Minkowski's 
community profoundly. Nevertheless, we want to mention these
aspects, and we hope that some science historians or sociologists
have interest to look deeper into these interesting relations 
and the historical development. As for us, today we can state 
only the occurrence of this notation in the three given book 
references before 1908/1910, but we have had no time so far to 
follow the occurrence of the coordinate notation $t$ for the 
fourth ordinary homogeneous point coordinate back to its origin(s) 
in mathematical literature.}. So to a certain extent, it is known
from letters exchanged between Lorentz and Poincar{\'{e}} \cite{LorPoinc},
that both discussed an additional quantity $t'$ introduced into 
physical equations and descriptions, however, originally not 
connected to a physical 'time' identification. In addition, 
at the beginning of the $20^{th}$ century, there existed a 
notation already in mathematical textbooks (see e.g.~\cite{hudson:1905},
\cite{staude:1905}, \S 47, or \cite{hilbert:1909}) to describe
the fourth component of {\it ordinary} homogeneous coordinates
by the letter $t$, following $x$, $y$, and $z$. This seems to 
be the common notational convention of that time when using $x$,
$y$, or $z$, instead of index notation, because on the other hand, 
the tetrahedral homogeneous coordinates were usually denoted by 
$x_{\beta}$, $1\leq\beta\leq 4$, or $0\leq\beta\leq 3$, dependent
on the author. In Hudson's case \cite{hudson:1905}, there is some
more evidence of such a common notion because whereas in the first
sections (see e.g.~\S 2 or \S 10) he uses 'homogeneous coordinates'
$x$, $y$, $z$, and $t$ to denote individual point coordinates, 
beginning with \S 11 he explicitly argues to switch to tetrahedral
coordinates which he also denotes by $x$, $y$, $z$, and $t$, and 
he requires to reinterpret this coordinate notation. Staude 
\cite{staude:1905} treats the notation much stricter, compare
e.g.~\S 47 on ordinary homogeneous coordinates versus \S\S 57
and 59 on tetrahedral homogeneous coordinates. Note with respect
to the interpretation of 'coordinates' and 'masses' especially, 
that he recalls implicitly \MOo's barycentric ideas of weights 
and masses of such coordinate definitions!

So according to our current understanding, it is at least worth 
to keep track of the parallel but {\it different} notations 
in physics and mathematics, and take care of the coordinate
interpretation and ambiguities. It is definitely {\it NOT} 
sufficient, to naively put $c=1$, neglect dimensional considerations
and hope for unique formal treatments in mathematics. So it 
is part of our ongoing work to separate the various, sometimes
overlapping coordinate interpretations in detail, and we see 
projective geometry on the one hand as an appropriate tool 
to handle the different coordinate interpretations versus 
the related Euclidean and non-Euclidean geometries. On the 
other hand, it allows to consistently complete point descriptions
of 3-space by plane descriptions and duality, or if we switch
instead to the equivalent line representation (by lines and 
their conjugated lines) we may describe physics and comprise
established formalism like the examples we'll discuss in 
the next section, or like in the case of gauge formulations 
\cite{dahm:MRST10}. % Prague 2019

\section{A Reinterpretation -- back to the roots!}
\label{sec:reinterpretation}
% falsche Koordinateninterpretation; falsche Interpretation der
% Zeit; Raum und Geschwindigkeit sind über die Zeit funktional
% verknüpft, im einfachsten Fall konstant oder auch direkt linear
Throughout the preceding section, we've argued with respect 
to selected aspects and by various selected examples, only. 
However, as there exists a well-defined and well-established
general context in terms of Lie's contact geometry \cite{lie:1896},
which -- at least in its original form and its deep connection 
to physical descriptions -- seems to be forgotten, we want to
recall briefly this framework in the next subsection. Afterwards,
we want to switch to this framework and invert the first two
sections in that we want to arrange and integrate the aspects
mentioned above into their appropriate places\footnote{We've
applied this background already to approach and discuss the 
treatment of gauge theories in terms of Complexe and $P^{5}$
geometry \cite{dahm:MRST10}.}.

\subsection{Historical Context and Null Systems}
\label{sec:nullsystems}
Lie, however, has pointed out (for details, see \cite{lie:1896},
ch.~6, \S3) in the context of null systems that Pfaff's equations
have ever played an important r\^{o}le. Using Euclidean coordinates,
he summarizes the historical development and derives -- using metric
conditions in 3-space -- that the most general infinitesimal motion 
in 3-space when expressed in terms of Euclidean coordinates $x$, $y$,
and $z$ is given by
\begin{equation}
\label{eq:spacedisplacement}
\begin{array}{rl}
\delta x =
& \left(Bz-Cy+D\right)\,\delta t\\
\delta y =
& \left(Cx-Az+E\right)\,\delta t\\
\delta z =
& \left(Ay-Bx+G\right)\,\delta t\,.
\end{array}
\end{equation}
This reflects the theorem attributed to Euler (1775) and D'Alembert (1780)
that every motion in 3-space can be replaced by appropriate infinitesimal
rotations by the coordinate axes and appropriate translations along the
axes. Later, it has been shown that such a description of motion can be
replaced by a single screw (see e.g.~\PLL \cite{pluecker:1865}, 
\cite{pluecker:1866}, Ball \cite{ball:1876}, Lie \cite{lie:1896}, Klein
\cite{klein:screws}, \S2, or Klein and Sommerfeld \cite{kleinso:1897}, 
\cite{kleinso:1898}), and others.

In the next step, Lie argues that for the 1-dim set of planes
\begin{equation}
\label{eq:complexplane}
Ax+By+Cz=\mathrm{const}
\end{equation}
due to the displacements given by eq.~(\ref{eq:spacedisplacement}),
the planes given by (\ref{eq:complexplane}) undergo a constant 
increment on the lhs of their respective equations as can be 
seen immediately by $A\delta x+B\delta y+C\delta z= (AE+BE+CG)\delta t$.
Thus, they are displaced in parallel\footnote{This is why we 
may use {\it the normal} by means of Hesse's plane rep to describe
translations, fluxes, currents by a 4-dim rep, or the '4-momentum',
respectively, as long as we choose a common point 'to attach' the 
normal to the plane, i.e.~a point being member of a common line. 
Note that besides the usual tangent constructions -- thus founding
on polar constraints and reciprocity as transfer principle intrinsically
-- we may also use null points and their associated linear Complexe 
e.g.~in terms of skew matrix reps or Lie algebras.} by the infinitesimal
motions given by eq.~(\ref{eq:spacedisplacement}).

For points moving {\it perpendicularly} to the planes (\ref{eq:complexplane}),
their displacements $\delta x$, $\delta y$, and $\delta z$, thus should
be proportional to $A$, $B$, and $C$, respectively, and Lie thus obtains
the equations
\begin{equation}
\label{eq:orthogonal}
\begin{array}{rl}
\rho A = & Bz-Cy+D\\
\rho B = & Cx-Az+E\\
\rho C = & Ay-Bx+G\,,
\end{array}
\end{equation}
in order to derive the important equation 
\begin{equation}
\label{eq:importantequation}
AD+BE+CG=\rho\left(A^{2}+B^{2}+C^{2}\right)\,.
\end{equation}
Whereas Lie uses this equation to determine the 'Factor 
$\rho$' and a certain 'reduced' geometry, we want to 
emphasize already here its connection to the Complex 
parameter\footnote{There exists also an 'infinitesimal
version' by means of partial derivatives \cite{lie:1896}.
Moreover, it is evident that the lhs of eq.~(\ref{eq:importantequation})
corresponds to the \PLL condition, and that we have 
to treat the case \protect{$A^{2}+B^{2}+C^{2}=0$} by
involving complex numbers, i.e.~by invoking absolute
elements and 'spinors' \cite{dahm:MRST7}. Last not 
least, rewriting (\ref{eq:spacedisplacement}) in terms
of four (ordinary) homogeneous coordinates, the {\it 
common} $\delta t$ on the rhs of the three equations
reflects the same coordinate in the absolute plane. 
Replacing this related {\it space} coordinate by $ct$,
and the respective coordinate displacements $\delta x$,
$\delta y$, and $\delta z$ by their appropriate 
counterparts in terms of velocities $v_{i}t$ yields
an alternative approach to our derivation of the 
Lorentz factor \cite{dahm:MRST3}, \cite{dahm:MRST5},
and emphasizes our proposal to understand Lorentz 
transformations as a line (or Complex) counterpart 
when working with point coordinates in 3-space.}. 
Lie then emphasizes the reduction of this equation set
from three to two equations by an appropriate choice 
of $\rho$, which in turn reduces the geometrical setup
to a straight line being orthogonal to the plane(s) 
given by eq.~(\ref{eq:complexplane}). This straight
line, as given by (\ref{eq:orthogonal}), is invariant 
and transformed (or translated) into itself under the
transformations (\ref{eq:spacedisplacement}) as long 
as $A^{2}+B^{2}+C^{2}\neq 0$.

In a third and last step to obtain the rep of a screw
(or the related Complex), Lie rotates the coordinate 
system in a manner that the new $\hat{z}$-axis is 
identical with the invariant line from above. He obtains
the simple analytical expression for the infinitesimal 
motion,
\begin{equation}
\label{eq:simplescrew}
\delta x=-y\delta t, 
\delta y=x\delta t, 
\delta z=k\delta t\,,
\end{equation}
however, it is important the understand that he had 
{\it to redefine} his 'time' coordinate $\delta t$ 
in that he had to introduce a new $\delta t$ by 
$C \delta t$, and a new parameter $k$ by the original
identification $k\delta t = G\delta t$. In other words,
due to the transformation to let the invariant line 
of the motion coincide with the new $\hat{z}$-axis 
in order to obtain the screw rep and infinitesimally
an orthogonal decomposition, he had 'to merge' the 
original constants $C$ and $G$ into the new 'time',
and alter the constant in the $z$-coordinate from 
$G$ to $k$. So according to our understanding, it 
is this picture with all its geometrical assumptions
which yields the formal picture today of Lorentz 
transformations when leaving $x$ and $y$ invariant
whereas $z$ and $t$ 'mix'. As such we propose to
supersede it right from the beginning by Complex
geometry, and to consider $P^{5}$ directly.\\

However, to enhance the historical remarks related to
null systems, Lie has related the appropriate Pfaff 
equation by considering line elements orthogonal to 
the axis of motion. As per point of 3-space $(x,y,z)$,
there exist $\infty^{2}$ 5-dim line elements\footnote{We
have used the old notation in order to preserve the 
dependence of the ratios, i.e.~the three coordinate 
differentials $\mathrm{d}x$, $\mathrm{d}y$, and $\mathrm{d}z$
are {\it not} independent, and the expression $\mathrm{d}x 
:\mathrm{d}y : \mathrm{d}z$ symbolizes that we have 
to treat only {\it two} independent quantities, usually
denoted as $p$ and $q$, instead of three!} $(x,y,z, 
\mathrm{d}x : \mathrm{d}y : \mathrm{d}z)$, one obtains
the equation
\begin{equation}
\delta x\cdot \mathrm{d}x+
\delta y\cdot \mathrm{d}y+
\delta z\cdot \mathrm{d}z=0\,.
\end{equation}
Using eq.~(\ref{eq:simplescrew}), this results in the 
Pfaff equation
\begin{equation}
\label{eq:pfaff}
x\mathrm{d}y-y\mathrm{d}x+k\mathrm{d}z=0\,,
\end{equation}
however, expressed in Lie's {\it NEW} coordinates which 
already are transformed with respect to the original 
coordinate system $x$, $y$, $z$, and thus implicitly 
respect the mixture emphasized above in the coordinates
and parameters.

\begin{figure}
\includegraphics{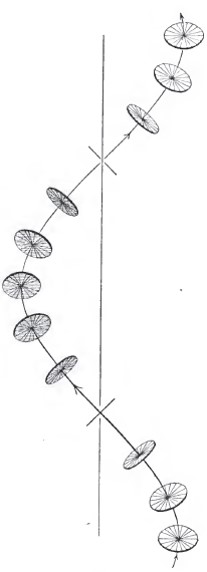}
\caption{Figure of the screw motion, taken from \cite{lie:1896},
p.~210.}\label{fig:complex}
\end{figure}

The last step to connect completely to linear Complexe
can use Lie's proof (see~\cite{lie:1896}, p.~210/211) 
that the integral curves of eq.~(\ref{eq:pfaff}) contain
$\infty^{3}$ straight lines, i.e.~integral lines as 
linear solutions, which results in the theorem (\cite{lie:1896},
p.~211, Satz~6) that related to every infinitesimal screw
there exist $\infty^{3}$ straight lines whose points move
{\it perpendicular} to the axis of motion. Per point of 
space $(x,y,z)$, this results in a $\infty^{1}$ set of 
lines, i.e.~a planar line pencil being orthogonal to the
axis of motion. The corresponding Figure~\ref{fig:complex}
is taken from \cite{lie:1896}, p.~210, figure~46.

Last not least, Lie presents the general Pfaff equation 
\begin{equation}
\begin{array}{cl}
\hphantom{+}A (y\mathrm{d}z -z\mathrm{d}y) & \\
+B (z\mathrm{d}x -x\mathrm{d}z) & \\
+C (x\mathrm{d}y -y\mathrm{d}z) & \\
+D \mathrm{d}x
+E \mathrm{d}y
+G \mathrm{d}z & =0\,,
\end{array}
\end{equation}
to describe the general infinitesimal motion in terms of 
perpendicular coordinates. So there exists always a set 
of $\infty^{3}$ straight lines as integral curves of this
equation which can be seen by using \PLu's rep of the 
straight line given by a running coordinate $z$ in terms
of its (linear) projections $x=rz+\rho$ and $y=sz+\sigma$
which we've used \cite{dahm:MRST7}, \cite{dahm:MRST9} to 
attach the spinor theory and 'quantum' notion in terms of
the Pauli algebra, too. The line projections and intersections
$r$, $s$, $\rho$, and $\sigma$, also known as \PLL coordinates
of the straight line when completed by the fifth line coordinate
$\eta=r\sigma-s\rho$, of the integral straight lines thus 
have to fulfil
\begin{equation}
\label{eq:complexinhomogen}
A \sigma -B \rho
+C (s\rho-r\sigma)
+D r
+E s
+G =0\,,
\end{equation}
which -- in other words -- constitute a line Complex. The
Pfaff equation selects the planar pencil of perpendicular
lines out of the lines through the point $(x,y,z)$ of 
3-space when using 4-dim line geometry as rep space. Note
also, that we may understand such a pencil as a flat planar 
(or projected) 'version' of a cone, obtained e.g.~by just
shifting its vertex to the plane!

For later use in conjunction with the tetrahedral Complex,
and for reference to \cite{dahm:MRST7} and \cite{dahm:MRST8},
we want to emphasize the special selection $\eta=0$. In 
this case, the definition yields $\eta=r\sigma-s\rho=0$, 
so $r\sigma=s\rho$, or $\frac{r\sigma}{s\rho}=1$. If we
rescale e.g.~$s$ and $\rho$ both by a parameter $\sqrt{\kappa}$,
the ratio $\frac{r\sigma}{s\rho}$ reads as $\frac{r\sigma}{s\rho}=\kappa$.
Now this ratio gains its interpretation and importance
as anharmonic ratio later with respect to the parameter
classifying tetrahedral Complexe.

Although Lie extends his discussion of null systems in 
\cite{lie:1896} much further, as with respect to our 
discussion of special relativity here, we think that 
we've given enough aspects of the background as well 
as of the geometrical context of the usual rep of 
Lorentz transformations. So after having mentioned 
few of Lie's aspects of transformation properties in
sec.~\ref{sec:remarkstransformations} and discussed 
some more 'well-known' relativity axioms in
sec.~\ref{sec:quadratic}, we'll use sections~\ref{sec:reinterpretation}
and \ref{sec:outlook} to assemble a guiding principle 
in terms of Complexe throughout topics in mechanics, 
optics, and relativity, and subsume typical formalism 
and 'axioms' into Complex geometry.\\

\subsection{Feynman Rules and Spin revisited}
\label{sec:weinbergNEW}
Before focussing on a further application of Complex geometry,
we want to discuss two more aspects briefly -- the rep theory
of the 'quantum' picture mentioned in sec.~\ref{sec:weinberg}
and the link to Onsager theory as discussed in \cite{dahm:MRST9}
4.3.\\

As we've extracted from  Weinberg's Hamiltonian formalism and 
his modeling in sec.~\ref{sec:weinberg}, one major issue is the
consistent (and 'overall' unique) treatment of phases. Whereas
from the practical viewpoint, we want to reduce phases and
ambiguities within rep theory, from the mathematical viewpoint
it would be nice to attach and use the enormous and extremely 
powerful machinery developed by Helgason \cite{helgason:1978},
\cite{helgason:GGA}, and moreover, to work with consistent 
phases throughout rep theory especially when treating group
chains, cosets, and algebras, and when identifying physics 
and measurements. So according to the notation in Helgason's
books, we want to rewrite some operators in sec.~\ref{sec:weinberg}
to use the exponential without additional '$i$'s, and extract 
a related operator rep. Meanwhile, we try to keep focus on 
physics and geometry.

As such, to recast Weinberg's operator structure in order 
to compare directly to Helgason, the first aspect is the 
exponential. Weinberg uses mappings $U\sim\exp\pm i\mathcal{O}$
of operators $\mathcal{O}$, $\mathcal{O}^{+}=\mathcal{O}$,
between Lie algebra and group. Formally, differentiation or 
series expansions of the exponential thus introduce '$i$'s 
into calculations. If, instead, we prefer to work directly 
with the operator action, there is no need to extract the 
$i$ by hand, but we can use the exponential directly to map
generators to group elements; the property we have to change
(and what we have to keep in mind throughout calculations) is
the Hermitean character of the generator which in conjunction 
with the additional $i$ has to be joined into an anti-Hermitean 
operator\footnote{For people using spectral theory, it is always
possible to introduce hermiticity by multiplying a commuting 
$-1=i^{2}$ to work with Hermitean operators relative to the
remaining phase $i$, however, this can be done at any time.}
$\mathcal{O}'=i\mathcal{O}$, ${\mathcal{O}'}^{+}=-\mathcal{O}'$,
and $U\sim\exp\pm\mathcal{O}'$. For us, this rep is closer 
to the logarithmic mapping \cite{lie:1896}, ch.~8, \S5, as 
well as closer to usual metric aspects by means of anharmonic
ratios (or 'W{\"{u}}rfe'), because we can now use the exponential 
directly to transfer products to sums (metric), and vice 
versa. Moreover, we think that we gain more control on 
symmetries because Hermitean conjugation consists of two 
operations, complex conjugation and transposition. Both 
have independent relevance, so we prefer the second picture
of mapping anti-Hermitean operators to unitary reps. Last 
not least, we can thus avoid superfluous and mostly meaningless 
'$i$'s like in perturbation theory which are sometimes 
introduced by 'magical' (or even mythical) rules or recipes.

With respect to Weinberg's Hermitean operator reps $J_{i}$ and 
$K_{j}$, we can rewrite the unitary rep $U$ (\cite{weinberg133:1964}, 
eq.~(2.20), or eq.~(\ref{eq:hermitisierung})) as
\[
\begin{array}{ccl}
	U[1+\omega] & = & 
	1+\frac{1}{2}\,\sum\limits_{\mu\neq\nu}\omega^{\mu\nu}\,iJ_{\mu\nu}\\
	& \sim & 1+\frac{1}{2}\,\sum\limits_{i}(\omega^{s}_{i}\,iJ_{i}+\omega^{0}_{i}\,iK_{i})
\end{array}
\]
where we've decomposed $\omega^{\mu\nu}$ into two triples of 
real parameters, $\omega^{0}$ and $\omega^{s}$. The notation
reflects the occurrence of $0-$ and purely spatial coordinates,
and using $\omega^{0}$ and $\omega^{s}$ helps to remember this 
background throughout calculations especially when having to 
identify contributions from rotations and 'boosts' at a later
stage within calculations. The operators $J_{i}$ and $K_{j}$ 
denote Weinberg's hermitean triples, $J^{+}_{i}=J_{i}$ and 
$K^{+}_{j}=K_{j}$. So the exponential maps $iJ_{i}$ and $iK_{j}$
to unitary $U$. The commutators derived by Weinberg via comparison
with \cite{weinberg133:1964}, eq.~(2.3) by use of~(2.1), yields
\begin{equation}
\begin{array}{cclclcl}
\left[J_{i}, J_{j}\right] & = & i\epsilon_{ijk}J_{k}\,, & & 
\left[J_{i}, K_{j}\right] & = & i\epsilon_{ijk}K_{k}\,,\\
\left[K_{i}, K_{j}\right] & = & -i\epsilon_{ijk}J_{k}\,. & & 
\end{array}
\end{equation}
Due to the additional '$i$'s in the commutation relations, however,
we define the 'new' anti-hermitean sets $\tilde{J}_{i}:=-iJ_{i}$ 
and $\tilde{K}_{i}:=-iK_{j}$, so that $\tilde{J}^{+}_{i}=-\tilde{J}_{i}$
and $\tilde{K}^{+}_{i}=-\tilde{K}_{i}$. Now, the commutators of
the anti-hermitean sets $\tilde{J}_{i}$ and $\tilde{K}_{i}$ read
as
\begin{equation}
\begin{array}{cclclcl}
\left[\tilde{J}_{i}, \tilde{J}_{j}\right] & = & \epsilon_{ijk}\tilde{J}_{k}\,, & &
\left[\tilde{J}_{i}, \tilde{K}_{j}\right] & = & \epsilon_{ijk}\tilde{K}_{k}\,,\\
\left[\tilde{K}_{i}, \tilde{K}_{j}\right] & = & -\epsilon_{ijk}\tilde{J}_{k}\,. & &
\end{array}
\end{equation}
The sets $\tilde{J}_{i}$ and $\tilde{K}_{i}$ with {\it real} coefficients
map to $U$, and the commutators do not introduce additional '$i$'s when
decreasing the operator grade by 1. In a last step, we modify the third
commutator only and want to get rid of the sign, so $\overline{K}_{i}:=i\tilde{K}_{i}$
is a suitable re-definition (i.e.~we revert to the original definition
${K}_{i}$), and we obtain (if we set in addition $\overline{J}_{i}:=\tilde{J}_{i}$) 
the commutation relations
\begin{equation}
\begin{array}{cclclcl}
\left[\overline{J}_{i}, \overline{J}_{j}\right] & = & \epsilon_{ijk}\overline{J}_{k}\,, & &
\left[\overline{J}_{i}, \overline{K}_{j}\right] & = & \epsilon_{ijk}\overline{K}_{k}\,,\\
\left[\overline{K}_{i}, \overline{K}_{j}\right] & = & \epsilon_{ijk}\overline{J}_{k}\,. & &
\end{array}
\end{equation}
Now we've reached the stage where we have a real, reductive algebra
\cite{helgason:1978}, and where only the product table of the algebra
determines via exponential mapping the respective special functions
while preserving a priori real coefficients. So it is purely the 
algebra itself which determines the reducibility of the result of
the mapping and the reality conditions.

Nevertheless, we have the choice to calculate either with Hermitean
$J_{i}$ and $K_{i}$ and the 'physical exponential' $\exp\pm i\,\cdot$,
or with anti-Hermitean operators $\tilde{J}_{i}$ and $\tilde{K}_{i}$
where we have to remember the sign in the third commutator within 
the exponential expansion or when summarizing the series by means
of special functions.

On the other hand, using $\overline{J}_{i}$ and $\overline{K}_{i}$
we may just apply and use the coset theory of the reductive 6-dim
Lie algebra above \cite{helgason:1978}, \cite{helgason:GGA}, and 
its decoupling into two commuting su(2) algebras. The relative,
intrinsic $i$ between the operator sets $\overline{J}_{i}=\tilde{J}_{i}$
and $\overline{K}_{i}=i\tilde{K}_{i}$ symbolizes the difference in 
the special functions between trigonometric and hyperbolic functions,
or in physical language the rotations and the boosts. In other words,
in both pictures it is visible that we are working with a non-compact
group and its Lie algebra. So far, the operator relations can be 
summarized by
\begin{equation}
\label{eq:operatorrelationsWeinberg}
\overline{J}_{i}=\tilde{J}_{i}=-iJ_{i}\,,\quad
\overline{K}_{i}=i\tilde{K}_{i}=K_{i}\,,
\end{equation}
so the additional '$i$'s are evident. Whereas the first set 
$\overline{J}_{i}$ and $\overline{K}_{i}$ is mapped by $\exp\,\cdot$,
the other sets are mapped by $\exp\pm i\,\cdot$ to obtain unitary 
reps of the group.

As side effects of Weinberg's publications on the Lorentz group, it is
noteworthy that in 1968, he applied almost the same approach to chiral 
symmetry of the pion-nucleon system, however, this time without the 
additional $i$ between the two su(2) algebras. So this time, instead 
of an su(2)$\oplus$i~su(2) algebra and SO(3,1) invariance, he could
use the same reasoning with respect to the p-wave coupling of the pion
on the nucleon according to the rules of almost the same Lie algebra.
This time the algebra su(2)$\oplus$su(2) corresponded to an SO(4) 
invariance, i.e.~the boosts of the Lorentz group corresponded to the
p-wave coupling character of the pion field according to the appropriate
rep identifications \cite{dahm:diss}. The linear rep theories emerged
as SO(4) invariant quadrics, the so-called $\sigma$-model, which 
comprised the power series of the Lie algebra parameters like above
in terms of special functions. Due to the exponential mapping above,
one could obtain also 'non-linear' reps, however, at the price to 
identify already the Lie algebra parameters as physical fields with
observable properties, and not the group reps. Practically, however,
Lagrangean expressions beyond quadratic terms were not really fruitful,
inclusion of other particles and resonances became ambiguous even 
with respect to electromagnetic coupling, and in general one had 
to introduce more and more assumptions and parameters into such 
effective Langrangeans to describe physics and experiments 
phenomenologically.\\

Now, before we switch over to aspects of line coordinates and line 
geometry, we want to summarize few central aspects of Weinberg's
description:
\begin{itemize}
	\item[-] He uses the Hamiltonian instead of a Lagrangean formalism
	because thus he doesn't need to eliminate superfluous field components,
	and because he knows the covariance and invariance properties which
	he applies to the field reps.
	\item[-] He has modeled the skew $4\times 4$ parameter set $\omega^{\mu\nu}$
	by associating (necessarily skew) operators $J_{\mu\nu}$, which 
	he re-expresses by two operator triples with su(2) commutation
	rules\footnote{Essentially, choosing the formalism of Lie symmetries
	instead of null systems, in this $4\times 4$-case we can just use
	the so(4) Lie algebra which can be rewritten in terms of so(3)$\oplus$so(3)
	(or su(2)$\oplus$su(2)), or even by reps of quaternions; all matrix 
	reps, of course, have to respect the skew rep property to relate to
	their background in null systems.} and complex parameters instead 
	of identifying and working with the 6-dim rep $\omega^{\mu\nu}$ 
	and its associated object(s).
	\item[-] He orientates himself with his modeling versus Wigner's boost
	reps and unitary rep theory, and the 6-dim rep is a special case 
	of representing the electromagnetic field by a special parameter 
	choice $j=1$, induced by the two su(2) algebras and their individual
	rep theory, however, here combined into $(j,0)\oplus(0,j)$.
\end{itemize}

For our further reasoning, we can use the case $j=1$ of the adjoint
reps in correspondence to the skew 6-dim rep. And as we know a 
geometrical identification of the 6-dim rep which was emphasized
much earlier already by the six-vector calculus of the electromagnetic
field rep, the real challenge is to keep track of the phases versus
physical identifications, and the correct association of the respective
transformation groups. As such, one has to consider very carefully
the association of the components $\vec{E}\pm i\vec{B}$ to the rep
$(1,0)\oplus(0,1)$ of the Lorentz group \cite{weinberg133:1964}, 
\cite{weinberg134:1964}, \cite{weinberg138:1965}. Although such
an identification is reasonable from the physical point of view,
and although we know of circularly polarized light, the components 
$\vec{E}\pm i\vec{B}$ are known to correspond to Klein coordinates
of the special Complex (see sec.~\ref{sec:quadratic}), and moreover,
this is related to the 'most compact' senary quadric (see e.g.~eq.~(\ref{eq:MinkInvariante1}) 
with invariance group SO(6)) as well as to the oval quadric in 
point space with SO(4) symmetry. So in both cases, we have formally
to work with compact symmetry groups. However, the real case using
the six \PLL coordinates parametrized by $(\vec{E},\vec{B})$ yields
SO(3,3) symmetry, and its associated point space symmetry group is
SO(3,1). Note, that we are talking of {\it real} parameters here! 
However, above we have seen, that one can easily redefine compact
to non-compact generators, and vice versa, by complexifying the 
parameters by a phase $i$. So in order not to get lost and to track
the correct phases between the generators, we have to start from 
scratch with line coordinates, so that at any time we can control
whether there is need to complexify real parameters by multiplying
with '$i$'s, and we can thus keep strict control on generators 
and coordinates.

As a phenomenological consequence (or result) of Weinberg's approach,
we can conjecture already here a relation between Lie algebra generators
and classical line coordinates, i.e.~we understand the differential 
rep of a su(2) generator as a certain rep of a line coordinate $p_{\mu\nu}$ on
function spaces. This can be read from eq.~(\ref{eq:hermitisierung})
if we understand the parametrization of the transformation as a
linear Complex with six real parameters $\omega^{\mu\nu}$.
So necessarily the corresponding 'operators' $i J_{\mu\nu}$ have
to represent the line coordinates in 'quantum' reps, or Hilbert 
space reps. In the appendix, we'll give a more precise derivation,
and we discuss the action of such generators on plane waves and 
especially on quadrics, however, for now this phenomenological 
identification yields a certain interpretation and guideline.

\subsection{Remarks on Transformations}
\label{sec:remarkstransformations}
So now, after having introduced null systems according to 
Lie's guideline \cite{lie:1896} in sec.~\ref{sec:nullsystems},
and having connected and discussed some differential operator
reps in sec.~\ref{sec:weinbergNEW}, it is time to have a 
deeper look on Lorentz transformations (here for short 'LT')
and their origin. As such, we'll concentrate not only as 
usually on point reps and an invariant quadric in point 
space, but we derive the background of LT from $P^{5}$ 
and line geometry.

As a side effect, we want to mention here only briefly 
Lie's research on the 10-dim symmetry group of Complexe
(more precise, of regular null systems, or dim 11 when 
treating special null systems). He has discussed this 
group as a subgroup of projective transformations in 
\cite{lie:1896}, ch.~6, exhaustively, and he has also 
related Pfaff's differential equations and discussed 
a lot of aspects. So later (if there is more time and
space to write things down) one can attach symplectic 
considerations, modern differential geometry and 
dynamics/kinematics here. Monge's equations enter in
his presentation as superset of Pfaff's equations, and
one can think (later on) to understand the cones by 
Complex geometry, and relate them to quadratic Complexe.
Such Complex cones automatically yield intersections 
with the absolute plane, so if we shift e.g.~the center/vertex
of the cone to the absolute plane, we can discuss 
Zindler's\footnote{Zindler summarized at the beginning
of the $20^{th}$ century some aspects of line geometry 
in two volumes \cite{zindler1:1902}, \cite{zindler2:1906}.}
rep at the beginning of vol.~I \cite{zindler1:1902},
and introduce there geometrically $\hbar=\frac{h}{2\pi}$,
$h$ describing the height of a full turn of the curve 
(or screw) with respect to the cylinder symmetry. 
Otherwise without absolute vertex/center, we can 
investigate and use the conic intersection of the
second order cone with the absolute plane throughout
'particle' motion, and thus we have immediate access
to metric properties by projective relations to the 
absolute conic, i.e.~by the Caley-Klein picture. 

Last not least, the ongoing discussion of 'entanglement'
in optics and 'quantum' physics is usually attributed 
to the intersection of second order ('light') cones, 
which -- using Monge's equations and Complex geometry
-- results automatically when relating line and 
differential geometry by considering Complex curves
(\cite{lie:1896}, ch.~7, \S5), and the Complex cones 
of two infinitesimal close curve points (see \cite{lie:1896},
p.~304, and figure~63).\\

However, we leave all those associated and interesting 
aspects and details, and stop here, because due to the 
topic and our focus on relativity, we want to address the 
origin of Lorentz transformations which we've mentioned 
briefly in \cite{dahm:MRST5} by presenting there few 
details only.\\

The usual setup is nowadays the view on point space and
point coordinate transformations, and the discussion of
transformations according to\footnote{For convenience, 
we've written transformations of the 0- and 1-coordinate,
however, it doesn't matter to change the coordinates to 
other values of $0,1,2,3$ as long as we leave a plane/two
of the point coordinates invariant, e.g.~$1$ and $3$ when
transforming $0$ and $2$ (compare also to figure~\ref{fig:complex}
in sec.~\ref{sec:nullsystems}).}
\begin{equation}
\label{eq:LTstd}
\begin{array}{ccl}
x_{0} & \longrightarrow & x'_{0}=\gamma x_{0}-\gamma\beta x_{1}\,,\\
x_{1} & \longrightarrow & x'_{1}=\gamma x_{1}-\gamma\beta x_{0}\,,\\
x_{2} & \longrightarrow & x'_{2}\,,\\
x_{3} & \longrightarrow & x'_{3}\,.\\
\end{array}
\end{equation}
Here, we've set $ct$, or $ct'$ as $x_{0}$, or $x'_{0}$, 
and we need the constraint 
$\gamma^{2}(1-\beta^{2})=\gamma^{2}-\gamma^{2}\beta^{2}=1$
as known from LT. The usual physical association is to 
define $\beta=\frac{v}{c}$, a ratio of velocities\footnote{See
our related discussions in \cite{dahm:MRST3}, \cite{dahm:MRST4},
\cite{dahm:MRST5} with respect to PG.}, and 
$\gamma=\sqrt{1-\beta^{2}}^{-1}=\sqrt{1-(\frac{v}{c})^{2}}^{-1}$.
A typical parametrization in terms of special functions 
is known to be $\beta=\tanh\eta=\frac{v}{c}$. Of course,
this rep is sufficient and works, but as we'll see below,
this explicit definition is not necessary as long as 
$\gamma^{2}(1-\beta^{2})=\gamma^{2}-\gamma^{2}\beta^{2}=1$.

Now before we start more detailed discussions, it is 
important to recall that we write the usual 'Lorentz 
invariant' in 4-vector notation and point coordinates
according to 
$x_{\mu}x^{\mu}\stackrel{\mathrm{\scriptscriptstyle LT}}{=}x'_{\mu}x'^{\mu}$.
In other words, we require that a quaternary point rep
remains on the invariant quadric $x_{\mu}x^{\mu}$. To 
gain some geometrical impression, we want to associate
the picture of a sphere with a certain radius (i.e.~a 
metric property) in order to use this picture later on
to visualize some of the geometrical aspects.

Now, usually people do not take care of spatial extension
but use models where $x$ represents a point only. But 
if we associate the sphere picture (or represent an 
extension by a second point $y$), we have to think about
representing and treating extension. So we need to know
the action of the Lorentz group on other points, located 
e.g.~close as well as far from the point $x$. The simplest 
approach is a second point $y$ on the sphere (or more
general on the quadric), because we know (or require) 
that the quadric is invariant. So one can discuss the 
transformation properties of this second point $y$ as
well by means of eq.~(\ref{eq:LTstd}). If we assume 
global transformations or if, with respect to the two 
points $x$ and $y$ we may assume them lying 'close 
together', then we can transform both points by the 
same transformation rules (\ref{eq:LTstd}).

At next, if we define second order objects according to 
$p_{\alpha\beta}:=x_{\alpha}y_{\beta}-x_{\beta}y_{\alpha}$
by using the coordinate enumeration of eq.~(\ref{eq:LTstd}),
we may ask for their transformation properties under LT
as well \cite{dahm:MRST5}, sec.~3. Of course, we can 
identify the origin and background of $p_{\alpha\beta}$
in line geometry, but we postpone synthetic discussions
for a while and follow algebra only. The easiest formal
approach (besides just calculating the transformation 
properties brute force in individual coordinates) is to
define $2\times 2$-determinants,
\begin{equation}
\label{eq:detLinien}
p_{\alpha\beta}\,=\,\left|
\begin{array}{cc}
x_{\alpha} & y_{\alpha}\\
x_{\beta}  & y_{\beta}
\end{array}
\right|\,,
\end{equation}
and use the standard rules on determinants. We have 
discussed in \cite{dahm:MRST5} the invariance properties
of two of the line coordinates. So it is evident, that 
the coordinate comprising the two point coordinates 
which are not affected by the LT do not change, i.e.~in
this case $p'_{23}=p_{23}$.

To understand why the line coordinate comprised of the 
two transformed point coordinates $0$ and $1$ doesn't
change, however, is not that obvious. In order to find
a general expression, we use the determinant rep, too. 
So with the transformation set
$x'_{\alpha}=\gamma x_{\alpha}-\gamma\beta x_{\beta}$, 
$x'_{\beta}=\gamma x_{\beta}-\gamma\beta x_{\alpha}$ 
for fixed $\alpha=0,\beta=1$, we have
\begin{eqnarray}
\label{eq:detLinieMain}
p'_{\alpha\beta}
& = & \left|
\begin{array}{cc}
x'_{\alpha} & y'_{\alpha}\\
x'_{\beta}  & y'_{\beta}
\end{array}
\right|\\
& = & \left|
\begin{array}{cc}
\gamma x_{\alpha}-\gamma\beta x_{\beta} & \gamma y_{\alpha}-\gamma\beta y_{\beta}\\
\gamma x_{\beta}-\gamma\beta x_{\alpha}  & \gamma y_{\beta}-\gamma\beta y_{\alpha}
\end{array}
\right|
\end{eqnarray}
Decomposing the last determinant by linearity and switching
rows, we obtain
\begin{equation}
\label{eq:bothtransformed}
p'_{\alpha\beta}=(\gamma^{2}-\gamma^{2}\beta^{2})
\left|
\begin{array}{cc}
x_{\alpha} & y_{\alpha}\\
x_{\beta}  & y_{\beta}
\end{array}
\right|
+(\gamma^{2}\beta-\gamma^{2}\beta)
\left|
\begin{array}{cc}
x_{\alpha} & y_{\beta}\\
x_{\beta}  & y_{\alpha}
\end{array}
\right|
\end{equation}
Now, independent of values of $\beta$ and $\gamma$, the second 
term on the rhs will always vanish. The remaining first term, 
however, can be considered also for transformations with other 
parameter $\beta$ and $\gamma$, so that in general there is a 
priori no need for the LT constraint,
$\gamma^{2}-\gamma^{2}\beta^{2}=1$. If, however, we require 
$\gamma^{2}-\gamma^{2}\beta^{2}=1$, this yields $p'_{01}=p_{01}$,
and we've identified a second line coordinate which remains 
invariant although we are applying the nontrivial (and 'non-local')
transformations (\ref{eq:LTstd}) (see also \cite{dahm:MRST5}). 

At this stage, we've identified already two out of six
line coordinates which behave not only irreducibly, but
each of them is individually invariant under LT.\\

Time to check the remaining four line coordinates $p'_{02}$,
$p'_{12}$, $p'_{03}$, $p'_{13}$, and their respective
transformation behaviour. With respect to the transformations
(\ref{eq:LTstd}), it is obvious that while one can leave
the lower row in the determinant without changes, we 
have to treat the two cases $p'_{02}$, $p'_{03}$, and 
$p'_{12}$, $p'_{13}$, separately according to the two 
different transformation rules of $x_{0}$ and $x_{1}$
in eq.~(\ref{eq:LTstd}).

The first case $p'_{0\delta}$, $\delta=2,3$, reads as 
\begin{equation}
\label{eq:0transformed}
p'_{0\delta}=
\left|
\begin{array}{cc}
\gamma x_{0}-\gamma\beta x_{1} & \gamma y_{0}-\gamma\beta y_{1}\\
x_{\delta}  & y_{\delta}
\end{array}
\right|=
\gamma p_{0\delta}-\gamma\beta p_{1\delta}\,,
\end{equation}
whereas the second case $p'_{1\delta}$ reads as
\begin{equation}
\label{eq:1transformed}
p'_{1\delta}=
\left|
\begin{array}{cc}
\gamma x_{1}-\gamma\beta x_{0} & \gamma y_{1}-\gamma\beta y_{0}\\
x_{\delta}  & y_{\delta}
\end{array}
\right|=
\gamma p_{1\delta}-\gamma\beta p_{0\delta}\,.
\end{equation}
As we see, the result can be still expressed in line
coordinates (which features once more the irreducibility
of the 6-dim set of line coordinates as an appropriate 
basis), however, this time, we have obtained additional
terms, or a mixture of two coordinates by the LT. But 
because each $\delta$-point coordinate remains invariant
for $\delta=2,3$, the transformation induces a linear 
combination only in terms of the point coordinates affected
by the LT. To summarize what we know so far with respect
to LT acting on $0$-/$1$-point coordinates, we can list:
\begin{equation}
\label{eq:LTonPcoordinates}
\begin{array}{lcl}
p'_{01}=p_{01} 
& \,, & 
p'_{23}=p_{23}\,,\\
p'_{02}= \gamma\,p_{02}-\gamma\beta\,p_{12}
& \,, &
p'_{12}= \gamma\,p_{12}-\gamma\beta\,p_{02}\,,\\
p'_{03}= \gamma\,p_{03}-\gamma\beta\,p_{13}
& \,, &
p'_{13}= \gamma\,p_{13}-\gamma\beta\,p_{03}\,.
\end{array}
\end{equation}
So although the two individual coordinates $p_{01}$ 
and $p_{23}$ remain invariant (and of course, their
combinations $p_{01}\pm p_{23}$, too), there is no 
similar {\it linear} mechanism if we consider individual
coordinates $p'_{02}$, $p'_{03}$, $p'_{12}$, $p'_{13}$,
or the combinations $p'_{02}\pm p'_{31}$, or $p'_{03}\pm p'_{12}$,
which both result in sums and an additional mixture
of the linear combinations\footnote{Although this 
gives rise to further interesting symmetries, we 
postpone the discussion here. Parts have already 
given in \cite{dahm:MRST9}, and \cite{dahm:MRST6}.}.

Now to understand the background, we have to recall 
that the line coordinates fulfil the \PLL condition.
In other words, we may ask what happens to this 
constraint on the line coordinates after having 
performed the LT on point coordinates. As such, 
if we consider at first the \PLL condition in 
transformed coordinates,
\begin{equation}
\label{eq:PLLconditionNew}
P'=p'_{01}p'_{23}+p'_{02}p'_{31}+p'_{03}p'_{12}\,,
\end{equation}
the first summand on the rhs remains trivially invariant
by (\ref{eq:LTonPcoordinates}), and we have to calculate
the last two summands. The result is
\begin{equation}
\label{eq:PLLconditionNewResult}
P'=p_{01}p_{23}+(\gamma^{2}-\gamma^{2}\beta^{2})\,
\left(p_{02}p_{31}+p_{03}p_{12}\right)\,,
\end{equation}
so that LT with $\gamma^{2}-\gamma^{2}\beta^{2}=1$ 
guarantee the invariance of the \PLL condition throughout
Lorentz transformations! In other words: lines remain
lines under LT. However, this explains an additional 
aspect because we know about the background of the 
\PLL condition in $P^{5}$. There, this condition was
used to map points of the \PLu-Klein quadric $M_{4}^{2}$ 
of $P^{5}$ to lines in $P^{3}$, and the \PLL condition 
has to be fulfilled by {\it each} line in $P^{3}$. 
Preserving this conditions under LT, i.e.~$P'\stackrel{\mathrm{\scriptscriptstyle LT}}{=}P$, 
therefore can be re-expressed by asking for the 
automorphism group of the \PLu-Klein quadric 
$M_{4}^{2}$ in $P^{5}$ (which is known to be a 
twofold 15-dim transformation group \cite{kleinHG:1926},
\S69). Vice versa, LT can be seen as the analogue
of such automorphisms in $P^{3}$.\\

As another interesting aspect for later use, it is 
noteworthy that eq.~(\ref{eq:PLLconditionNewResult})
allows for an alternative identification. Therefore,
we have to mention the general form of a tetrahedral
Complex in line coordinates,
\begin{equation}
\label{eq:tetrahedralComplex}
a\,p_{01}p_{23}\,+\,b\,p_{02}p_{31}\,+\,c\,p_{03}p_{12}\,=\,0
\end{equation}
(see e.g.~\cite{kleinHG:1926}, p.~172, or p.~288, or 
\cite{lie:1896}, p.~319). The anharmonic ratio of 
the lines intersecting the planes of the coordinate
tetrahedron corresponds to $\kappa=\frac{a-c}{b-c}$,
where the $\infty^{3}$ lines\footnote{Remember that
Lie uses another orientation of the coordinate axes.}
$x=rz+\rho$, $y=sz+\sigma$ fulfil $\kappa=\frac{s\rho}{r\sigma}$.
The Complex cones are elementary cones of the differential
equation
\begin{equation}
\label{eq:mongetypeeqn}
(b-c)x\mathrm{d}y\mathrm{d}z
+(c-a)y\mathrm{d}z\mathrm{d}x
+(a-b)z\mathrm{d}x\mathrm{d}y=0
\end{equation}
of Monge-type \cite{lie:1896}, p.~319.

So we can trace and understand the action of Lorentz
transformations also as modification of the tetrahedral
complex (corresponding to the modification of the
coordinate tetrahedron) by $a=1$, and 
$b=c=\gamma^{2}-\gamma^{2}\beta^{2}$. In the same
manner, also the \PLL condition can be seen as a 
special case of a tetrahedral Complex with $a=b=c=1$,
i.e.~related to a special choice of the coordinate
tetrahedron in point space. Note, however, that in 
both cases due to $b=c$ the denominator of $\kappa$
vanishes, i.e.~$\kappa\longrightarrow\infty$, so that
we can append our discussion given in \cite{dahm:MRST7}
to derive the spinorial picture. There, we have discussed
and used the fifth inhomogeneous \PLL coordinate 
$\eta=r\sigma-s\rho$ with the constraint $\eta=0$
(see the discussion with respect to $\kappa$ before).\\

So far, we've established a direct trail with respect
to linear transformation behaviour of the line coordinates
under LT (see eq.~(\ref{eq:LTonPcoordinates})) and 
the invariance of the \PLL condition. This invariance
guarantees that we can rely on our line picture 
throughout calculations and LT in $P^{3}$. However, if 
we take a closer look on the transformation laws in
eq.~(\ref{eq:LTstd}) and in eq.~(\ref{eq:LTonPcoordinates}),
we notice that these laws in the first two rows of
(\ref{eq:LTstd}) are similar to those in the last
two rows of (\ref{eq:LTonPcoordinates}) if we replace
$x \longleftrightarrow p$. Therefore, recalling 
Minkowski's 'two invariants' given in eqns.~(\ref{eq:MinkInvariante1})
and (\ref{eq:MinkInvariante2}), we may calculate\footnote{Background
details will be given in sec.~\ref{sec:quadratic}.}
the coordinate squares, too. Straightforward algebra
yields:
\begin{eqnarray}
\label{eq:squarepairs1}
{p'_{02}}^{2}-{p'_{12}}^{2}
& \stackrel{\mathrm{\scriptscriptstyle LT}}{=} &
\left(\gamma^{2}-\gamma^{2}\beta^{2}\right)
\left(p_{02}^{2}-p_{12}^{2}\right)\,,\\
\label{eq:squarepairs2}
{p'_{03}}^{2}-{p'_{13}}^{2}
& \stackrel{\mathrm{\scriptscriptstyle LT}}{=} &
\left(\gamma^{2}-\gamma^{2}\beta^{2}\right)
\left(p_{03}^{2}-p_{13}^{2}\right)\,,
\end{eqnarray}
so that the senary quadric yields under LT\footnote{We
do not discuss the additional freedom to switch the
signs of the squares which is due to the LT being given
only by point transformations and, as such, being not 
more restrictive. LT are a special case only, and the
assumptions on transformations of single points in 
3-space are not sufficient to separate a unique invariant
senary quadric. As such, we find a couple of symmetry 
groups SO($m$,$n$), $n+m=6$, discussed throughout 
literature.}
\begin{equation}
\begin{array}{ll}
& \pm{p'_{01}}^{2}\pm{p'_{23}}^{2}\\
& \pm({p'_{02}}^{2}-{p'_{12}}^{2})
\pm({p'_{03}}^{2}-{p'_{13}}^{2})\\
\stackrel{\mathrm{\scriptscriptstyle (\ref{eq:LTonPcoordinates})}}{\longrightarrow} & \\
& \pm p_{01}^{2}\pm p_{23}^{2}\\
& \pm\left(\gamma^{2}-\gamma^{2}\beta^{2}\right)\left(p_{02}^{2}-p_{12}^{2}\right)\\
& \pm\left(\gamma^{2}-\gamma^{2}\beta^{2}\right)\left(p_{03}^{2}-p_{13}^{2}\right)\,.
\end{array}
\end{equation}
For LT, i.e.~$\gamma^{2}-\gamma^{2}\beta^{2}=1$, we 
have
\[
\begin{array}{ll}
& \pm {p'}_{01}^{2}\pm {p'}_{23}^{2}
\pm({p'}_{02}^{2}-{p'}_{12}^{2})
\pm({p'}_{03}^{2}-{p'}_{13}^{2})\\
\stackrel{\mathrm{\scriptscriptstyle LT}}{\longrightarrow} 
& \pm p_{01}^{2}\pm p_{23}^{2}
\pm\left(p_{02}^{2}-p_{12}^{2}\right)
\pm\left(p_{03}^{2}-p_{13}^{2}\right)\,.
\end{array}
\]
So by LT alone, we still have the freedom to absorb
the one or other sign in an additional phase in point 
space, or in a transferred rep, e.g.~by projections or
special mappings. We cannot overemphasize that with each
mapping or transfer, the respective 'objects' change 
physically, and we have to adjust the physical picture
and its relevance appropriately. So pure axiomatization,
or Bourbaki-style technocracy, may help with formal rep
theory and yield a consistent 'grammar' of the language,
however, to stay in this picture, there is still the
need to use the correct words and pictures, and to 
write the epic works and poems of physics.

Last not least, to begin writing the poems, it is 
worth spending some thoughts on the geometrical 
setup we've introduced above, and on the associated 
picture\footnote{Before discussions start, please 
note that using homogeneous coordinates, we can 
apply the formalism to general quaternary quadrics 
also without using this geometrical picture in depth. 
As a further example of a quadric different from 
the sphere, we can e.g.~discuss the 'Schmiegtetraeder'
of an hyperboloid, see also \cite{lie:1896}, p.~348ff,
with respect to generating lines and the associated 
tetrahedral Complex. Recall also, that we may change 
easily from \PLL to Klein coordinates in a controlled
manner.} of the sphere, or quadric. If we understand 
the invariant quadric in point space as a sphere, then
the invariance requirement of the quadric allows to 
shift points on the sphere (or in general along the 
quadric). If we look closer to the fundamental 
tetrahedron, we can define coordinates (e.g.~by 
shifting the unit point appropriately) where the 
vertices lie on the circumscribed unit sphere, just 
in order to realize the invariant quadric. In this 
picture, the line coordinates correspond to edges
of the tetrahedron which connect the vertices 
appropriately.

Now by fixing the two line coordinates $p_{01}$ and 
$p_{23}$, we have identified two opposite edges of 
the tetrahedron with their endpoints each representing
two of the four vertices of the tetrahedron (which 
by construction lie on the invariant sphere). So in
order to fix these two edges/line coordinates, we 
can think of two stiff (or solid) edges whereas 
their endpoints can't leave the sphere (or quadric),
and as such the ('Lorentz') transformations\footnote{In
the general case the transformations leaving the quadric
invariant.} shift the endpoints of one edge/stick $p_{01}$
along the quadric whereas the second one, $p_{23}$, 
remains fixed. So in essence, while the endpoints of
one\footnote{By more general transformations, the 
endpoints of both edges can be moved.} of the stiff
edges can be moved along the quadric, the four remaining
edges according to their non-trivial transformations
under LT (or even in the case of more general 
transformations) can be seen as realized by rubber 
band, however, the sphere (quadric) will remain 
invariant.\\

Closing this section, it is necessary to integrate our 
aspects above, and to mention some more background and 
an additional picture used by the old geometers. As such, 
we can identify the skew edges/line coordinates $p_{01}$ 
and $p_{23}$ with two skew lines $l_{1}$ and $l_{2}$. 
Now by the definition of a 'ray system'\footnote{German:
Strahlensystem erster Ordnung und erster Klasse.} which
we've used already above, we have exactly one spatial 
line per general space point $x$ which intersects both
lines $l_{1}$ and $l_{2}$ (see \cite{lie:1896}, p.~187,
or p.~291ff, for details and background). In order to 
repeat not essential parts of Lie's book, it is sufficient
to reference to \cite{lie:1896}, ch.~6, which yields 
the context of Lie's line and area elements, of null
systems and various related aspects of Monge's and 
Pfaff's differential equations. Moreover, having 
identified the 'ray systems' above, we can connect 
to Kummer's work on such ray systems \cite{kummer:1866},
and their general relevance in optics as well as in
geometry. The line geometrical approach can be equally
well applied by means of Complexe or Congruences, 
and it relates to \PLu's work.

So for now, we think that we've attached special 
relativity sufficiently to classical line geometry,
and it is up to generalize to regular linear as well
as higher order Complexe.

\subsection{Quadratic Relations}
\label{sec:quadratic}
So far in this section, we have discussed some basic 
properties and examples of {\it linear} Complexe as
well as their relation to Pfaff's and Monge's equation,
to Lorentz transformations in subsection \ref{sec:nullsystems}
as well as aspects of related transformation groups in 
subsection \ref{sec:remarkstransformations}. Most of the
discussion was related to special linear Complexe and
lines\footnote{We discuss aspects of regular linear
Complexe in \cite{dahm:MRST6} in conjunction with (null) 
planes when representing 4-momentum and Dirac spinors,
i.e.~instead of second {\it order} surfaces one can
address as well second {\it class} surfaces, or transformation
properties of planes with respect to null systems, and 
especially under additional tangential constraints.}.
However, with respect to $P^{3}$ and known physics and
geometry, we have to consider also briefly {\it quadratic}
Complexe. There are multiple reasons to do this:
\begin{enumerate}
\item[-] The \PLL condition 
$P=p_{01}p_{23}+p_{02}p_{31}+p_{03}p_{12}=0$
(or equivalently $p_{1}p_{4}+p_{2}p_{5}+p_{3}p_{6}=0$)
is quadratic in line coordinates, so it is inevitable
to consider quadratic Complexe besides just using linear
Complexe and $P^{5}$ like in the last section. Moreover,
if we work practically with lines, the coordinates of
a special linear complex $\sum a_{\alpha\beta}p_{\alpha\beta}$
have to fulfil the \PLL condition, too. Intersection of
lines can be attributed to incidence of two lines with
a (common) plane, so if we take four points (two points
per line) the $4\times 4$-determinant of the four points
has to vanish. Rewriting this determinant in terms of
line coordinates $p_{\alpha\beta}$ and $p'_{\alpha\beta}$
of the two lines yields
\begin{equation}
\label{eq:lineintersection}
\begin{array}{cl}
& p_{12}p'_{03}+p_{31}p'_{02}+p_{01}p'_{23}\\
+ & p_{23}p'_{01}+p_{02}p'_{31}+p_{03}p'_{12}=0\,,	
\end{array}
\end{equation}
or using 'polarity', we may introduce partial derivatives
and write 
$\sum \frac{\partial\,P}{\partial p_{\alpha\beta}}p'_{\alpha\beta}=0$
where the sum has to be taken over all six coordinates.
As Cayley has mentioned, due to linearity of the linear
Complex $\sum a_{\alpha\beta}p_{\alpha\beta}$ in $p_{\alpha\beta}$
(which is essentially a 3-dim line set in 3-space), this 
intersection condition with respect to a line $(p'_{\alpha\beta})$
can be fulfilled by all member lines of the Complex, if
they fulfil the quadratic relation
$P_{a}=a_{01}a_{23}+a_{02}a_{31}+a_{03}a_{12}=0$. This
describes the special linear Complex of the electromagnetic 
field $(\vec{E},\vec{B})$, and we may associate the picture
given in \cite{dahm:MRST3}, sec.~2.2 or appendix~A, and 
\cite{dahm:MRST4}, sec.~3 by a moving point light (if we
simplify light emission to lines). In other words, the axis
of the special linear Complex (i.e.~the trajectory of the 
moving point) serves to collect those lines of 3-space 
into a Complex which hit this axis. This, of course, can
be understood in a second picture as observers located 
throughout arbitrary points of 3-space and watching a
moving point (once more, if we simplify the 'observation',
or information transfer, by lines between observer location
and moving point \cite{dahm:MRST3}, \cite{dahm:MRST4}.
\item[-] Another major aspect is the need to introduce
coordinate systems into this world as soon as we want 
to talk about physics, measures and real world observations.
Although we can use line geometry of $P^{3}$ and the 
intersections and unions of lines, it is often necessary
to have additional possibilities to use individual point
or plane reps in $P^{3}$, and compare to observations. 
However, introducing the fundamental coordinate tetrahedron
in $P^{3}$, i.e.~4 non-planar points or 4 appropriate 
planes as tetrahedron sides, yields immediately a 
classification scheme of lines in $P^{3}$ because each
of the lines hits the tetrahedron planes in four distinct
points. This scheme almost automatically yields the 
definition of a quadratic Complex, the so-called
tetrahedral Complex (see e.g.~\cite{dahm:MRST3}, 
sec.~3.2, or \cite{dahm:MRST4}, sec.~3.3, and references 
there), and the associated anharmonic ratio $\kappa$ 
from above.
\item[-] We've addressed the usual interpretation of 
the 4-dim rep of the '4-momentum' already a couple of
times. Especially in \cite{dahm:MRST6}, we address the
interpretation in terms of Hesse coordinates of plane
reps, and to keep the 6-dim rep for forces and momentum
like given in \cite{klein:1871}. This is based on the 
regular linear Complex as transfer mapping, or a correlation,
of points to planes and vice versa \cite{dahm:MRST6}.
As such, we have to consider normals of planes in the
Hesse rep, and a system of normals related\footnote{This
is a consequence of the mechanism/formalism how we 
treat dynamics and describe points (see \cite{study:1905}).}
to second order surfaces. In this environment, however, 
we benefit from the 'natural occurrence' of the tetrahedral
Complex when treating normals (see \cite{lie:1896}, 
ch.~7, \S2).
\item[-] Working with Complexe as base elements, it is
natural to ask for common lines or line sets e.g.~in 
Congruences or Configurations, i.e.~we have to handle
families or sets of Complexe, and related assemblies
thereof. This results analytically in typical structures
like 
$(\textfrak{A})_{\alpha\beta}\sim a_{\alpha\beta}
=\lambda_{1}a^{1}_{\alpha\beta}
+\lambda_{2}a^{2}_{\alpha\beta}$,
or
$(\textfrak{B})_{\alpha\beta}\sim b_{\alpha\beta}
=\lambda_{1}b^{1}_{\alpha\beta}
+\lambda_{2}b^{2}_{\alpha\beta}
+\lambda_{3}b^{3}_{\alpha\beta}$
etc. Now the quest for lines common to the pencil or
the triple set of Complexe may be re-expressed by means
of the quadratic \PLL condition, because the lines $l$ 
in $P^{3}$ have to fulfil the condition 
$P_{l}=a_{01}a_{23}+a_{02}a_{31}+a_{03}a_{12}=0$ from
above. As such, we have to resolve quadratic equations
in the coefficients $a_{\alpha\beta}$ or $b_{\alpha\beta}$
of the Complexe $\textfrak{A}$ or $\textfrak{B}$ which
results in quadratic equations to determine the parameters
$\lambda_{i}$ from above (or their ratios, respectively).
In the case of the pencil\footnote{We have two binary 
parameters $\lambda_{1}$ and $\lambda_{2}$, i.e.~the 
assembly is dependent from {\it one} parameter and 
constitutes a pencil.} $\textfrak{A}$, this yields for
example
$(\lambda_{1}a^{1}_{01}+\lambda_{2}a^{2}_{01})
(\lambda_{1}a^{1}_{23}+\lambda_{2}a^{2}_{23})
+\ldots
+\ldots=0\,.$
In the same manner, without restricting the products
on the rhs to 0, one can calculate the invariant of
the pencil and relate the invariants of the original
Complexe. This also justifies investigations of the
relative position of such Complexe in $P^{5}$, and 
it gives rise to the notion of 'involutions' of 
Complexe \cite{klein:1872a}. For now, this aspect is
sufficient to justify a deeper look onto quadratic 
Complexe in general.
\item[-] Last not least, because some aspects above
already mention relations to the tetrahedral Complex,
and because there are further contexts where this
Complex emerges, it is worth keeping focus on this
object (or the family of such objects). The tetrahedral
Complex is a quadratic Complex (see eq.~(\ref{eq:tetrahedralComplex})),
but although it is deeply connected to projective geometry
and especially, due to the fundamental tetrahedron, 
with the coordinate definition, we found only few of
its aspects discussed analytically in literature. 
Most of the discussion took place in synthetic geometry,
originated by von Staudt and in parts by Reye, and there
is a historical survey by Lie in \cite{lie:1896}, ch.~7,
\S2, which addresses some of its historical aspects and
contexts. However, because our discussion of Lorentz 
transformations and automorphisms of the \PLu-Klein 
quadric $M_{4}^{2}$ throughout the last subsection, 
sec.~\ref{sec:remarkstransformations}, lead us to the
tetrahedral Complex, it is worth to start a discussion
of quadratic aspects, too.
\end{enumerate}

So as a first step to approach analytic calculations with
Complexe as base elements, we can follow Klein \cite{kleinHG:1926},
\S23, and start right from the beginning with $P^{5}$. As 
such, if we recall the basic check for lines to fulfil the
\PLL condition, we can ask as well for rules and properties
to work with the coefficients of such elements. More general,
we have to construct an invariant theory and appropriate
forms and rules. As Klein remarks in \cite{kleinHG:1926},
\S23, one has to address quadratic forms already when working
with linear Complexe, and one can consider in a first stage
the transformation behaviour of quadratic (senary) forms
when substituting the coefficients of the linear elements.

We skip Klein's discussion of inertia indices (\cite{kleinHG:1926},
p.~97) here and focus on special substitutions to rearrange
the \PLL condition. Due to the quadratic character and our
results from the last subsection, it is evident that we can
use binomial expressions to transfer the products to plain
squares of a quadric, i.e.~if we rename the coefficients 
$a_{\alpha\beta}$ by a senary index $\alpha$ into $x_{\alpha}$,
the invariant of the linear Complex in $P^{5}$ reads 
as\footnote{To keep track with Klein and his book \cite{kleinHG:1926},
we adopt to his special naming convention \cite{kleinHG:1926}, 
p.~96, of the subscripts. Take care, that this notation differs
from senary indices used elsewhere, and here Klein's 'even' 
coefficients $x_{2}$, $x_{4}$, and $x_{6}$ are related to 
absolute elements (i.e.~to the index 0, or 4) in point space.
In order to recall the appropriate coordinate set of the 
respective rep of $\Omega$, we have attached the related 
subscript to $\Omega$.}
$\Omega_{x}=x_{1}x_{2}+x_{3}x_{4}+x_{5}x_{6}$.

By {\it real} and {\it linear} substitutions
\begin{equation}
\label{eq:kleinY}
\begin{array}{lll}
x_{1} = y_{1}+y_{2}, & x_{3} = y_{3}+y_{4}, & x_{5} = y_{5}+y_{6},\\
x_{2} = y_{1}-y_{2}, & x_{4} = y_{3}-y_{4}, & x_{6} = y_{5}-y_{6},
\end{array}
\end{equation}
it is evident, that the quadratic form $\Omega_{x}$ from
above now reads as
$\Omega_{y}=y^{2}_{1}-y^{2}_{2}+y^{2}_{3}-y^{2}_{4}+y^{2}_{5}-y^{2}_{6}$,
so with respect to {\it real} and {\it linear} 
substitutions, we have reached the final stage of
a sum of perfect squares and can read off their 
'inertial indices'. The senary quadric $\Omega_{y}$
now shows formally its SO(3,3) symmetry which we've
mentioned before as resulting from line geometry. 
Now a naive interpretation of this quadric in terms
of (abstract) {\it point} coordinates allows to treat
this SO(3,3) symmetry with formal arguments, however,
as is obvious, in such a case the interpretation 
against the background of line and Complex geometry
is completely lost.

However, in Minkowski's and Klein's papers 1910, the 
authors have left real substitutions and transformed 
the 0- (or 4-)component of the point coordinates by
multiplying it with a phase $i$ in order to represent
the quaternary quadric as sum of squared components,
and {\it in addition}, Minkowski introduces an 'imaginary'
angle $i\psi$ as argument of trigonometric functions,
i.~e.~in essence hyperbolic transformations with $\psi$.
To sort things, Minkowski associates the real, hyperbolic
transformations to the set $x,y,z,t$ whereas $x_{4}$ and
$t$ have the relative imaginary phase $i$ by \cite{mink:1910},
eq.~(3). So we have a mixture of non-compact SO(3,1) 
transformations and compact SO(4) transformations, 
acting on point reps with real coefficients or point
reps with 3 real and one purely imaginary coefficient
which can only implicitly resolved because in addition
Minkowski attaches the notion of a space-time point
to both reps $x,y,z,t$ AND $x_{\alpha}$ \cite{mink:1910},
\S1, p.~475, last line\footnote{The symmetry of his eqns.~(A)
and (B) \cite{mink:1910}, p.~477, and his eq.~(17) on
p.~481 associate the groups SO(4) to $x_{\alpha}$, and
SO(3,1) to $x,y,z,t$.}. Due to the character of \PLu's 
line coordinates, we have to address this change by 
'absorbing' the additional $i$ of the fourth point 
coordinate in a redefinition of the related coefficients
of three of the line coordinates. So to keep the old 
definition of the line coordinates, one can shift 
the $i$ to the coefficient, which results in sign 
changes of the quadric.\\

Klein starts the other way around and rewrites $\Omega$
by introducing imaginary transformations\footnote{On 
demand, one can also symmetrize the normalization by 
using $\sqrt{\frac{1}{2}}$ in both sets, however, this
in turn can be reabsorbed by the homogeneous coordinates
definition. Nevertheless, for calculations using coordinate
transformations it is helpful to perform the 'bookkeeping' 
correctly.} of the $y$, i.e.\\
\begin{equation}
\begin{array}{lll}
z_{1} = y_{1},  & z_{3} = y_{3},  & z_{5} = y_{5},\\
z_{2} = iy_{2}, & z_{4} = iy_{4}, & z_{6} = iy_{6},
\end{array}
\end{equation}
so that we now have 
$\Omega_{z}=z^{2}_{1}+z^{2}_{2}+z^{2}_{3}+z^{2}_{4}+z^{2}_{5}+z^{2}_{6}$
with formal SO(6) symmetry. This basically reflects Hermite's
trick to represent the sum of two squares of real numbers by 
the product of two conjugate complex numbers\footnote{Moreover,
it leads to reps by 3 complex numbers, i.e.~$\mathbbm{C}$ as 
used by Study in various contexts which we'll discuss elsewhere.}.
Rewriting the set~(\ref{eq:kleinY}), we obtain
\begin{equation}
\label{eq:kleinZ}
\begin{array}{lll}
x_{1} = z_{1}-iz_{2}, & x_{3} = z_{3}-iz_{4}, & x_{5} = z_{5}-iz_{6},\\
x_{2} = z_{1}+iz_{2}, & x_{4} = z_{3}+iz_{4}, & x_{6} = z_{5}+iz_{6}.
\end{array}
\end{equation}
So with respect to Minkowski's '{\it two} invariants'
given in eqns.~(\ref{eq:ComplexInvariante1}) and 
(\ref{eq:ComplexInvariante2}) here, it is now obvious
that we are talking on $\Omega_{x}$ and $\Omega_{z}$
of the same linear Complex $f$, however, related by 
the linear substitutions above. Therefore, we consider
it more appropriate to discuss the Complex invariant 
with respect to the $P^{5}$ background instead of 
'two invariants', or 'quantum numbers', related to 
SO(4), SO(3,1), or glued to assemblies of two su(2)
'spin' algebras with or without relative complexification.
As such (see also \cite{dahm:MRST3}, sec.~2) the source-free
electromagnetic case featuring $\vec{E}\cdot\vec{B}=0$ 
can of course be attached to vanishing values of invariants,
however, geometrically it is more fruitful and constructive
to discuss linear Complexe in general, i.e.~in $P^{5}$, 
and points on the \PLu-Klein quadric as special cases 
thereof. Besides relating 'quantum numbers', we expect
this to be the framework to discuss further some known
relations and aspects like summarized in \cite{biedenharnlouck8:1981},
ch.~6, \S22, or by various aspects of physical applications
in the following ch.~7, or in \cite{biedenharnlouck9:1981}
e.g.~with respect to the W-algebra in ch.~4, and the 
'Special Topics' in ch.~5, especially the Regge trajectories
('Topic 5'), and SU(1,1).\\

To proceed further in our current context of quadratic 
Complexe, the simultaneous invariant of the linear 
Complexe in the case $\Omega_{y}$ then reads as
\begin{equation}
\label{eq:invariante_y}
y_{1}y'_{1}-y_{2}y'_{2}+y_{3}y'_{3}-y_{4}y'_{4}+y_{5}y'_{5}-y_{6}y'_{6}
\end{equation}
which Klein uses to discuss the $\pm 1$-handedness
of the six fundamental Complexe and their grouping
into $3\oplus 3$. With respect to $\Omega_{z}$,
one has to discuss Complexe with general complex
coefficients, too, however due to the geometric
interpretation of $i$ and hypercomplex numbers as
an algebraic symbolism, right here, we do not feel
the need to do so. Here, in order to close this 
section on quadratic relations, it is important 
to mention Klein's subsequent arguments (\cite{kleinHG:1926}, 
p.~100) that working with quadratic Complexe one
automatically has to consider linear substitutions
of {\it two} quadratic forms to result in pure squares
as summands, because in addition to the quadratic 
Complex one has to respect the \PLL condition $\Omega$,
usually as $\Omega_{y}$ or $\Omega_{z}$. In essence,
also with respect to quadratic Complexe one can use
a system of six {\it linear} Complexe, either purely
real or imaginary \cite{kleinHG:1926}, p.~100, and
introduce concepts like confocal Complexe in analogy
to confocal cyclids, etc.

In physics, because quadratic Complexe are related 
to second order cones ('Complex cones' \cite{plueckerNG:1868}),
this framework generically yields access to the 
requirements discussed with respect to relativity
by Ehlers et al.~\cite{ehlers:1972}, and we find 
immediately applications in entanglement discussions
usually performed by intersecting two second order 
cones\footnote{We have already mentioned the background
of Monge's cones discussed by Lie \cite{lie:1896} 
with respect to dynamics and contact transformations,
and we want to mention with respect to our older 
work discussing the relation of quadratic Complexe 
and the 'light cone' in point, that throughout the 
discussion of confocal quadratic Complexe one can 
also find one special/singular quadratic Complex.}.

\subsection{Discrete Transformations}
\label{sec:discrete}
There are further beautiful relations to discrete 
transformations as well as to mathematics which 
is nowadays usually discussed in conjunction with
'quantum' physics, namely the Heisenberg group and 
Kummer surfaces, or K3 theory in general, which 
in our current context is off-topic. Nevertheless,
it is worth to recall these relations briefly as 
an 'interlude', or 'stint', as they are attached 
to ongoing discussions and to $P^{5}$ geometry 
which we'll propose later as a suitable framework
to generalize and unify some of the {\it physical}
ideas discussed above.

Now, here it is sufficient to follow Hudson \cite{hudson:1905}
and his discussion of permutation symmetries of points in the
(projective) point/plane incidence equation\footnote{According
to our notion, we denote the point and plane by $x$ and $u$, 
respectively, thus changing Hudson's notation in \cite{hudson:1905},
ch.~I, \S3.} $x\cdot u=0$ of 3-space. He then derives the 
$16_{6}$ configuration, some dimensional identifications 
from line and Complex geometry which are interesting with
respect to our earlier work\footnote{As mentioned before, 
we've used SU(4) and related non-compact groups to discuss
a linear covering of chiral and Lorentz group reps, linear
as well as nonlinear. The su(4) root diagram when rotated 
to 3-space yields geometrically a tetrahedron and is associated
to the fundamental rep $\underline{\bf 4}$. As such, SU(4) 
can serve to treat spatial geometry whereas SU(3) and the 
reps can serve to treat planar geometry by means of three 
homogeneous coordinates.}, and he discusses apolarity of 
Complexe. In his preface to Hudson's book \cite{hudson:1905},
Wolf Barth translates this apolarity condition into modern 
terminology by means of tensors $S$ and $T$ representing 
anti-selfadjoint maps 
$\mathbbm{C}^{4}\longrightarrow(\mathbbm{C}^{4})^{*}$ while 
the apolarity can be represented by $S^{-1}T=T^{-1}S$ with 
$S^{-1}T=(T^{-1}S)^{-1}$, thus being an involutory map of 
$\mathbbm{C}^{4}$. Whereas the rich content of the book 
yields various aspects of algebraic geometry, here we can
focus on chapters IV - VI. Ch.~IV can be used with respect
to linear Complexe, apolarity when considering the fundamental
Complexe, and the parity decomposition. Ch.~V discusses the 
simultaneous diagonalization of quadratic forms in $P^{5}$ 
and the hyperplane geometry there with respect to Kummer's 
and \PLu's work in optics. The main topic, however, is the
Kummer surface for the quadratic Complex and its model as
a complete intersection of three quadrics in $P^{5}$: the 
\PLL quadric, the quadratic Complex itself, and a third 
quadric attached to both (\cite{hudson:1905}, preface,
p.~xvi). \PLu's surface is comprised as a special case, 
however, especially with respect to optics, Hamilton theory
and QFT, the further chapters of Hudson's book (or their
content in modern terminology, respectively) will be part 
of upcoming papers after we've discussed more of the 
physical foundations to be settled more deeply in $P^{5}$.

\subsection{Classical PG and Dirac}
We want to address briefly\footnote{Details are discussed
in \cite{dahm:MRST6} with respect to null systems and as
well with respect to polar systems and surfaces of second
order and of second class. This yields also details with
respect to the non-local 'Hilbert' rep of Lie generators
which we've mentioned in appendix~\ref{app:polarity} to
find operator counter parts of line generators vs. second 
order surfaces.\label{fn:mrst6} Last not least, we address
there \cite{dahm:MRST6} Dirac's papers on wave equations
\cite{dirac:1935}, \cite{dirac:1936} in de~Sitter and
conformal space where -- similarly to Cartan's theory of
spinors (see \cite{dahm:MRST7}) -- reference to geometrical
background and to existing relevant references are {\it NOT}
given by the author.} another aspects of null systems as
correlation, i.e.~realizing a special mapping (or a 
transfer) between points and planes in $P^{3}$. As such,
the rep of such a system (because either acting on point
or plane reps of $P^{3}$ which are both quaternary) is 
necessarily a $4\times 4$ matrix rep $M$, and in order to
avoid singular cases, the determinant of a regular null
system is not zero. On the other hand, the null point $x$
is member of the null plane $u$ (or incident with the null
plane), so $x\cdot u=0$, and for regular null systems
(representing a correlation by e.g.~$u_{\mu}=M_{\mu\nu}x_{\nu}$,
or $x_{\mu}=(M')_{\mu\nu}u_{\nu}$), we have a constraint
$x\cdot u=xMx=0$ (or $x\cdot u=uM''u=0$) which in essence
is fulfilled for skew $M$ (or $M''$) because
$x\cdot u=M_{\alpha\beta}x_{\alpha}x_{\beta}=0$ (or
$x\cdot u=M''_{\alpha\beta}u_{\alpha}u_{\beta}=0$) by
symmetry and commutation of the two quaternary reps of 
$x$ (or $u$)\footnote{Note that in case of symmetric 
$M$ or $M''$, this yields reps of second {\it order} 
surfaces or second {\it class} surfaces.}.

However, misinterpreting $Mx$ as collineation, i.e.~using
$x\longrightarrow x'=Mx$ and interpreting the skew matrix 
rep $M$ as rep of appropriate Lie generators acting on 
space coordinates, gives rise to a transformation theory
of coordinates, or 'infinitesimal transformations' of a 
'4-dim space'. So 'by accident'\footnote{Essentially, this
rises the quest of an adequate identification of $Mx$ with
the adjoint $x^{T}$, or $x^{+}$ which has to be answered.},
because this also fulfils the basic formal procedures of 
Lie theory in this special $4\times 4$ case, and may be 
interpreted in terms of the related different geometry, 
the doors are wide open to discuss SO(4) or SO(3,1) 
transformations as well, or subgroups and subalgebras,
or even complex coverings like SU(2)$\times$SU(2), or 
SU(2)$\times$i~SU(2), etc., see above\footnote{To provide
a {\it phenomenological} association, one can map one 
(classical) line to one su(2) generator in mind, linear
independence of the 3 generators maps to (pairwise) 
skew lines, so su(2)$\oplus$su(2), or su(2)$\oplus$i~su(2)
map to two families of lines, and we may even understand
$i$ by orthogonality, i.e.~in case of the hyperboloid 
the two line families are orthogonal.}.

Now, if we rewrite in this context the 'Lorentz invariant'
in point space, $x_{\mu}x^{\mu}$, which we might also fix to 
the 'light cone' $x_{\mu}x^{\mu}=0$, by means of a regular
null system, we have several possibilities to 'insert' a 
six-dimensional skew matrix $M$ with an appropriate 
normalization by $M^{T}M=\mathbbm{1}$, or $M^{+}M=\mathbbm{1}$,
or even $M^{2}=\mathbbm{1}$. So the related second {\it class}
surface may be written as $u^{T}M^{T}Mu$, and in the general
case, we may also treat additional symmetric transformations
instead of using only skew '$M$'s related to null systems 
if we consider duality and treat the singular tangential 
case of the plane carefully with respect to the surface.

We have discussed in \cite{dahm:MRST6} the interpretation
of the Dirac 'spinors' $u$ and $v$ in terms of 4-dim Hesse
forms of planes, which reflects the idea to consider the
correlations as projective mappings in 3-space, either by
null systems or by polar systems (duality/reciprocity). 
There we can address easily well-known properties of Dirac
theory and the $g-2$ discussion, and we find a consistent
interpretation because we have switched from second 
{\it order} surfaces to second {\it class} surfaces. 
Moreover, we are consistent with Study's discussion of
normal systems and Hamiltonians \cite{study:1905}, and
the four parameters of the '4-momentum' gain immediately
natural explanations if we consider e.g.~tangent planes 
to a sphere in the Hesse form which have distance $\pm p_{0}$
from the center. So also the wave picture of spherical 
waves vs.~plane waves can be related easily. The general
correlation mapping can be decomposed into a 10-dim 
symmetric/polar part $g_{\mu\nu}$ (which can be further
decomposed into 6+4, or 6+(3+1) of the $4\times 4$
symmetric matrix rep) and an antisymmetric 6-dim part 
$\omega_{\mu\nu}$ of the null system. This addresses 
also some of the questions and problems raised by Sexl
and Urbantke in \cite{sexlurbantke:1992}, ch.~4, with 
respect to a generalization of classical mechanics to
'4-momentum vector'-calculus and related approximations.
We discuss few more aspects on the '4-vector' and the rep
by partial differential operators with respect to linear
rep theory on Hilbert spaces in appendix~\ref{app:polarity}.
All cases, however, show that by means of a '4-momentum'
we discuss a plane rep and no 'vector' as part of the usual
line rep (an orientation) of 3-space, and in order to use 
geometrical lines, we have to invoke the 6-dim homogeneous
rep.

\section{Outlook}
\label{sec:outlook}
It is evident that we have discussed only few aspects
of the general framework here, however, we think that
the major idea to use different, but equivalently suited
geometrical reps instead of the point picture has enough
relevance to describe physics in 3-space. As such, we
want to emphasize as a first stage what we've called 
in \cite{dahm:MRST3} a 'programmatic approach'. So in
analogy to the figure given by Ehlers et al. \cite{ehlers:1972},
featuring certain requirements and grouped relations
of 'relativity', we propose to start even earlier by
pursuing two stages, and within the stages even two
parallel tracks each of which should be compared with
respect to each aspect within one track versus the other
track: On the major level in stage~I, to describe 3-space
by PG, we should have one track using points and planes
(to initially respect {\it linear} reciprocity/duality)
in the context of usual physical/mathematical descriptions.
Within a {\it parallel} track in stage~I, one should 
address the same problems while focusing on reps in 
terms of line and Complex geometry. In both tracks, 
one can perform the steps from projective geometry 
and appropriate reps 'down to' affine and Euclidean 
space which allows on one hand to trace the results 
and their origin, on the other hand, we can compare 
both descriptions and relate them analytically to 
compare them to physical descriptions and reps. Both 
tracks can be confined to Klein's 'Erlanger Programm'
so that we can identify objects and transformations, 
i.e.~we can relate to known concepts and identifications,
and we have strong transfer principles at hand to relate
'objects' like e.g.~lines and spheres \cite{lie:1872}, 
\S7, or properties, like e.g.~special tangents and 
curvatures \cite{lie:1872}, \S12.

In a second stage~II, we can go beyond, and that's what
we understand when talking of the 'future of relativity'.
Having located e.g.~Lorentz transformations in Complex
geometry as above, we have the possibility to use other
viewpoints and generalize aspects of the first stage
with respect to e.g.~rep theory of a more general object
or a more general superset of transformations. So it is
this stage~II, where we see advanced concepts of projective
and algebraic geometry, however, for our part we follow
\PLL and Lie with respect to higher dimensions in that
we interpret them as base elements of a different 'new'
geometry, respectively, and not in the nowadays usual
fashion to generalize formalism to arbitrary $n$, mostly
following Grassmann and Riemann. In the example of Lorentz
transformations, by starting the original invariance 
discussion of the quadric $x^{2}+y^{2}+z^{2}-t^{2}$ 
in point space, one can -- of course -- discuss the
standard form of LT.

However, even with point space reps, one has the relation
to Amp{\`{e}}re's, or to Monge's equation like discussed
in (\cite{lie:1896}, see also sec.~\ref{sec:nullsystems}
and figure~\ref{fig:complex}) which yields more background,
or one can also think of exchanging quadratic subquadrics 
by mappings, or discussing the invariance e.g.~of $x^{2}+y^{2}$
with respect to conics, or even to absolute points by
$x^{2}+y^{2}=0$, etc. More interesting considerations
appear, however, if we enter the $P^{5}$ discussion. Having
started in line geometry from an originally 4-dim theory,
\PLL has introduced his fifth inhomogeneous line coordinate
$\eta$ to preserve the grade of the coordinates with respect
to {\it linear} transformations. As such, one has further
introduced senary homogeneous line coordinates e.g.~according
to eq.~(\ref{eq:inhomoglinecoord}) to work with a {\it linear}
transformation theory. One possibility to check for alternatives
accordingly is to work 'in the other direction' by loosening 
the conditions, especially with respect to non-linear mappings
and applications. Another possibility using $P^{5}$ and the 
\PLu-Klein quadric is a more geometrical one: If we restrict
general $P^{5}$ to the \PLu-Klein quadric by a quadratic 
constraint, we can select three special points on the quadric
to represent a family of generating lines, and construct a 
line-based (i.e.~{\it linear}) rep of a quadric in 3-space
with rep counterpart in the point picture. Now whereas we
can understand and represent the transformations of the 
respective quadric in point space (e.g.~LT with respect 
to an 'hyperboloid') by transforming with respect to the
generating linear elements ('lines'), we can in parallel
remember the triangle of points on the \PLu-Klein quadric
associated to the generating lines, and discuss transformations
in $P^{5}$, also with respect to the different 'duality' 
situation. Last not least, we have thus reached the regime
of general transformations acting on $P^{5}$ and of general
non-linear mappings, which opens up for further research
with respect to physical applications.\\

We want to close for now with two physical pictures which 
we've discussed this year \cite{dahm:MRST10} and which yield
some insight into geometry vs.~rep theory, however, here
we just want to present briefly the synthetic part of both
pictures.

The first picture relates to the construction of the
one-sheeted hyperboloid, so we have a real case which can
be visualized. We know that the hyperboloid can be generated
by three lines of a family, or we can use the $P^{5}$ picture
above, and select three Complexe, or a Configuration \cite{plueckerNG:1868},
to extract such lines by a linear constraint. Now, if we 
understand the lines as axes of three individual special 
linear Complexe, we may, of course, use the appropriate
reps as well, i.e.~we now have three 6-dim objects 
$F^{a}_{\mu\nu}$, $a=1, 2, 3$, but completely equivalent 
to the electromagnetic case which we've discussed above. 
Now we know also, that the hyperboloid constructed in this
way, has a symmetry axis with respect to a rotation around
this line, i.e.~an SO(2), or U(1) dependent on the rep. 
So we can introduce a fourth special linear Complex
$F^{0}_{\mu\nu}$, accompanied by some parameters to qualify
either a single hyperboloid (or we can admin even a family 
of hyperboloids with this symmetry axis). Last not least,
we can even include an observer moving linearly, i.e.~we
have an additional line, and to make things easy and preserve
the rotational symmetry of the hyperboloid\footnote{This is
no restriction, because otherwise (like in the theory of the
massive top) we can use Staude's reasoning and select the
axis of the Staude rotation \cite{staude:1894}.}, we assume
that the line of the observer movement intersects the symmetry
axis of the hyperboloid $\sim F^{0}_{\mu\nu}$. So we can 
construct our Langrangean in terms of line coordinates by
the standard scheme of intersecting lines (see \cite{klein:1871},
or \cite{dahm:MRST3}, or \cite{dahm:MRST5}), and thus end
up with what is nowadays called a 'gauge theory'.

The second aspect relates to planar coordinates (or as well
as differential operators) with respect to null systems.
Although being used to discuss transformations or movements
and dynamics in terms of points and (Euclidean) displacement,
we can also take to points of the trajectory e.g.~of a curve
or from a tangential plane. But in this plane, we can associate
a null point with the first point, and with point and plane
fixed, we'll find a null system relating both. Now for the
same Complex, we'll find no second null point in the same
plane\footnote{Using the same Complex and shifting the point
will result in a rotated plane at the other point. Although
(due to the skew rep of the Complex), we can interpret this
in terms of 'skew generator reps' and Lie theory, or in 
special 'derivatives' or some kind of precession, we are in
the mind setting of (singular) tangential transformations
and polar theory, i.e.~using second order surfaces and 
symmetric reps.}. As we've required to connect the points
{\it in the same plane}, we have necessarily to introduce
a second Complex in order to associate its null point to 
the second point. This introduces immediately a second, 
skew matrix to describe the same plane, but as the null
plane of the second point. So a point translation in the
same plane (or its rep by 'derivatives') alters the rep
in terms of classes because we have to consider the two
skew reps for one and the same plane $u$. To preserve the
skewness of the underlying null system(s), this can not be
realized by simple matrix multiplication, but we need the
commutator, so the whole description once more seems to
require Lie algebras and groups like known from covariant
derivatives. In addition, as with each null point, we may
associate a line pencil, we may apply elementary PG (see 
e.g.~Doehlemann \cite{doehlemann:1905}) which yields 
projective relations and a planar conic generated by 
mapping the two pencils. But moreover, we may as well
switch to $P^{5}$ and associate Congruences and the 
related physical reps, or treat it completely with 
$P^{5}$ transformation theory.

\appendix

\section{Coordinate Systems and Velocities}
\label{app:coordinatesystems}
Now, in order to discuss some aspects of line geometry
versus the point picture especially with respect to 
velocities and applications of Lorentz transformations,
we do not want to perform all the -- almost trivial 
and straightforward -- algebra, but just recall few
geometrical aspects and pictures.

\begin{figure}[h]
\includegraphics[width=\columnwidth]{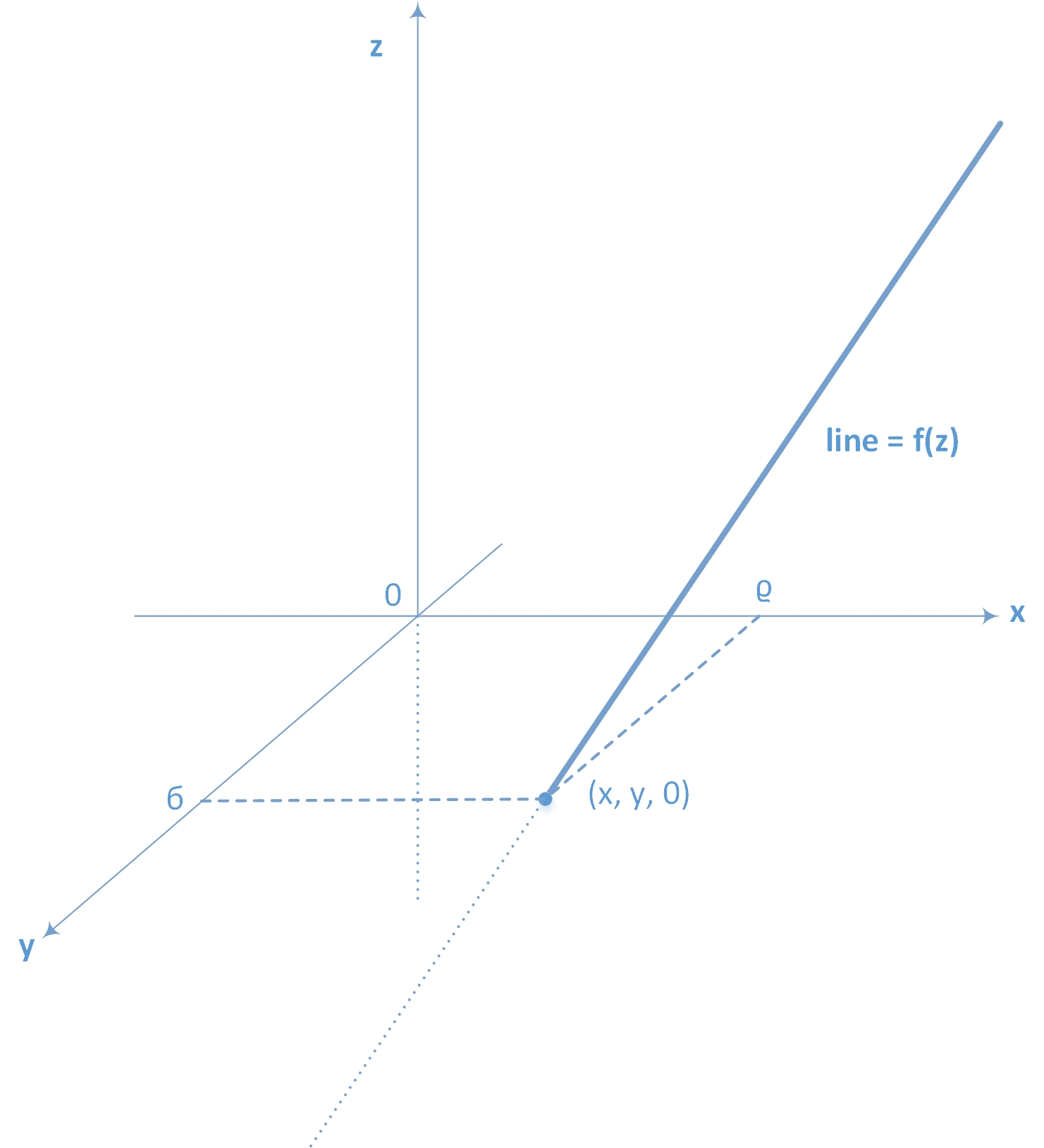}
\caption{Notation used by \PLu, or Lie.}\label{figA1}
\end{figure}

As such, it is noteworthy, that the original analytical
approach to line reps used a 'running coordinate' z in
Euclidean 3-space, and described the two remaining space
coordinates\footnote{Also at that time, people often used
-- like in \PLu's and Lie's case -- a different orientation
of the coordinate system.} $x$ and $y$ by $x=rz+\rho$ and 
$y=sz+\sigma$ \cite{plueckerNG:1868} (see figure~\ref{figA1}).
Of course, this is sufficient to describe the intersection
point $\rho,\sigma$ at $z=0$, and due to the slopes 
$r$ and $s$ (i.e.~the slopes of the planar projections
of the line onto the respective coordinate planes) the
line is determined throughout space and $\forall z$. 
One can also generalize this description versus analysis
by setting $x=f_{1}(z)$ and $y=f_{2}(z)$ where 
$f_{1}(z)=rz+\rho$ and $f_{2}=sz+\sigma$ describe the 
simplest, linear case, and one can think about other 
functional dependencies $f_{i}(z)$, also about switching
to $z\in\mathbbm{C}$ and performing analysis, relating 
geometrical pictures like the stereographic projection,
etc. We have modified this approach using homogeneous
coordinates to discuss spin and spinors, as well as 
aspects of special relativity with respect to the 
exceptional value $\eta=0$ (\cite{dahm:MRST7} and 
\cite{dahm:MRST8}, or sec.~\ref{sec:nullsystems}).

As we want to keep our focus on lines and different reps
only, we do not follow these aspects. Here, it is more
interesting with respect to relativity -- which can be 
understood also to make claims on velocities with respect
to our surrounding space-time -- to introduce a point on 
the line and discuss linear motion. This can be seen from
two sides:
\begin{itemize}
\item[-] Given a point, acting with a force on the point
will put the point into motion. Instead, we can assume 
also existing linear uniform motion and ask for the 
trajectory rep throughout {\it time}, which of course
should yield a line rep in 3-space. So today, the usual 
rep in 3-space will consist of a 3-vector pointing from 
the coordinate origin to a point on the line, and a second
3-vector pointing in the direction of the force or the 
movement\footnote{This  picture shows impressively that
the notion of a 'translation' in time is attached to this
vectorial picture only. In essence, we consider different
possibilities to define a parameter on the line, and there
are other parameters in different geometrical setups suited
as well.}. In order to describe the point on the line, we 
thus have a 3-vector at 'time' $t_{0}$ when we've 'started'
our process, and we can track the point using a parameter 
$t-t_{0}$, or $\Delta t$. So for all 'times' $t\geq t_{0}$,
we 'know' the position of the point, however, we also know 
the integrated line as trajectory. So this picture 'lives'
from the fact that a line is a point set, and that we can
tag the points on the line individually by an appropriate
(linear) parametrization if in addition we accept 
'velocities' of the point on the line.
\item[-] An alternative geometrical view can be given by
understanding a tagging of points on the line e.g.~by its 
intersection with other known lines of a planar pencil, 
or even by known planes of a spatial pencil. So whereas 
on the one hand, we can construct lines as point sets 
(with at least two points), on the other hand the intersection 
of two lines, or a line by a plane, is by geometrical 
definition a point. In this picture, the movement of a
point on a given line can be equally well parametrized
by intersecting the line with a pencil of lines or planes,
and individual points on the line are tagged by tagging 
the respective intersecting lines or planes, and by 
ordering the pencil elements appropriately. Instead of a 
'coordinate system' on the original line e.g.~by a 3-vector
and 'time', we may use as well the coordinate system of 
the pencil and an additional rule (i.e.~a mapping) related
to the metric (if necessary). In the planar case, the line
pencil (which we can e.g.~take from a null system/Complex 
with the null point as center) and the original line should
be coplanar, and the center of the pencil (in our example
the null point) should not be incident with the original
line (or analytically: $x_{i}l_{i}\neq 0$, where $x_{i}$
are ternary homogeneous point coordinates and $l_{i}$ are
planar line coordinates). In other words, in this scenario
the line $l$ shouldn't be a null or Complex line. In case 
of the pencil of planes in 3-space, the axis of the pencil
(i.e.~the intersection of the planes) should be skew to 
the line of movement, so we can apply Congruences and null 
systems, too, or when introducing 'conjugation', besides 
using null systems, we can introduce second order 
surfaces/quadrics, and apply polar theory and duality/reciprocity
(i.e.~symmetric matrix reps and 'anti-commutators'). Now, 
the points on the lines are annotated by an 'angle' within
the respective pencil related to fundamental rays or planes
of the pencil, which we can describe by an anharmonic ratio
(or better with von Staudt: as 'Wurf'). PG then yields 
several mechanisms to adopt this view to analytical and
algebraical reps, typically by 'W{\"{u}}rfe' or anharmonic
ratios, and by deriving metrical properties
%DOEHLEMANN? III. Abschnitt/Transfer Doppelverhaeltnis Punkte
%auf Strahlbueschel\\
thereof, see e.g.~Doehlemann \cite{doehlemann:1905}, 
III.~Abschnitt.
\end{itemize}

Last not least, we may introduce and discuss 'velocities'
to denote the motion of the point on the line by different
'coordinate systems', or simply by different reps. Whereas 
the original static picture $x=f_{1}(z)$ and $y=f_{2}(z)$
fails without further assumptions on a parameter dependence
of $z$ and enhanced notation, we are used to think of a 
linearly moving point (i.e.~a tagged point on a line!) in
vector notation using point coordinates.
\begin{figure}[h]
\includegraphics[width=\columnwidth]{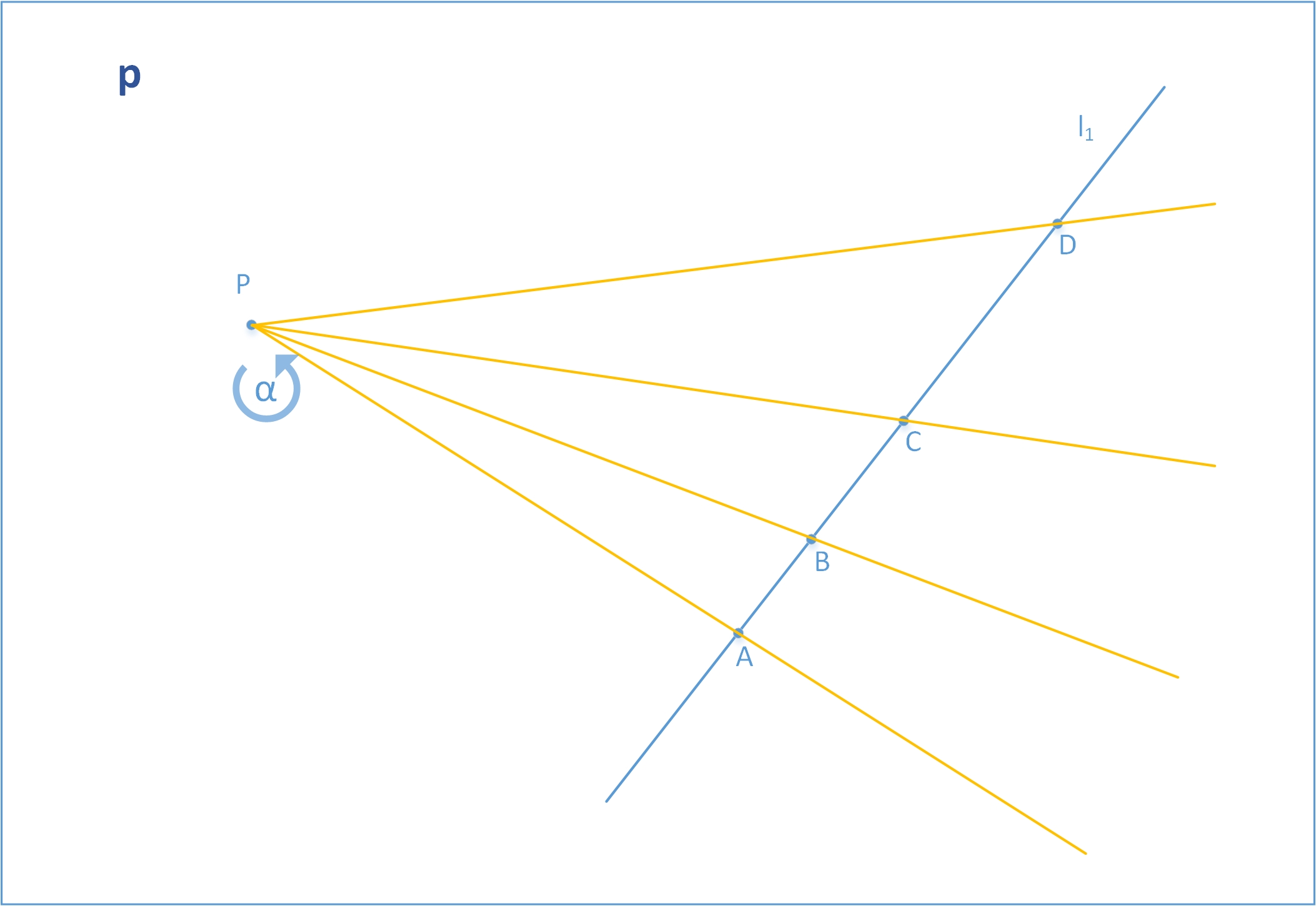}
\caption{Points on a line $l_{1}$ versus a pencil coordinate
$\alpha$.}\label{figA2}
\end{figure}
So the adequate picture (see e.g.~figure~\ref{figA2}) -- 
as soon as we {\it interpret} the direction of the line
by associating it with a velocity vector\footnote{It 
is important to understand that this is an intrinsic 
identification of one of the two vectors of the vector
description above with a physical picture. In the linear
case, this is well expressed and justified by the linear
dependence of the parameter on the line which we interpret,
relative to the velocity, as 'time'. There are, however, 
alternatives, e.g.~by discussing points in the absolute 
plane.}. In terms of the line pencil, however, switching
to the different parametrization, we have {\it to interpret}
the coordinate of the planar line pencil as an angular 
velocity related to a rotation so that 'over time' we 
measure different angles of $\alpha$. If we interpret 
the original motion as continuous, we have introduced,
of course, once more a dependence of a 'time' $t$, 
i.e.~$\alpha(t)$. The same picture holds for null 
systems or Congruences if we use the angle between the
planes\footnote{With respect to covariant derivatives
and skew transformation terms, or effects like Thomas
precession, see \cite{dahm:MRST6}.}.\\

Not that this switch is a problem, however, notation and
the physical pictures change although they are, of course,
related, and the relation between $\alpha$ and the position
measured on the line -- although nonlinear -- can be given 
immediately because we know the respective functional 
dependencies from planar (or spatial) Euclidean geometry
nowadays by heart, and we have even introduced names or 
naming of their groups like 'trigonometric functions', 
'hyperbolic functions', 'HuppelDuppel' functions or 
polynomials, etc.

Using this simple setup, we can already discuss {\it various} 
possibilities describing point velocities on the line $l_{1}$ 
versus pencil parameter $\alpha$:
\begin{itemize}
\item[-] $\delta\alpha$ between pencil lines is constant
per 'time' interval, i.e.~$\omega=\alpha t$. Switching
to metric properties on $l$ and assuming common 'time'
$t$ for line and pencil yields that related distances 
on $l$ between intersections in the line metric are a 
priori NOT equal. If we introduce 'physics' on the line
(i.e.~a velocity $v$ on the line $l$ with respect to the
metric and synchronize clocks at the intersection
points), we find an accelerated motion $\dot{v}\neq 0$.
\item[-] We can change the usual metric to a 'non-Euclidean
metric' on the line in that we require the distances
between the intersection points on the line to be
equidistant. Then in the linear picture, we can
recover linear space and velocity, however, we have
to pay the price with a strange (and for our common
picture unusual) 'time definition'. Our understanding
can be restored if we recall that a projective line
can be understood as a special circle closing through
'infinity' (i.e.~by enhancing the Euclidean picture
by an absolute element). Then, the line behaves as
a circle around the center of the planar pencil.
However, we have to switch to projective instead of
Euclidean geometry only\footnote{We do not want to
discuss the Cayley-Klein mechanism and associating
different polar systems here in general.}.
\item[-] As an alternative, we can absorb this 'strange' 
behaviour by using different 'times' on $l$ and on 
the rotating line of the pencil, so dependent on the 
metric on $l$ whether we require the distance or the
velocity to be constant, this can be absorbed as well
by different non-constant rotations, or even by 
non-linear 'times'.
\end{itemize}
So in general the notion of velocities is generated 
by identifying and tagging a special (sub-)element 
of a lower stage -- the point on the line or the 
line or plane of the respective pencil -- and recalling
projective construction schemes. The 'velocity' is 
introduced by the observation that the tagged elements
evolve 'in time' through the allowed (or a priori 
existent) positions of the superior element, i.e.~the
mapping itself can be understood to change 'in time'.
In essence, we thus obtain different mappings between
the line and the pencil by different identifications
and transfer rules.

\begin{figure}[h]
\includegraphics[width=\columnwidth]{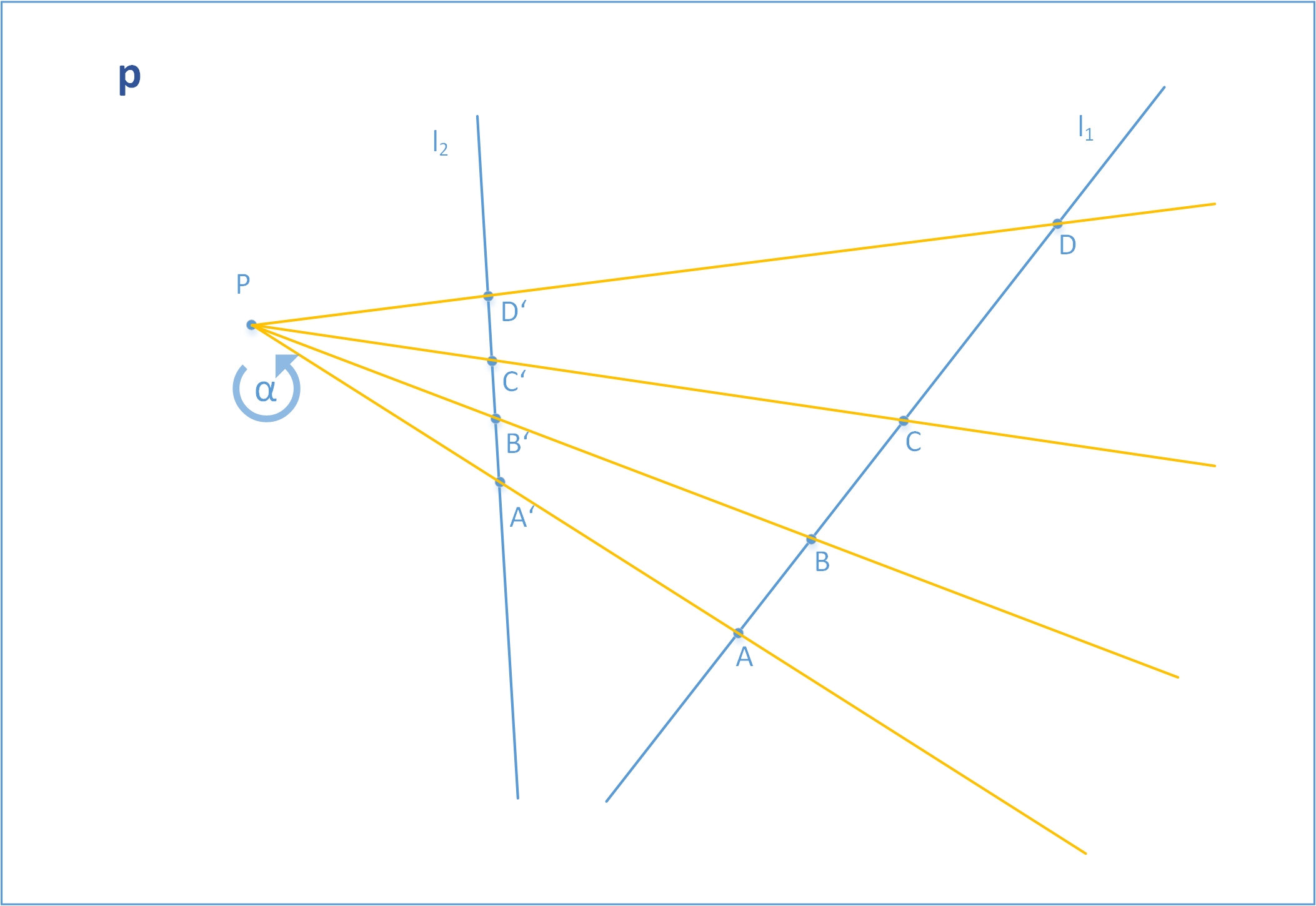}
\caption{Points on two lines $l_{1}$ and  $l_{2}$ with different
velocities versus the same pencil coordinate $\alpha$.}\label{figA3}
\end{figure}

\begin{figure}[h]
\includegraphics[width=\columnwidth]{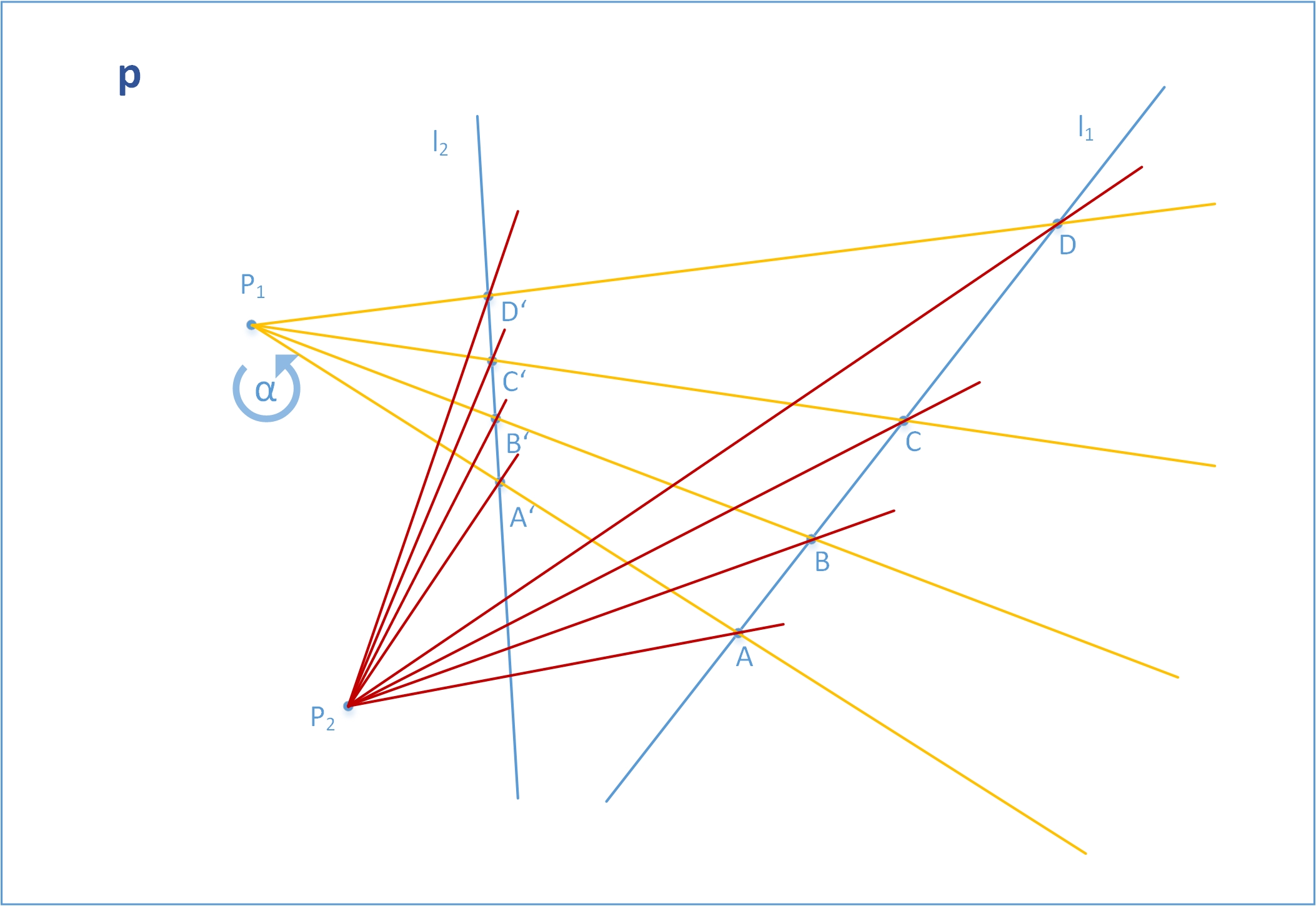}
\caption{The same setup with two 'observers'.}\label{figA4}
\end{figure}

In order to gain some more insight into such constructions,
we may assume a second line $l_{2}$, and if -- by chance 
-- it happens (see figure~\ref{figA3}, or the discussions 
in \cite{dahm:MRST3}, \cite{dahm:MRST4}) that the motion
of a point on $l_{2}$ can be tagged in the same manner
like the point motion on $l_{1}$, we can apply some fundamental
theorems of planar projective geometry.
The points are determined by linear sections of the same
pencil which relates them, i.e.~we can write a well-defined
mapping down by means of the anharmonic ratios of point sets
and lines although the individual 'velocitities' on $l_{1}$
and $l_{2}$ are different. It is obvious, that this situation
changes, if the points at the respective angles $\alpha_{i}$
do not coincide, or if -- like in figure~\ref{figA4} --
there is a second 'observer' at $P_{2}$. Nevertheless, we
can discuss this planar setup by projective, or perspective
mappings \cite{doehlemann:1905}, and use anharmonic ratios
to describe and relate the setup parameters.

Last not least, leaving the planar setup and switching to
lines in space, we have even two scenarios at hand:

The general setup we have discussed so far, are two skew
lines where points are moving (see \cite{dahm:MRST3}, or
\cite{dahm:MRST4}). In this scenario, to connect (or 
relate) points, we cannot in general use only {\it one}
planar pencil, but we have to discuss Congruences, or 
'ray systems', i.e.~we need to consider at least two 
linear Complexe and their common lines. This illustrates
why we can use the description of line geometry given 
above to treat Lorentz transformations and relativity.

The second (and related) scenario can be developed right
from figure~\ref{figA4}. With respect to 'the second observer'
at $P_{2}$, we may switch the lines $l_{1}$ and $l_{2}$
'out of plane' into different spatial positions. But now
we can apply a construction scheme which \PLL has published
\cite{pluecker:1847} as an alternative to Monge's generation
of a one-sheeted hyperboloid. Understanding $P_{2}$ as 
center/vertex of the original (planar) line pencil, the
plane transforms into a one-sheeted hyperboloid while moving
the lines out of the plane $p$. The two lines $l_{1}$ and 
$l_{2}$ (where their intersection points with the pencil 
remain fixed in the sense of a mechanical model/construction
e.g.~by filaments, or fibres), now positioned in space, 
belong to one of the family of generators whereas the 
intersecting lines belong to the second family. \PLL in
\cite{pluecker:1847} discussed explicitly this filament/fibre
model, and he categorizes his {\it second class} construction
versus Monge's construction as well as the r\^{o}le of 
duality. Especially with respect to cones, he emphasizes 
their singular/special character and comments on the 
relevance of duality.

As such, the notion of 'relativity' in general can be associated
to construction schemes in PG, and as such to the respective 
transfer principles. Moreover, it can be subsumed under Klein's
'Erlanger Programm'. And it is obvious, that the parameter (or 
'coordinate') 'time' has a different meaning by its tagging 
function than point reps in terms of pure spatial coordinates.

\section{Projective Generation and Fields}
\label{app:fieldrep}
In addition, we want to recall a construction scheme 
of field reps in terms of circle pencils which is 
automatically introduced by the tetrahedral Complex
and some transfer principles\footnote{We have omitted 
here the relation to W-curves and W-surfaces, because
this is an interesting topic in itself with respect to
W{\"{u}}rfe and metric considerations. If of interest,
please see the introductory remarks in \cite{kleinHG:1926}
\S41. We have postponed also Hesse transfer and rational
curves, although by means of the tetrahedron, we should 
discuss cubics in general as well as twisted cubics. We've
made first use with respect to reps of the quadric and 
spinors throughout \cite{dahm:MRST7}, but it is evident
that with respect to our old work on SU(4) and the 
tetrahedral symmetry of the symmetric reps ${\bf\underline{4}}$,
${\bf\underline{10}}$, and ${\bf\underline{20}}$, we may
investigate appropriate cubics passing through the vertices
and their relations. We've used ${\bf\underline{20}}$ to
treat the $N\Delta$-reps ('3 quarks') in order to complete
vector and axial charge commutators in pion interactions,
and PG seems to yield the appropriate background if we
associate ${\bf\underline{4}}$ with the fundamental tetrahedron
of point reps, ${\bf\underline{4}}^{*}\sim(1,1,1)$ as 
conjugate with plane coordinates, ${\bf\underline{15}}$
with the adjoint rep, and construct reps as usual. Moreover,
the transformations of points on the line using binary reps 
has a couple of interesting applications and identifications
by means of Clebsch \cite{clebsch:1872} and Hesse transfer
\cite{hesse:1866}, besides exchanging vertices of the 
tetrahedron of the tetrahedron as points of a $C_{3}$
when using the 3-dim transformation group \cite{kleinHG:1926}
\S51.}. We have mentioned above the relation of the 
tetrahedral Complex and the fundamental coordinate 
tetrahedron. It is also evident that lines in 3-space
hit the planes of the coordinate tetrahedron in four 
points (the four planes opposite of the vertices of 
the tetrahedron, or within a self-polar fundamental 
coordinate tetrahedron the planes even dual to the 
vertices). Now by recalling that the tetrahedral 
Complex, a quadratic Complex, consists of all lines
with the same anharmonic ratio $\kappa$ of the four
intersection points with respect to the tetrahedron,
we find an associated set of $\infty^{1}$ such ratios
depending on different intersections of lines with the
fundamental coordinate tetrahedron. In other words, 
given a fundamental tetrahedron of coordinates and 
an additional parameter, interpreted as the anharmonic
ratio of the intersection points, one tetrahedral 
Complex is selected. Note that the anharmonic ratio
is invariant under projective transformations of 
projection and intersection!

Now, if in the next step we take a line with its four 
intersection points, and if -- in addition -- we intersect
the line with a plane, i.e.~the line with its four points
are all elements of the plane (i.e.~incident with the plane),
we can apply an interesting geometrical construction scheme 
\cite{doehlemann:1905} in order to attach pencils of circles 
and their orthogonal pencils\footnote{From the physical 
viewpoint, we enter the discussion of potential lines and
fields. Here, we leave the discussion open on whether the
intersection points of the conics should be identified 
with physical charges or masses.}. Because Doehlemann's 
original purpose to investigate point involutions on the 
line by identifying the two carriers of the points is 
comprised, identifying point sets on the remaining common
carrier, the line, yields an additional projective mapping,
and a birational mapping of the individual 'parameters'
\cite{doehlemann:1905} \S\S 17ff. Moreover, we thus can 
apply immediately the original framework of Hesse's transfer 
principle to the point sets, use binary form reps and the 
associated differential equations \cite{clebsch:1872} which
occur in various contexts in physics and especially 'quantum'
theories.

\begin{figure}[h]
\includegraphics[width=\columnwidth]{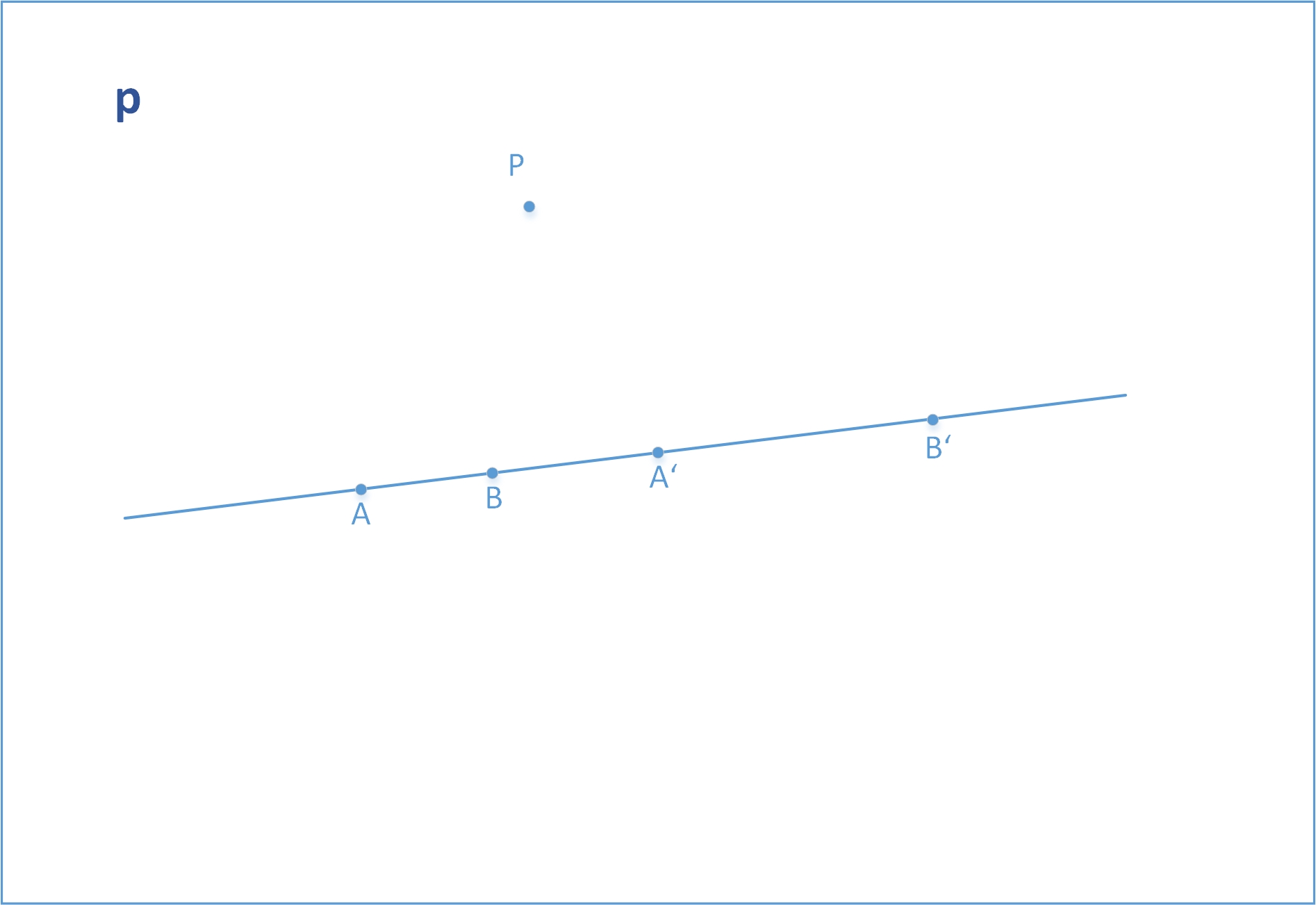}
\caption{Step 1: Define point $P$ and plane $p$ with respect
to the line, the four points $A$, $A'$, $B$, $B'$, and their
anharmonic ratio $\kappa$.}\label{figB1}
\end{figure}

Here, we focus on the construction of circles and pencils of
circles by using the line from the tetrahedral Complex with 
its four intersection points $A$, $B$, $A'$, and $B'$, with 
the fundamental tetrahedron\footnote{To emphasize the point
involution mentioned above, one can map $A\longrightarrow A'$,
and $B\longrightarrow B'$, and follow Doehlemann's outline.}.
In addition, we associate a point $P$ off-line in the plane 
$p$ to avoid analytical problems with degeneracies for the 
moment (see figure~\ref{figB1}). Physically, the point $P$ 
may be associated with a probe or an observer, and it is 
evident that with respect to the fundamental tetrahedron, 
we can either associate quaternary homogeneous space-coordinates
of $P^{3}$, or switch to affine or Euclidean coordinates, 
i.e.~we will find linear (or of order 1) reps of this point.

\begin{figure}[h]
\includegraphics[width=\columnwidth]{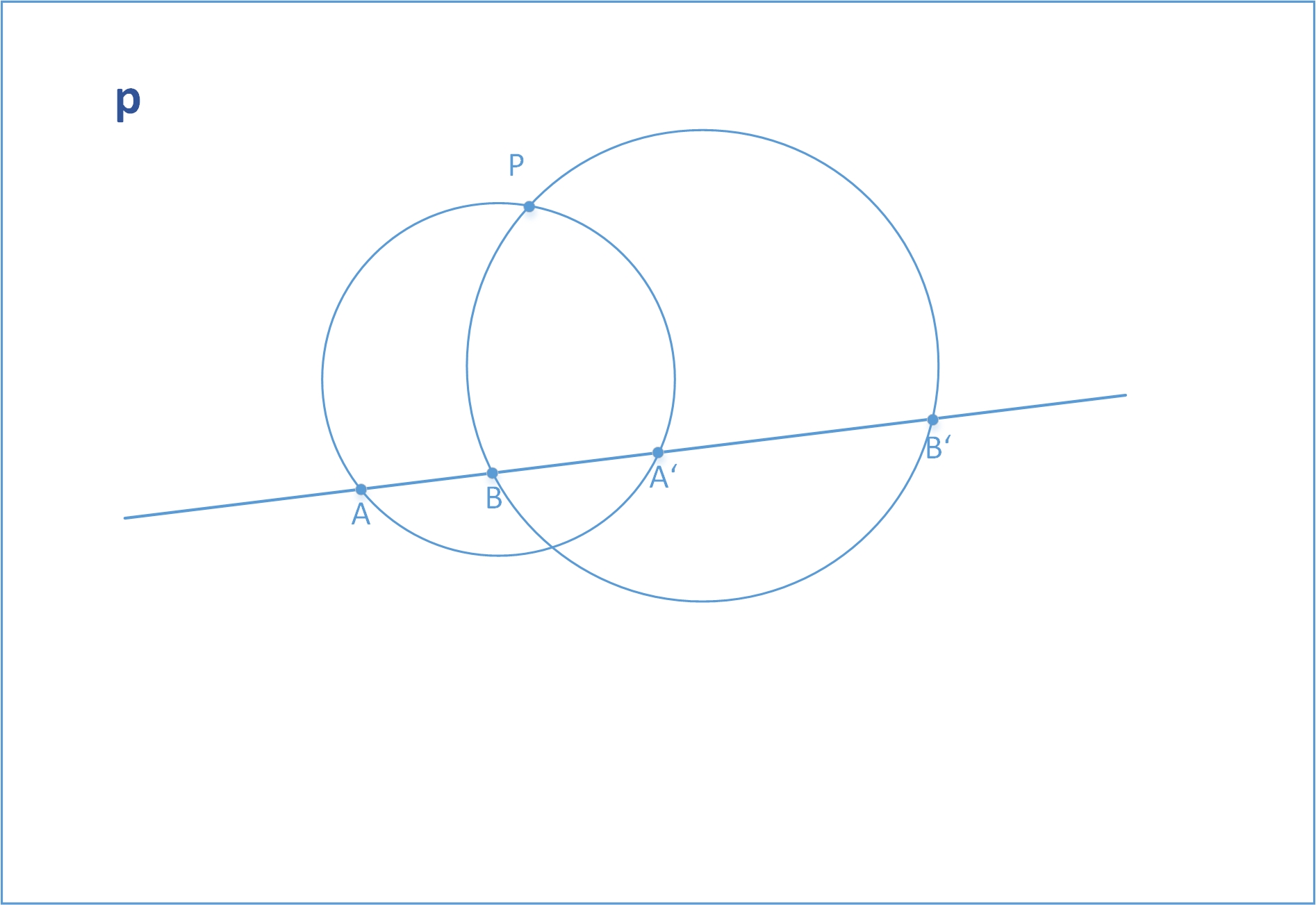}
\caption{Step 2: Connect $P$ with the two point sets ($A$, $A'$)
and ($B$, $B'$) by circles.}\label{figB2}
\end{figure}

\begin{figure}[h]
\includegraphics[width=\columnwidth]{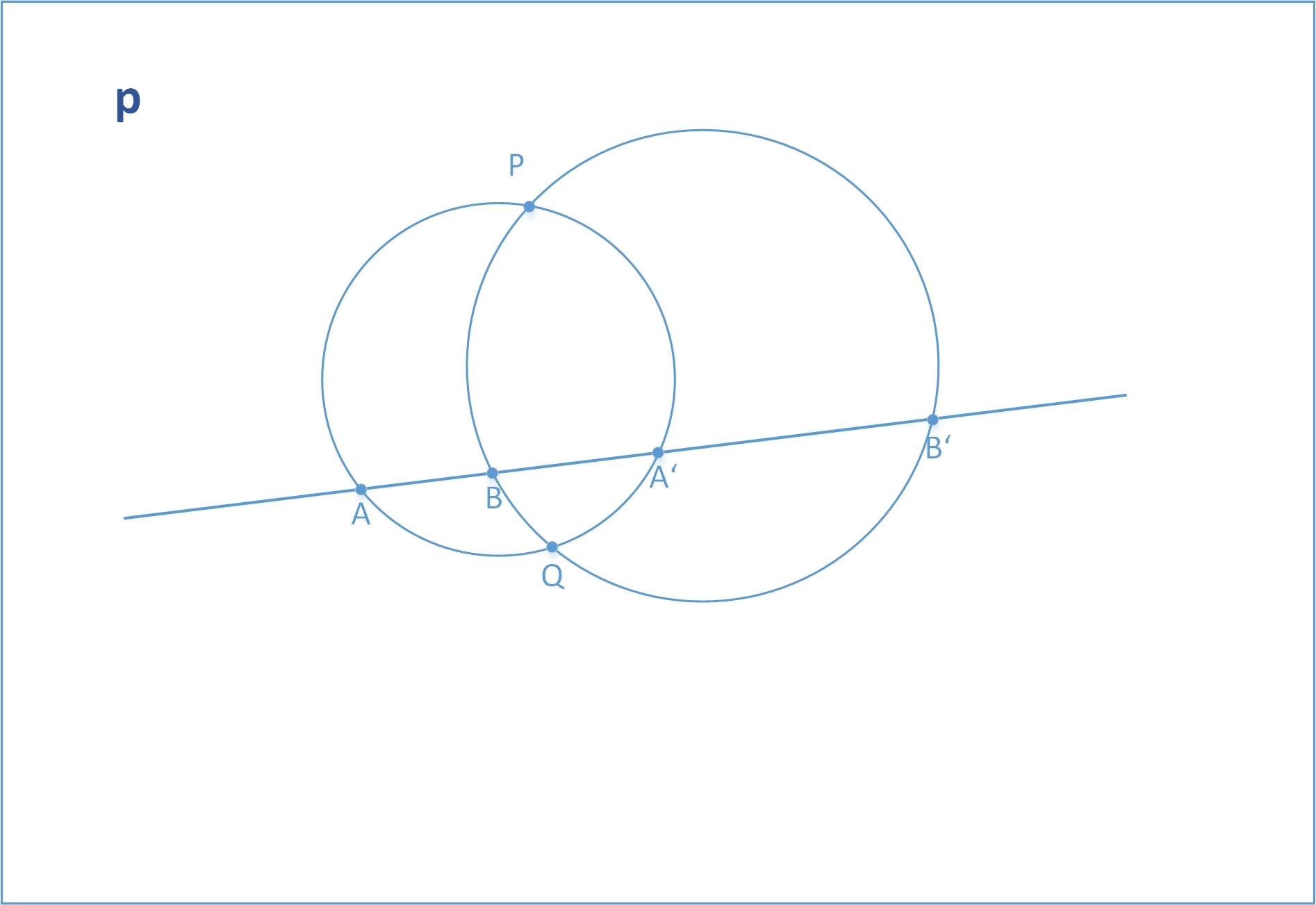}
\caption{Step 3: Identify $Q$ as second intersection 
of the two circles.}\label{figB3}
\end{figure}

\begin{figure}[h]
\includegraphics[width=\columnwidth]{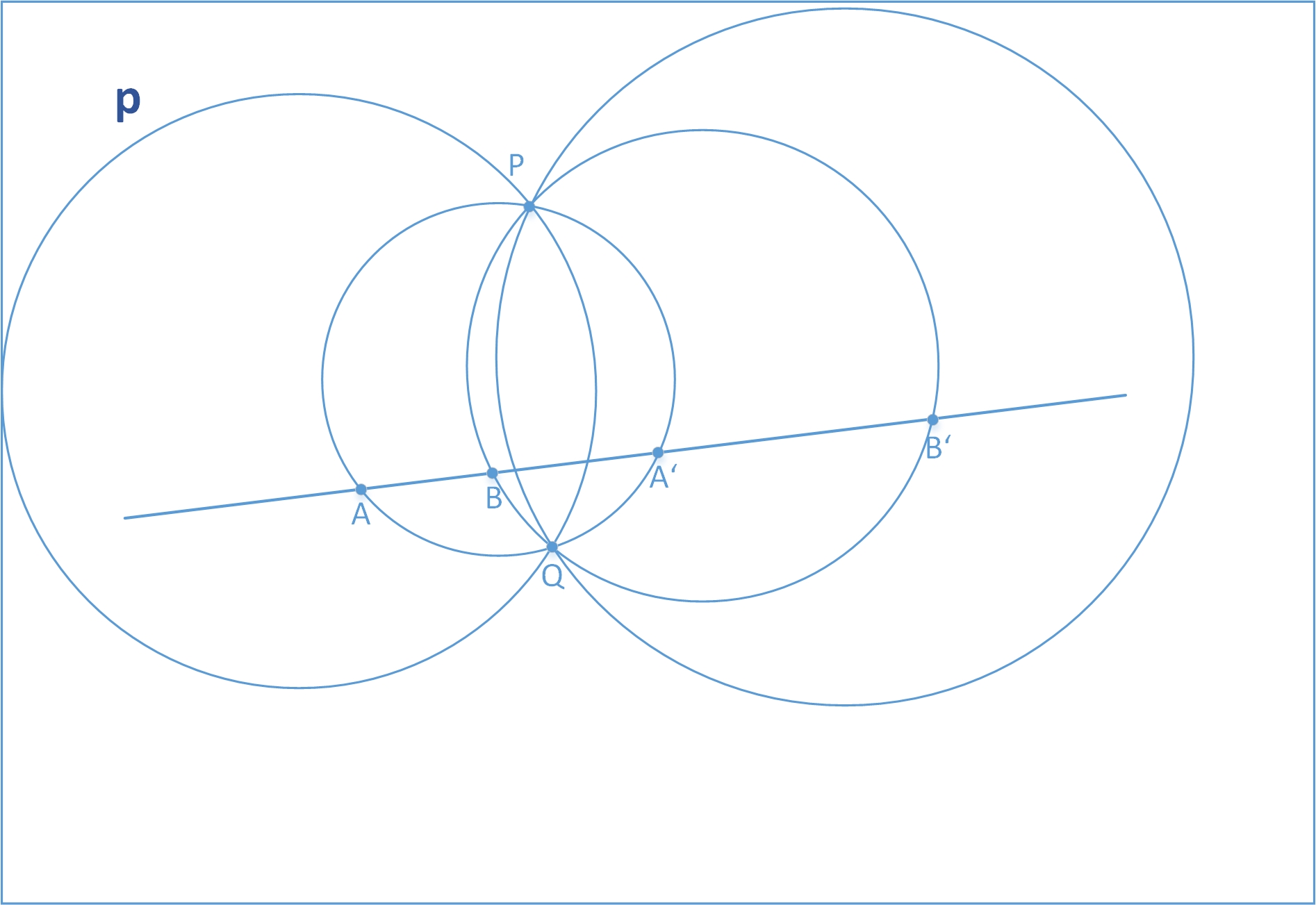}
\caption{Step 4: Visualize the pencils of circles through 
$P$ and $Q$.}\label{figB4}
\end{figure}

However, if in the next step we intersect the point sets
($P$, $A$, $A'$) and ($P$, $B$, $B'$) by circles like in
figure~\ref{figB2}, we can denote the intersection point
of the circles by $Q$ (see figure~\ref{figB3}).
For the upcoming steps, the line with the points  $A$, $B$,
$A'$, and $B'$ is no longer of direct use, but we draw all
circles through the points $P$ and $Q$ like in figure~\ref{figB4}.
In this figure, we have drawn only few lines to recall the
principle but it is evident that one can find planar circles
connecting each planar point with $P$ and $Q$, i.e.~we may
attach (at least) one quadric rep to each planar point\footnote{Note 
the relation to second order partial differential equations.}.
So in \PLu's sense, we may as well switch to a geometry 
using the circle as base element of this geometry\footnote{With
respect to spheres in 3-space, we may use symmetry arguments 
for now, however, the analytic rep has to base on Lie transfer.}.

\begin{figure}[h]
\includegraphics[width=\columnwidth]{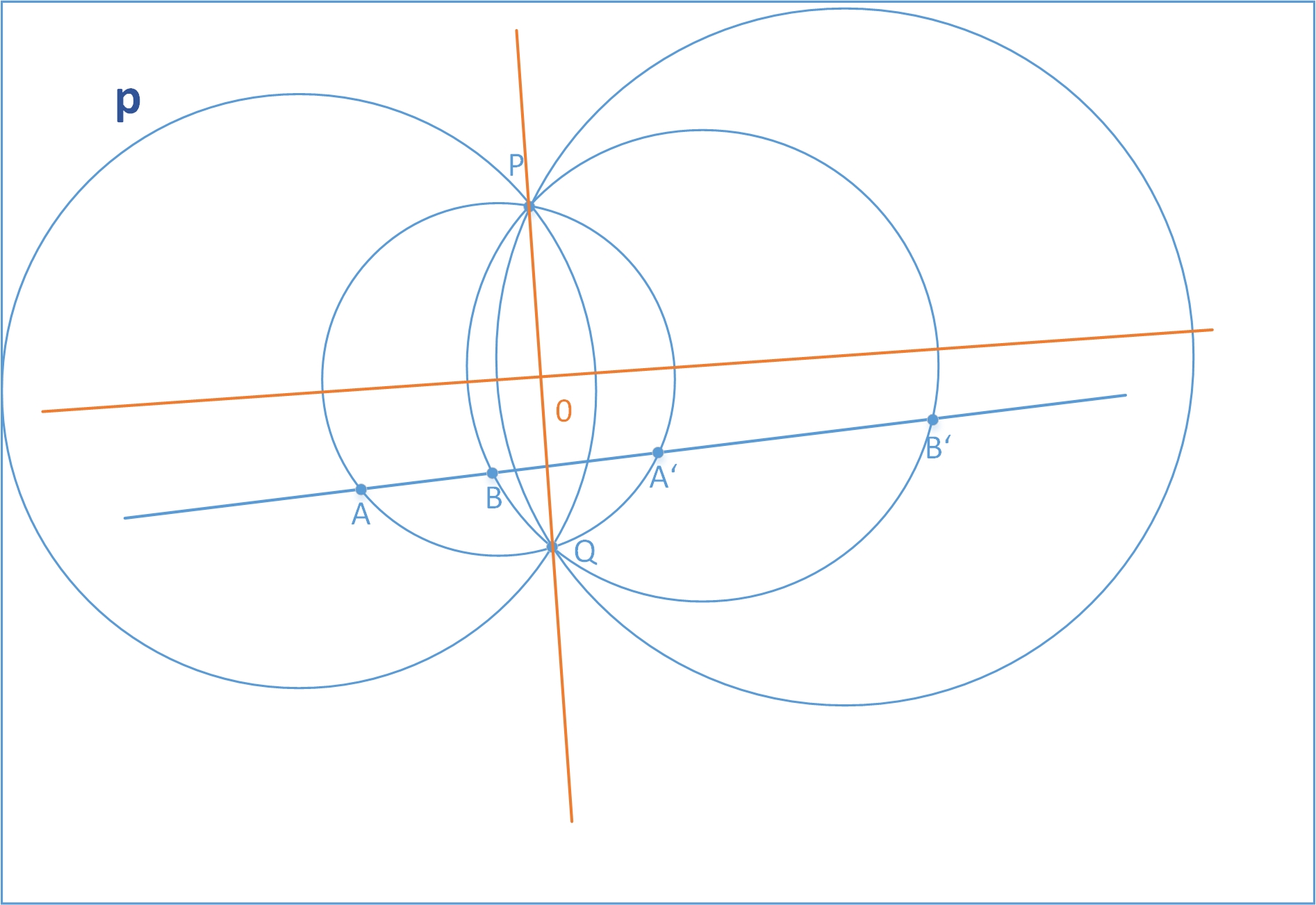}
\caption{Step 5: Draw line through $P$ and $Q$, and an orthogonal
line at $0$.}\label{figB5}
\end{figure}

\begin{figure}[h]
\includegraphics[width=\columnwidth]{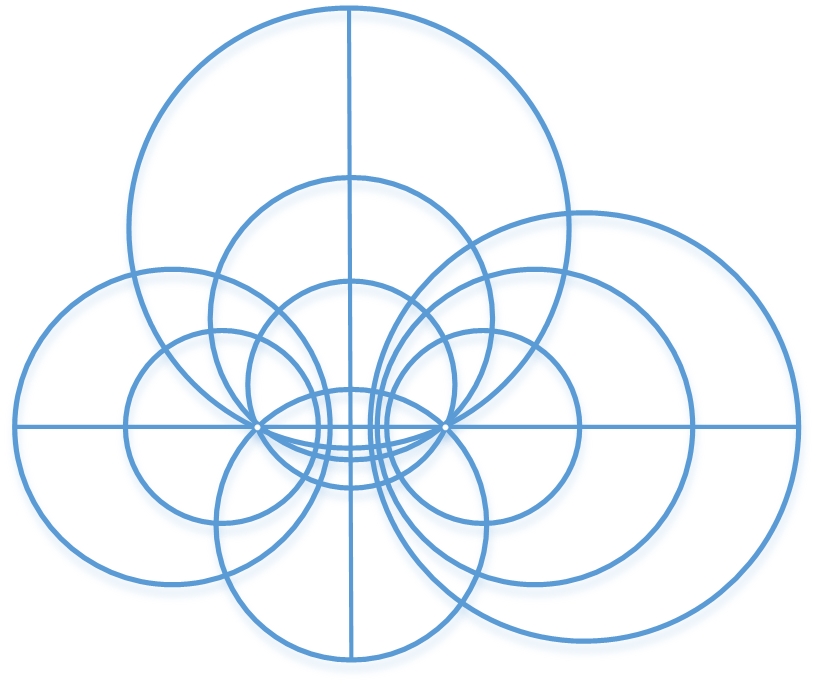}
\caption{Comparison to FIG.~\ref{figB5}: Pencils of original 
and orthogonal circles given by Klein \cite{kleinHG:1926}, p.~42.}\label{figBComparison}
\end{figure}

In a last step to get closer to usual descriptions\footnote{In
order to remove the circular restriction, one can even switch 
to general conics and 'deform' the description which complicates
the setup considerably.}, it is helpful to change the planar 
coordinate system like in figure~\ref{figB5} where the first 
axis is a line incident with $P$ and $Q$, and we may choose 
the origin by an orthogonal axis half between $P$ and $Q$, 
i.e.~in addition we can fix the orientation by the anharmonic
ratio ($Q$,$0$,$P$,$\infty$)=-1.
As a comparison, we want to refer to the picture in figure~\ref{figBComparison}
as given in \cite{kleinHG:1926} when presenting pencils of 
circles and of orthogonal circles\footnote{We have omitted the
orthogonal circles here, but like in the case of line generators
of quadrics, one can of course consider the orthogonal system, 
too, which then usually appears as 'rotated' by 90 degrees, or
$i$.}.

Now if we recall $P$ representing originally a probe or an
observer, we have introduced a description where the 'forces'
(or field line orientations') point towards $Q$, but we can 
introduce in addition orthogonal sets of circles or conics 
to introduce the notion of 'potentials'. As most of this is
classical analytical geometry, we leave it to the reader to
play with conics and partial differential operators using 
incidence relations (due to tangential conditions) in the 
linear and the quadric case. Moreover, it is interesting to
play with the 'two infinities' of the rectangular axes and
relate them to the scenario where the plane is the projection
plane of a stereographic projection and both axes meet in 
a second point. Vice versa, one can think to replace $P$ 
and $Q$ by $\pm i$, and perform some algebra also with their
anharmonic ratio versus $0$, and $\infty$.

\section{Complex Numbers and Geometry}
\label{app:complexnumbers}
Although it sounds silly to discuss nowadays complex numbers
because everybody is used to work with them from scratch,
it is worth to briefly recall their use by means of a simple
geometrical example\footnote{Which -- by the way -- enforced
the introduction of spinors and their special properties\ldots}.

As example, one can write down the intersection of a circle
and a line in planar coordinates where it doesn't matter
whether we use two Euclidean coordinates or three homogeneous
planar coordinates. We obtain -- of course -- three cases to
be distinguished:
\begin{enumerate}
	\item[a)] the line hits the circle in two real points
	('secant'),
	\item[b)] the line is a tangent to the circle (one real
	double-point), or
	\item[c)] the line misses the circle, i.e.~there is no 
	intersection in the Euclidean plane.
\end{enumerate}
The analytic calculation to find the two intersection points
by finding the two roots is well-known.

Now it is noteworthy, that exactly the physically not so
interesting case where there is no intersection in the
Euclidean plane requires to complex roots to treat the
analytical rep in terms of coordinates. In other words, 
the complex numbers serve as an algebraic or analytic 
unification tool by introducing the symbol $i$ and 
appropriate rules to work with this symbol. In the planar
case, this extension serves to treat two 'conjugate' 
points on the absolute line, whereas in 3-space, we
have to introduce additional symbols to treat a line or a
plane vs.~a second order surface. Based on Lie transfer
and Study's work in line geometry, we have discussed the 
construction of spinors in \cite{dahm:MRST7} by means of
homogeneous coordinates while respecting the absolute
elements and treating 'null' elements. So it is justified
to understand complex numbers (and quaternions, and in 
general hypercomplex numbers as well) as symbols to serve
for algebraic/analytic completion when representing geometry
and associating absolute elements in homogeneous coordinates
to the respective geometries.

However, being physicist, when being asked to vote for
the interesting case(s), I'd vote for a) and b) where 
the line meets the circle at least in one point and
where the probability that 'something interesting 
happens' seems to be much higher than in the 
non-intersecting case c). It seems justified to regard
and understand the case c) as being necessary to 
complete the algebraic/analytic description formally,
and to see whether the mathematical tool sets work 
appropriately or not -- the more, as we can apply
a lot of more unphysical ideas to this picture like
shrinking the circle or even the line to a point, 
etc.~which introduces ambiguities or even singularities.

\section{Einstein and 'Six-Vectors'}
\label{app:einstein}
As brief and quick access to Einstein's publications in 
order to guarantee a certain overview of the original 
concepts of special and general relativity, we've used
publications of his talks in the transactions of the 
Prussian Academy of Science \cite{einstein:1914-32} 
during his time in Berlin as well as his publications
in 'Annalen der Physik' which we'll discuss briefly 
with focus on the use of the 'six-vector' in order to
attach Complex geometry to special and general relativity
as discussed nowadays by 4-vector calculus and differential
geometry.

\subsection{Prussian Academy of Science}
Here, we've found three publications which we want to 
mention for later geometrical use.

In \cite{einstein:1914} Einstein mentions in a very small
paragraph briefly 'six-vectors' as an alternative notation
in the context of $2^{nd}$-rank antisymmetric tensors 
(p.~1037/1038) without physical context. Only after having
introduced the 'antisymmetric fundamental tensor of Ricci
and Levi-Civita' yielding\footnote{Here, Einstein uses 
$\delta$ instead of the nowadays common notation $\epsilon$
for the totally antisymmetric tensor with 4 indices.}
\begin{equation}
\begin{array}{lcl}
G_{iklm} & = & \sqrt{g}\,\delta_{iklm},\\
G^{iklm} & = & \frac{1}{\sqrt{g}}\,\delta_{iklm},\quad\mathrm{and}\\
G^{lm}_{ik}
& = & \sum_{\alpha\beta}\sqrt{g}\,\delta_{ik\alpha\beta}\,g^{\alpha l}g^{\beta m}\\
& = & \sum_{\alpha\beta}\frac{1}{\sqrt{g}}\,\delta_{lm\alpha\beta}\,g_{\alpha i}g_{\beta k}
\end{array}
\end{equation}
in eqns.~(19), (21a), and (22), Einstein discusses the 
case of dual six-vectors in eq.~(24),
\begin{equation}
(F^{\mu\nu})^{*}=\frac{1}{2}\sum_{\alpha\beta}G^{\mu\nu}_{\alpha\beta}F^{\alpha\beta}
\end{equation}
with respect to the electromagnetic field rep. However, 
the whole discussion seems to focus mainly on tensor 
algebra. The major part of discussing $2^{nd}$-rank 
tensors throughout this contribution is related to 
symmetric tensors which yield 10 components and which
in \S 9, p.~1055, is related as $T_{\sigma\nu}$ to the
energy tensor. Because we do not want to discuss details
here due to the focus on six-vectors (and Complexe or 
null systems), we just want to emphasize the connection
of the symmetric part to polar systems and second order
surfaces for later use.\\

In \cite{einstein:1916}, Einstein starts from Minkowski's
identification of the six electromagnetic field components
as six-vector and discusses Maxwell's equations. Einstein
defines a 'new' six-vector $\mathcal{F}^{\mu\nu}=\sqrt{-g}
\sum_{\alpha\beta}g^{\mu\alpha}g^{\nu\beta}F_{\alpha\beta}$
which in case of special relativity reduces to the usual 
field components $\vec{h}$ and $-\vec{e}$ in order to
incorporate electromagnetism into general relativity by
means of the tensor $\mathcal{T}^{\nu}_{\sigma}$,
\begin{equation}
\mathcal{T}^{\nu}_{\sigma}=\sum_{\alpha\beta}
\left(
-\mathcal{F}^{\nu\alpha}F_{\sigma\alpha}
+\frac{1}{4}\mathcal{F}^{\alpha\beta}F_{\alpha\beta}\delta^{\nu}_{\sigma}\,.
\right)
\end{equation}
From our point of view, it is important to note that
throughout his calculations, he uses only the field
tensor $F^{\alpha\beta}$, even in his generalized form
$\mathcal{F}^{\alpha\beta}$ and in derived equations,
i.e.~{\it one} six-vector (or Complex).
% - Minkowski-Sermon
% - wichtig: nur EIN Sechser-Vektor! singulär linear -> Feld,
%   keine Materie

In \cite{einstein:1932b}, Einstein and Mayer decompose in 
\S 2 the antisymmetric $4\times 4$-tensor rep of $R^{4}$.
By means of 'the totally antisymmetric tensor $\eta^{iklm}$'
they exercise an algebra which basically represents working
with pairs of four-indices, and which in line geometry is
related to switching ray versus axis coordinates. $\eta$
can be understood as a rep of the $4\times 4$-determinant
when working with point or plane coordinates. As such, 
one obtains a rep of line coordinates and in consequence
naturally Lorentz transformations (see the discussion 
above or \cite{dahm:MRST8}).

% 6-dim reps -> Klein-Koordinaten, fast Liniengeomtrie

% noch erwähnen: 5-dim Grütze, P^4 (Klein, Weyl), Kaluza-Klein
% refs SB1931 und SB 1932a mit Mayer

\subsection{Annalen der Physik}
Within his publications throughout this journal, we've 
found only two places which seem to be related to Complexe,
or to six-vectors. However, in both cases this seems to 
be a naming convention, and Einstein has made no real use
of this geometry or related formalisms.

To be complete, we've found in his paper of 1905, titled
'Zur Elektrodynamik bewegter K{\"{o}}rper', \S8, p.~913f,
five times the expression 'Lichtkomplex', i.e.~a complex
of light, however, without explanation or definition. So 
there is no evidence that Einstein referenced a line 
Complex, or whether he used some contemporary notion of 
describing light in a certain manner.

The other occurence, we've found in his paper on general
relativity of 1916, titled 'Die Grundlagen der allgemeinen
Relativit{\"{a}}tstheorie'. There in \S20, he discusses 
briefly the Maxwell equations of the electromagnetic field
based on Minkowski's rep theory, as he mentions.

\section{Remarks on Representation Theory and Lie Symmetries}
\label{app:polarity}
\subsection{Remarks}
Throughout the text and the other appendices, we've switched 
'on demand' between classical and 'quantum' reps of objects
and operators. We have prepared in \cite{dahm:MRST6} a more
detailed account, however, as we've used various aspects so
far, we want to provide a brief overview(or summary) at least
on the aspects in 3-space. From the viewpoint of classical
geometry, the central aspects we've used are points, lines 
and planes, and we've discussed mappings where (besides a 
lot of other 'transfer principles') duality is a central
aspect. As such, within classical geometry of 3-space, we
have to quaternary reps $x$ and $u$ with respect to points
and planes, and besides mapping points to points, and planes
to planes, we know correlations mapping points to planes, 
and vice versa. Lines map under duality to lines, we may 
use senary reps as above (or nowadays 'Grassmannians' 
Gr(2,4)), so on the one hand, we have only a single line
formalism, on the other hand, we automatically have to 
take care and work with quadratic forms a priori. As a
sufficient framework, we can invoke classical projective
and advanced geometry like discussed e.g.~in \cite{kleinHG:1926}
or literature cited there.\\

The classical approach, however, doesn't provide a rep for
what people started to discuss as 'quantum' theories in 
terms of differential operators on Hilbert spaces, i.e.~using
function reps. First of all, we want to attach linear rep 
theory (or 'plane waves'), and in order to associate physical
pictures, we need to find appropriate reps on function spaces,
too. With respect to relativity, a suitable starting point 
in classical geometry of 3-space is the plane rep either in 
general form $Ax+By+Cz-D=0$ or in terms of Hesse's implicit 
representation, i.e.~if we define the plane as usual by all 
$\vec{x}$ which are normal to a given {\it normal} vector 
$\vec{n}$ in Euclidean 3-space, the plane $\textfrak{p}$ 
reads as $\textfrak{p}=\vec{n}\vec{x}-d$, where $d$ denotes
an {\it oriented distance} of plane and origin. Now, in this
picture we can replace the Euclidean coordinates, of course,
by the known fractions of homogeneous coordinates above, so
this plane can be rewritten either as 
$\textfrak{p}=Ax_{1}+Bx_{2}+Cx_{3}-Dx_{0}=0$ for the general
form, or by $\textfrak{p}=n_{1}x_{1}+n_{2}x_{2}+n_{3}x_{3}-dx_{0}$
in Hesse form, which both formally resemble a 'Lorentz invariant'
product $\textfrak{p}=p_{\mu}x^{\mu}$. Now the '4-momentum' 
$p_{\mu}$ comprises the components $(p_{0}=d, n_{1}, n_{2}, n_{3})$,
however, it is a quaternary plane rep which in classical
geometry we can treat also as $u_{\alpha}$, and as such the
$0$-component $p_{0}$ representing the oriented distance
changes sign appropriately. As an example, in the case of 
a sphere with radius $r$, we have two parallel tangential
planes at distance $\pm r$ at antipodal points of the sphere,
and we have to take care only whether we discuss spherical
or elliptical geometry, but we do not have to take care of
'anti-particles' or 'time-reversal'. Note, however, that 
we have attached a special interpretation to the coordinates,
and especially the $0$-coordinate denotes a {\it metrical}
property. So the calculus itself has to be attached to 
normals and normal congruences which relates this rep theory
on the one hand to Study's summary paper \cite{study:1905},
on the other hand we are back again on the original ground 
of maps of normal Complexe (see \cite{lie:1896} and references
therein).

If we specialize our coordinatization further with respect
to a planar point in 3-space, in order to define the metric
we have mentioned the usual procedure to use logarithms of
anharmonic ratios of points. So using the above plane rep
as argument of an exponential, and adding a phase $i$ to
compensate the phase in the metric, yields reps according 
to $\exp \pm i p\cdot x$. As check, if we take the tangential
point as origin, the vector $\vec{x}$ of planar points in 
3-space is always perpendicular to the normal $\vec{p}$, 
so we the logarithm of the metric mapping yields the
remaining $0$-component $p_{0}$ as metric distance. Acting
with $\partial_{\alpha}$ on this rep yields $p_{\alpha}$,
and we may use standards rules like known from QFT to
multiply exponentials. As such, we can see $i\partial_{\alpha}$
as a phenomenological (however, restricted) replacement
of quaternary plane reps on function spaces, especially
since the plane exponential remains but produces a linear
quaternary rep $i\partial_{\alpha}\exp(-i p\cdot x) \sim
p_{\alpha}\exp(-i p\cdot x)$, and moreover, it yields 
according to its later processing e.g.~in QFT the necessary
information on 'momentum conservation' within this rep.
Formally, we can remember the geometrical relation of 
order and class, so we understand this 'plane wave'
rep as class view, however, one has to be careful with 
the physical interpretation of the components. So the
axiomatic approach by first quantization can be understood
as well by transferring classical geometry to function
spaces, and integer numbers or multiples of $2\pi$ emerge
automatically for certain geometric conditions of Complexe
(or 'screws') (see e.g.~\cite{zindler1:1902}, ch.~1) or
cones (second class surfaces), especially if the vertex
is a point in the absolute plane, and we obtain a cylinder.
The interpretation of 'currents' or 'lines' by means of
such '4-momenta', however, is coincident with the senary
line definition only if we respect Lie's picture discussed 
in sec.~\ref{sec:nullsystems}, and visualized by 
figure~\ref{fig:complex}, i.e.~if we understand the normal
as the direction of the senary line rep, or in terms of
line elements to follow Lie's presentation of differential
geometry in 3-space.\\

Here, we've discussed this phenomenology with respect to
identifications and linear reps used in various physical
models and discussions. So far, we have reps for points
and planes, however, with respect to Lie's and Study's
refs cited above with respect to the normal systems of
surfaces, we want to find also appropriate reps of lines
and second order surfaces to gain control and to reproduce
the construction schemes of projective geometry. So besides
representing the line family sets of second order surfaces,
it is interesting to have possibilities to investigate 
or even gain control on non-local behaviour. Wigner's 
'relativistic angular momentum operator $M_{\mu\nu}$' 
in \cite{wigner:1962} shows dependence only with respect
to one point coordinate rep, and Gilmore's generator 
definitions (see e.g.~\cite{gilmore:1974}, p.~450/451,
eqns.~(1.28s) or (1.28h)), given for the case SO($n$,1)
according to $X_{ij}=x^{i}\partial_{j}-x^{j}\partial_{i}$,
$X_{i,n+1}=x^{i}\partial_{n+1}+x^{n+1}\partial_{i}$, 
depend (like elsewhere when performing e.g.~angular 
momentum algebra) only from one point rep $x$. However,
right from the definitions above we see that the action
of $\partial_{n+1}$ in this case produces an additional
sign, and if we recall that we act on second order 
surfaces, we obviously perform the algebra of a special
or even singular rep\footnote{It is not that we want 
to abolish standard quantum mechanics exercises or 
standard reps of Lie generators on function spaces,
because due to what is known from force systems or 
e.g.~from Staude with respect to the Staude rotation
of the massive top \cite{staude:1894} which can be 
represented by one special axis, we can find effective
reps attached to one point of space.}.

As such, we've defined for our use with respect to 
second order surfaces and (senary) line reps an a 
priori 'non-local' operator
\begin{equation}
L_{\alpha\beta}=\frac{1}{2}\left(
x_{\alpha}\frac{\partial}{{\partial y}_{\beta}}
-x_{\beta}\frac{\partial}{{\partial y}_{\alpha}}
\right)
\end{equation}
which we can justify generally for two points $x$ 
and $y$ of a quadric $S(x)=a_{\alpha\beta}x_{\alpha}x_{\beta}$
because acting with this operator (on a canonical/'diagonal'
form of S) yields 
$L_{\alpha\beta}S=
x_{\alpha}y_{\beta}
-x_{\beta}y_{\alpha}$
which can be identified as $p_{\alpha\beta}$. So we 
have gained operator reps which by acting on second
{\it order} surfaces behave as line coordinate reps
$p_{\alpha\beta}$.

The typical differential discussion of taking two 
'close' points $x$ and $y$ on the quadric $S$ has 
to be modified due to the quaternary {\it homogeneous}
coordinates used above, however, we can write at 
least symbolically 
$M_{\alpha\beta}=\lim\limits_{y\longrightarrow x} L_{\alpha\beta}$
and relate to Wigner's 'relativistic angular momentum' 
and to usual angular momentum algebra. As such, we 
discuss details in \cite{dahm:MRST6} but right now,
one can try to understand Schr{\"{o}}dinger equations
or Klein-Gordon equations by the class view and 
planar reps in 3-space.

If -- as an example -- we calculate the commutator
$[L_{\alpha\beta}, L_{\gamma\delta}]$ of the 'line
operators', we obtain
\begin{equation}
\label{eq:commMegaLine1}
\begin{array}{ll}
4 \left[L_{\alpha\beta}, L_{\gamma\delta}\right]\,=
& \hphantom{+}\delta_{\beta\delta}\left(
x_{\gamma}\frac{\partial}{{\partial x}_{\alpha}}
-x_{\alpha}\frac{\partial}{{\partial x}_{\gamma}}
\right) \\
& +\delta_{\alpha\gamma}\left(
y_{\delta}\frac{\partial}{{\partial y}_{\beta}}
-y_{\beta}\frac{\partial}{{\partial y}_{\delta}}
\right)\,.
\end{array}
\end{equation}
For $y\longrightarrow x$, this yields 
\begin{equation}
\label{eq:commMegaLine2}
\begin{array}{rl}
4 \left[L_{\alpha\beta}, L_{\gamma\delta}\right]\,=
& \left(\delta_{\beta\delta}\,\delta_{\gamma\rho}\delta_{\alpha\sigma}
+\delta_{\alpha\gamma}\,\delta_{\delta\rho}\delta_{\beta\sigma}
\right)\cdot\\
& \cdot\left(x_{\rho}\frac{\partial}{{\partial x}_{\sigma}}
-x_{\sigma}\frac{\partial}{{\partial x}_{\rho}}\right)\\
\,= & 2\left(\delta_{\beta\delta}\,\delta_{\gamma\rho}\delta_{\alpha\sigma}
+\delta_{\alpha\gamma}\,\delta_{\delta\rho}\delta_{\beta\sigma}
\right)L_{\rho\sigma}\,,
\end{array}
\end{equation}
and finally
\begin{equation}
\left[L_{\alpha\beta}, L_{\gamma\delta}\right]\,=
\,\frac{1}{2}
\left(\delta_{\beta\delta}\,\delta_{\gamma\rho}\delta_{\alpha\sigma}
+\delta_{\alpha\gamma}\,\delta_{\delta\rho}\delta_{\beta\sigma}\right)
L_{\rho\sigma}\,.
\end{equation}
The remaining operators on the rhs show the form of 
standard Lie generators of compact groups (see also
Gilmore's operator definitions cited above), and due
to the quaternary indices, we are concerned with so(4).
However, the intricate factor in front of the operators
depends on Kronecker symbols of indices and restricts 
the algebra, and we have to resolve the result and its
background by line geometry. This, however, is ongoing,
and so far we do not have appropriate results from 
'classical' line geometry to compare to.

\end{document}